\documentclass[a4paper,11pt]{article}
\usepackage{jcappub} 
\usepackage{lineno}

\usepackage{dcolumn}
\usepackage{verbatim}

\usepackage{breakurl}

\usepackage{lmodern}

\pdfoutput=1
\usepackage{graphicx}
\usepackage{bm}
\usepackage{amsmath}
\usepackage{amsfonts}
\usepackage{amssymb}
\usepackage{multirow}
\usepackage[capitalise]{cleveref}
\usepackage{acro2}
\usepackage[utf8]{inputenc}

\usepackage{tikz}
\usepackage{pgfplots}

\crefname{figure}{Fig.}{Figs.}
\def\({\left(}
\def\){\right)}
\def\[{\left[}
\def\]{\right]}

\newcommand{\Hz}{{\rm Hz}}
\newcommand{\mHz}{{\rm mHz}}

\newcommand{\be}{{\begin{eqnarray}}}
\newcommand{\ee}{{\end{eqnarray}}}

\newcommand{\overbar}[1]{\mkern 1.5mu\overline{\mkern-1.5mu#1\mkern-1.5mu}\mkern 1.5mu}

\newcommand{\mpl}{m_\mathrm{Pl}}

\newcommand{\fnl}{f_\mathrm{NL}}
\newcommand{\abs}[1]{{\left \vert #1 \right \vert}}

\newcommand{\cH}{\mathcal{H}}
\newcommand{\cS}{\mathcal{S}}
\newcommand{\cO}{\mathcal{O}}

\newcommand{\bk}{\mathbf{k}}
\newcommand{\bq}{\mathbf{q}}

\newcommand{\bx}{\mathbf{x}}
\newcommand{\bn}{\mathbf{n}}

\newcommand{\ud}{\mathrm{d}}
\newcommand{\uin}{\mathrm{in}}
\newcommand{\uRD}{\mathrm{RD}}

\newcommand{\uGW}{\mathrm{gw}}
\newcommand{\uc}{\mathrm{c}}
\newcommand{\Beq}{\begin{align}}
\newcommand{\Eeq}{\end{align}}

\DeclareAcronym{GW}{
  short = GW,
  long = gravitational wave ,
  short-plural = s ,
}
\DeclareAcronym{LIGO}{
  short =LIGO ,
  long = Laser Interferometer Gravitational-Wave Observatory ,
  short-plural = ,
}
\DeclareAcronym{LVK}{
  short = LVK ,
  long = {Advanced LIGO, Virgo, and KAGRA Collaborations} ,
  short-plural = ,
}

\DeclareAcronym{SGWB}{
  short = SGWB ,
  long = stochastic gravitational-wave background ,
  short-plural = s ,
}
\DeclareAcronym{CGWB}{
  short = CGWB ,
  long = cosmological gravitational-wave background ,
  short-plural = s ,
}
\DeclareAcronym{CBC}{
  short = CBC ,
  long = compact binary coalescence ,
  short-plural = s ,
}
\DeclareAcronym{BH}{
  short = BH ,
  long = black hole ,
  short-plural = s ,
}
\DeclareAcronym{BBH}{
  short = BBH ,
  long = binary black hole ,
  short-plural = s ,
}
\DeclareAcronym{PBH}{
  short = PBH ,
  long = primordial black hole ,
  short-plural = s ,
}
\DeclareAcronym{SNR}{
  short = SNR ,
  long = signal-to-noise ratio ,
  short-plural = s ,
}
\DeclareAcronym{IMRPPv2}{
  short = ,
  long = {\normalsize IMRP}{\footnotesize HENOM}{\normalsize P}v2 ,
  short-plural = ,
}
\DeclareAcronym{PTA}{
  short = PTA ,
  long = pulsar timing array ,
  short-plural = s ,
}
\DeclareAcronym{SFR}{
  short = SFR ,
  long = star formation rate ,
  short-plural =  ,
}
\DeclareAcronym{FRW}{
  short = FRW ,
  long = Friedmann-Robertson-Walker ,
  short-plural =  ,
}
\DeclareAcronym{IMR}{
  short = IMR ,
  long = inspiral-merger-ringdown ,
  short-plural =  ,
}
\DeclareAcronym{LISA}{
	short = LISA ,
	long  = Laser Interferometer Space Antenna,
  short-plural =  ,
}
\DeclareAcronym{ET}{
	short = ET ,
	long  = Einstein Telescope,
  short-plural =  ,
}
\DeclareAcronym{CE}{
	short = CE ,
	long  = Cosmic Explorer,
  short-plural =  ,
}
\DeclareAcronym{BBO}{
	short = BBO ,
	long  = Big Bang Observer,
  short-plural =  ,
}
\DeclareAcronym{DECIGO}{
	short = DECIGO ,
	long  = Deci-hertz Interferometer Gravitational wave Observatory,
  short-plural =  ,
}
\DeclareAcronym{ABH}{
	short = ABH ,
	long  = astrophysical black hole,
  short-plural = s ,
}

\DeclareAcronym{PNG}{
	short = PNG ,
	long  = primordial non-Gaussianity ,
  short-plural =  ,
}

\DeclareAcronym{CMB}{
	short = CMB ,
	long  = cosmic microwave background ,
  short-plural =  ,
}
\DeclareAcronym{LSS}{
	short = LSS ,
	long  = large-scale structure ,
  short-plural =  ,
}

\DeclareAcronym{PGW}{
	short = PGW ,
	long  = primordial gravitational wave ,
  short-plural = s ,
}
\DeclareAcronym{SIGW}{
	short = SIGW ,
	long  = scalar-induced gravitational wave ,
  short-plural = s ,
}

\DeclareAcronym{RD}{
	short = RD,
	long  = radiation-dominated ,
  short-plural =  ,
}
\DeclareAcronym{MD}{
	short = MD,
	long  = matter-dominated ,
  short-plural =  ,
}
\DeclareAcronym{eMD}{
	short = eMD,
	long  = early-matter-dominated ,
  short-plural =  ,
}
\DeclareAcronym{SW}{
	short = SW,
	long  = Sachs-Wolfe ,
  short-plural =  ,
}
\DeclareAcronym{ISW}{
	short = ISW,
	long  = integrated Sachs-Wolfe ,
  short-plural =  ,
}

\DeclareAcronym{DM}{
	short = DM,
	long  = dark matter ,
  short-plural =  ,
}

\DeclareAcronym{NANOGrav}{
	short = NANOGrav ,
	long  = North American Nanohertz Observatory for Gravitational Waves ,
  short-plural =  ,
}

\DeclareAcronym{PDF}{
	short = PDF ,
	long  = probability distribution function ,
  short-plural = s ,
}

\arxivnumber{2305.19950 } 
\title{\boldmath Primordial Non-Gaussianity $f_{\mathrm{NL}}$ and Anisotropies in Scalar-Induced Gravitational Waves}

\author[a,b]{Jun-Peng Li,}
\author[a,b]{Sai Wang, }
\author[c]{Zhi-Chao Zhao,}
\author[d,e,f,g]{Kazunori Kohri}

\affiliation[a]{Theoretical Physics Division, Institute of High Energy Physics, Chinese Academy of Sciences, Beijing 100049, China}
\affiliation[b]{University of Chinese Academy of Sciences, 
Beijing 100049, China}
\affiliation[c]{Department of Applied Physics, College of Science, China Agricultural University,
Qinghua East Road, Beijing 100083, China}

\affiliation[d]{Theory Center, IPNS, KEK, 
1-1 Oho,Tsukuba, Ibaraki 305-0801, Japan}
\affiliation[e]{International Center for Quantum-field Measurement Systems for Studies of the Universe and Particles (QUP, WPI), KEK, 1-1 Oho, Tsukuba, Ibaraki 305-0801, Japan}
\affiliation[f]{The Graduate University for Advanced Studies (SOKENDAI), 
1-1 Oho, Tsukuba, Ibaraki 305-0801, Japan}
\affiliation[g]{Kavli Institute for the Physics and Mathematics of the Universe (WPI), UTIAS, The University of Tokyo, Kashiwa, Chiba 277-8583, Japan}

\emailAdd{wangsai@ihep.ac.cn}

\abstract{Primordial non-Gaussianity encodes vital information of the physics of the early universe, particularly during the inflationary epoch. To explore the local-type primordial non-Gaussianity $f_{\mathrm{NL}}$, we study the anisotropies in gravitational wave background induced by the linear cosmological scalar perturbations during radiation domination in the early universe. We provide the first complete analysis to the angular power spectrum of such scalar-induced gravitational waves. The spectrum is expressed in terms of the initial inhomogeneities, the Sachs-Wolfe effect, and their crossing. It is anticipated to have frequency dependence and multipole dependence, i.e., $C_\ell(\nu)\propto [\ell(\ell+1)]^{-1}$ with $\nu$ being a frequency and $\ell$ referring to the $\ell$-th spherical harmonic multipole. In particular, the initial inhomogeneites in this background depend on gravitational-wave frequency. These properties are potentially useful for the component separation, foreground removal, and breaking degeneracies in model parameters, making the non-Gaussian parameter $f_{\mathrm{NL}}$ measurable. Further, theoretical expectations may be tested by space-borne gravitational-wave detectors in future. }

\begin{document}
\maketitle
\flushbottom

\section{Introduction}

The \acl{PNG} refers to deviations from Gaussian statistics in the linear cosmological perturbations originated from quantum fluctuations during the inflationary epoch of the early universe \cite{Maldacena:2002vr,Bartolo:2004if,Allen:1987vq,Bartolo:2001cw,Acquaviva:2002ud,Bernardeau:2002jy,Chen:2006nt}. 
A quantity of mechanisms related to the generation of \acl{PNG} have been proposed (see Ref.~\cite{Meerburg:2019qqi} for reviews), for example, nonlinear couplings between the inflaton and other fields \cite{Lyth:2002my,Bartolo:2003jx,Zaldarriaga:2003my,Lyth:2005qk,Linde:2012bt,Torrado:2017qtr,Frazer:2011br,McAllister:2012am,Bjorkmo:2017nzd}, non-standard inflation models \cite{Namjoo:2012aa,Martin:2012pe,Chen:2013aj,Huang:2013oya,Mooij:2015yka,Bravo:2017wyw,Finelli:2017fml,Cai:2018dkf,Passaglia:2018ixg}, and so on. 
The primordial non-Gaussian parameter $\fnl$ represents a higher or lower probability of large overdensities, depending on its sign. 
Therefore, the study of \acl{PNG} is not only important for understanding the underlying physics of the early universe, but also the formation and evolution of cosmic structures.

There are several observational constraints on the \acl{PNG}, but they are limited to cosmological curvature perturbations on large scales comparable to the whole scale of the observable universe. 
Via measurements of anisotropies and polarization in the \ac{CMB}, the Planck collaboration \cite{Planck:2019kim} has reported highly Gaussian curvature perturbations that are compatible with anticipations of canonical single-field slow-roll inflation \cite{Mukhanov:1981xt}. 
Constraints have also been provided via measurements of \ac{CMB} spectral distortions \cite{Chluba:2015bqa,Rotti:2022lvy}, galaxy formation \cite{Stahl:2022did}, and UV luminosity function \cite{Sabti:2020ser}, etc., but all of them are less precise than the Planck results. 
We should note that the above measurements are only sensitive to the large-scale curvature perturbations, which are related to the dynamics of inflation during the 50-60 e-foldings before its end.

Detection of \acp{GW} can provide a new observational window to the nature of cosmological curvature perturbations on smaller scales, which were generated during later stages of inflation. 
It is well known that only the physics imprinted on the last-scattering surface of \ac{CMB} can be measured, due to the tightly coupled limit before the free streaming of photons \cite{Dodelson:2003ft}. 
In contrast, the \ac{GW} probe overcomes such a defect and thereby has potentials to directly measure the physics playing significant roles on more remote distances \cite{Maggiore:1999vm,Inomata:2018epa,Hajkarim:2019nbx,Domenech:2020kqm,Yu:2023lmo,Zhang:2022dgx,Chang:2022vlv,Chang:2020tji,Inomata:2019ivs,Inomata:2019zqy}, 
which are corresponded to higher redshifts, because \acp{GW} propagate almost freely after production \cite{Bartolo:2018igk,Flauger:2019cam}. 
Smaller-scale modes exited the Hubble horizon later during inflation, but reentered the Hubble horizon at higher redshifts after the end of inflation. 
Therefore, we expect the \ac{GW} probe to be sensitive to the primordial non-Gaussianity of small-scale perturbations and thereby the dynamics of inflation at the late stage.

After reentering into the Hubble horizon, the small-scale curvature perturbations nonlinearly produced a \ac{CGWB}, 
and the \acl{PNG} left significant imprints on the background \cite{Garcia-Bellido:2017aan,Domenech:2017ems,Nakama:2016gzw,Cai:2018dig,Unal:2018yaa,Yuan:2020iwf,Atal:2021jyo,Adshead:2021hnm,Ragavendra:2021qdu}, making the background to be a potential probe to the \acl{PNG}. 
Conventionally, such a \ac{CGWB} is also called the \acp{SIGW} \cite{Ananda:2006af,Baumann:2007zm,Mollerach:2003nq,Assadullahi:2009jc,Espinosa:2018eve,Kohri:2018awv}, since it was induced at second order by the linear scalar perturbations in the early universe. 
Depending on values of the local-type non-Gaussian parameter $\fnl$, the contribution of \acl{PNG} to the energy-density fraction spectrum of \acp{SIGW} could be two orders of magnitude larger than the Gaussian contribution, as was shown in Ref.~\cite{Adshead:2021hnm}. 
If the perturbativity conditions are required during inflation, some viable models have been considered in Ref.~\cite{Ragavendra:2021qdu}, where the authors studied \acp{SIGW} in the Starobinsky's model with a dip \cite{Atal:2019cdz,Mishra:2019pzq} and the model of critical-Higgs inflation \cite{Ezquiaga:2017fvi,Bezrukov:2017dyv,Drees:2019xpp}. 
For simplicity, we would not be concerned with such concrete scenarios in our current work. 
Recently, a common-spectrum process reported by the \ac{NANOGrav} collaboration \cite{NANOGrav:2020bcs} was speculated to be evidence for \acp{SIGW} in the literature \cite{DeLuca:2020agl,Vaskonen:2020lbd,Kohri:2020qqd,Domenech:2020ers,Atal:2020yic,Yi:2022ymw,Zhao:2022kvz,Dandoy:2023jot,Cai:2021wzd}, though not confirmed until now \footnote{In late June of 2023, four \ac{PTA} collaborations further reported strong evidence for the Hellings-Downs correlations that indicate a gravitational-wave background in the nano-Hertz frequency band \cite{Xu:2023wog,Antoniadis:2023ott,NANOGrav:2023gor,Reardon:2023gzh}.}.

Besides the vital contribution to \acp{SIGW}, the \acl{PNG} also impacts the formation of \acp{PBH} and particularly alters the mass distribution function of \acp{PBH} \cite{Bullock:1996at,Byrnes:2012yx,Young:2013oia,Franciolini:2018vbk,Passaglia:2018ixg,Atal:2018neu,Atal:2019cdz,Taoso:2021uvl,Meng:2022ixx,Chen:2023lou,Kawaguchi:2023mgk}.   
Since \acp{PBH} could be formed due to gravitational collapse of enhanced small-scale curvature perturbations \cite{Hawking:1971ei} and the \ac{PDF} of the latter is deformed by the \acl{PNG}, the abundance of \acp{PBH} would be significantly enhanced or suppressed compared with results for the Gaussian perturbations, depending on the sign of the non-Gaussian parameter (e.g., see Refs.~\cite{Byrnes:2012yx,Young:2013oia}). 
On the other hand, in the early universe, \acp{SIGW} were also produced as an accompaniment to the production of \acp{PBH}, making \acp{SIGW} a potential probe to \acp{PBH} \cite{Bugaev:2009zh,Saito:2009jt,Wang:2019kaf,Kapadia:2020pnr} and then the \acl{PNG} correspondingly.

In summary, the study of \acp{SIGW} is important for determination of the \acl{PNG}. 
The energy-density fraction spectrum of \acp{SIGW} has been used for this aim in the literature \cite{Nakama:2016gzw,Cai:2018dig,Unal:2018yaa,Yuan:2020iwf,Atal:2021jyo,Adshead:2021hnm,Ragavendra:2021qdu}. 
However, we will show that such a spectrum, i.e., the monopole, has a sign degeneracy in the non-Gaussian parameter. 
We will further show that there are degeneracies in the non-Gaussian parameter and other model parameters, indicating that the non-Gaussian contribution can be mimicked by these parameters. 
In addition, other gravitational-wave backgrounds originating from astrophysical processes would be foregrounds that may contaminate the signal (see Ref.~\cite{LISACosmologyWorkingGroup:2022kbp} and references therein). 
Due to the above reasons, it is particularly challenging to measure the \acl{PNG} with the monopole in \acp{SIGW}. 
Therefore, it is necessary to develop some new probes.

In this work, we propose that the anisotropies in \acp{SIGW} could be a powerful probe to the local-type \acl{PNG} on scales that can not be probed otherwise (e.g., via \ac{CMB}). 
We will provide the complete analysis to the angular power spectrum of \acp{SIGW} for the first time. 
We will also show its frequency dependence and multipole dependence, which could be useful for breaking the aforementioned degeneracies of model parameters and the foreground removal as well as component separation \cite{Chung:2023rpq}. 
Before our present work, the line-of-sight method for the study of anisotropies in a \ac{GW} background has been developed in Ref.~\cite{Contaldi:2016koz}, analogue to that for the study of anisotropies and polarization in \ac{CMB} \cite{Seljak:1996is}. 
Subsequently, it was adopted to study the anisotropies and non-Gaussianity in \acp{CGWB} in Refs.~\cite{Bartolo:2019oiq,Bartolo:2019yeu}. 
Assuming the local-type \acl{PNG} upon the squeezed limit, the anisotropies in \acp{SIGW} as well as implications of them for \acp{PBH} were studied for the first time in Ref.~\cite{Bartolo:2019zvb}. 
However, such a study is incomplete, as will be demonstrated in our present work. 
Following Ref.~\cite{Bartolo:2019zvb}, other related works can be found in Refs.~\cite{ValbusaDallArmi:2020ifo,Dimastrogiovanni:2021mfs,Schulze:2023ich,LISACosmologyWorkingGroup:2022kbp,LISACosmologyWorkingGroup:2022jok,Unal:2020mts,Malhotra:2020ket,Carr:2020gox}.
One of the leading aims of our present work is to establish the first complete analysis.

The remaining context of this paper is arranged as follows. 
In \cref{sec:Theory}, we will briefly review formulae of the inhomogeneous energy density of gravitational waves as well as the Boltzmann equation for the distribution function of gravitons. 
In \cref{sec:SIGW}, we will summarize the generic theory of \acp{SIGW}. 
In \cref{sec:mono}, we reproduce the theoretical results of the monopole in \acp{SIGW}, and show the degeneracies in model parameters. 
In \cref{sec:multi}, we provide the complete analysis of multipoles in \acp{SIGW}, including the formulae of angular power spectrum and its properties. 
In \cref{sec:Conclusion}, we make concluding remarks.

\section{Basics of cosmological gravitational wave background}\label{sec:Theory} 


We consider a spatially-flat \ac{FRW} metric in the conformal Newtonian gauge, with perturbations characterized by the linear scalar perturbations $\Phi(\eta,\bx)$ and $\Psi(\eta,\bx)$, and the transverse-traceless tensor perturbations $\chi_{ij}(\eta,\bx)$, i.e., the \acp{GW}. 
We disregard the vector perturbations due to inflation. 
The perturbed metric is given by 
\begin{equation}\label{metric} 
    \ud s^2 
    =  a^2 \left\{ 
            - (1 + 2 \Phi) \ud \eta^2 
            + \left[ (1 -2 \Psi) \delta_{ij} + \chi_{ij} \right] \ud x^i \ud x^j 
    \right\}\ ,
\end{equation} 
where $a(\eta)$ is the scale factor of the universe at conformal time $\eta$. 
It is convenient to expand $\Phi(\eta,\bx)$ (we expand $\Psi(\eta,\bx)$ in the same way) and $\chi_{ij}(\eta,\bx)$ in Fourier space, i.e.,  
\begin{eqnarray}
\Phi(\eta,\bx) & =& \int \frac{\ud^3 \bq}{(2\pi)^{3/2}} e^{i\bq\cdot\bx} \Phi(\eta,\bq)\ , \\
\chi_{ij}(\eta,\bx) & =& \sum_{\lambda=+,\times} 
\int \frac{\ud^3 \bq}{(2\pi)^{3/2}} e^{i\bq\cdot\bx} \epsilon_{ij}^{\lambda}(\bq) \chi_\lambda(\eta, \bq)\ ,
\label{eq:h-Fourier}  
\end{eqnarray}
where we define two polarization tensors $\epsilon^+_{ij}(\bq) = \left[\epsilon_i(\bq) \epsilon_j(\bq) - \bar{\epsilon}_i(\bq) \bar{\epsilon}_j(\bq)\right] /\sqrt{2}$ and $\epsilon^\times_{ij}(\bq) =  \left[\epsilon_i(\bq) \bar{\epsilon}_j(\bq) + \bar{\epsilon}_i(\bq) \epsilon_j(\bq) \right]/\sqrt{2}$, with  $\epsilon_{i}(\bq)$ and $\bar{\epsilon}_{i}(\bq)$ being a set of orthonormal basis which is perpendicular to the wavevector $\bq$. 
We further define the power spectrum of \acp{GW} as the two-point correlator of $\chi_{\lambda}$, i.e., 
\begin{equation}\label{eq:chi-cor} 
    \langle \chi_\lambda (\eta,\bq) \chi_{\lambda'} (\eta,\bq')\rangle 
    = \delta_{\lambda\lambda'} \delta^{(3)} (\bq+\bq') P_{\chi_\lambda} (\eta,q) \ , 
\end{equation} 
which characterizes the statistical property. 
In the following, we will introduce several useful definitions and conventions of \acp{CGWB}, as well as the Boltzmann equation of gravitons. 
In fact, most of them are analogue to those for \ac{CMB} \cite{Seljak:1996is}, and we would follow Refs.~\cite{Contaldi:2016koz,Bartolo:2019oiq,Bartolo:2019yeu}. 

\subsection{Energy density with inhomogeneities}\label{sec:ed}

At a conformal time $\eta$ and spatial location $\bx$, the energy density of \acp{GW} on subhorizon scales is defined as \cite{Maggiore:1999vm} 
\begin{equation}\label{eq:rho-def}
    \rho_\uGW (\eta,\bx) 
    = \frac{\mpl^2}{4 a^2(\eta)} \overbar{\partial_l \chi_{ij}(\eta,\bx) \partial_l \chi_{ij}(\eta,\bx)}\ ,
\end{equation}
where the overbar denotes a time average over oscillations, and $\mpl=({8\pi G})^{-1/2}$ is the Planck mass. 
Throughout this paper, we use $\partial_\eta$ and $\partial_i$ to denote $\partial / \partial\eta$ and $\partial / \partial x^i$, respectively.
The energy density spectrum $\Omega_\uGW(\eta,\bx,q)$ is defined as \cite{Maggiore:1999vm} 
\begin{equation}\label{eq:rho-Omega}
    \rho_\uGW (\eta,\bx) 
    = \rho_\uc \int \ud \ln q\, \Omega_\uGW (\eta,\bx,q)\ ,
\end{equation}
with the critical energy density of the universe defined in terms of the conformal Hubble parameter $\cH(\eta)={\partial_{\eta}{a}}/{a}$ as $\rho_\uc=3\mpl^2 \cH^2 / a^2$. 
We further introduce the energy-density full spectrum $\omega_\uGW (\eta,\bx,\bq)$, which is direction-dependent, as 
\begin{equation}\label{eq:omega-def}
    \Omega_\uGW (\eta,\bx,q) 
    = \int \ud^2 \bn\, \omega_\uGW (\eta,\bx,\bq)\ , 
\end{equation}
where $\bq$ denotes the comoving momentum of GWs and $\bn$ denotes the propagation direction of GWs, i.e., $\bq = q \bn$ with $q=|\bq|$. 
Therefore, we get an explicit expression of it to be 
\begin{equation}\label{eq:omega-chi}
     \omega_\uGW(\eta,\bx,\bq)= - \frac{q^3}{12 \cH^2} \int \frac{\ud^3 \bk}{(2\pi)^{3}} e^{i\bk\cdot\bx} \left(\bk-\bq\right) \cdot \bq  \sum_{\lambda,\lambda'} \epsilon_{ij}^{\lambda}(\bk-\bq) \epsilon_{ij}^{\lambda'}(\bq) \overbar{\chi_\lambda(\eta, \bk-\bq) \chi_{\lambda'}(\eta, \bq)}\ .
\end{equation}
It is crucial to note that $\bk$ is associated with the Fourier modes of overdensities in \ac{CGWB}. 

The full spectrum $\omega_\uGW(\eta,\bx,\bq)$ can be decomposed into a homogeneous and isotropic background $\bar{\omega}_\uGW(\eta,q)$ and superimposed fluctuations $\delta\omega_\uGW(\eta,\bx,\bq)$. 

The former is also called the monopole. 
It can be obtained from the definition of $\omega_\uGW(\eta,\bx,\bq)$ in \cref{eq:omega-def}, i.e., 
\begin{equation}\label{eq:omegabar}
    \bar{\omega}_\uGW (\eta,q) = \frac{\bar{\Omega}_\uGW (\eta,q)}{4\pi}\ .
\end{equation}
Here, $\bar{\Omega}_\uGW (\eta, q)$ stands for the energy-density fraction spectrum defined by the spatial average of $\Omega_\uGW(\eta,\bx,q)$ as follows \cite{Inomata:2016rbd}
\begin{equation}\label{eq:Omegabar} 
    \bar{\Omega}_\uGW (\eta, q) 
    = \left\langle\Omega_\uGW (\eta,\bx,q) \right\rangle_\bx
    = \frac{q^5}{24\pi^2 \cH^2} 
        \sum_{\lambda=+,\times} \overbar{P_{\chi_\lambda} (\eta,q)}\ ,
\end{equation}
where the angle brackets with a suffix $_\bx$ denote the spatial average that is equivalent to the ensemble average. 
Besides \cref{eq:omega-def}, we also have used \cref{eq:chi-cor} and \cref{eq:omega-chi} during the derivation process of \cref{eq:Omegabar}.

The inhomogeneities $\delta\omega_\uGW$ on top of the background, leading to the multipoles in a \ac{CGWB} discussed in the following, can be written as 
\begin{equation}\label{eq:delta-omega-def} 
    \delta\omega_\uGW (\eta,\bx,\bq) 
        = \omega_\uGW (\eta,\bx,\bq) - \bar{\omega}_\uGW (\eta,q)\ ,
\end{equation}
which can be recast to be the density contrast of the form  
\begin{equation}\label{eq:deltaGW-def}
    \delta_\uGW (\eta,\bx,\bq) 
        = \frac{\delta\omega_\uGW (\eta,\bx,\bq)}{\bar{\omega}_\uGW (\eta,q)}
        = 4\pi \frac{\delta\omega_\uGW (\eta,\bx,\bq)}{\bar{\Omega}_\uGW (\eta,q)}\ .
\end{equation}
To study the statistics of the fluctuations, we use the two-point correlation of $\delta_{\uGW}(\eta,\bx,\bq)$. 
It is useful to expand $\delta_{\uGW}$ in spherical harmonics along the direction $\bn$, i.e., 
\begin{equation}
    \delta_\uGW (\eta,\bx,\bq) =
        \sum_\ell\sum_{m=-\ell}^\ell \delta_{\uGW,\ell m} (\eta,\bx,q) Y_{\ell m}(\bn)\ ,
\end{equation}
where we have used the relation $\bq=q\bn$, and the multipole coefficients are given by 
\begin{equation}\label{eq:delta-lm}
    \delta_{\uGW, \ell m} (\eta,\bx,q)
    = \int \ud^2 \bn\, Y_{\ell m}^\ast (\bn)  
        \delta_\uGW (\eta,\bx,\bq)\ .
\end{equation}

Assuming the statistical isotropy on large scales, we define the reduced angular power spectrum as a two-point correlator of the multipole coefficients $\delta_{\uGW, \ell m} (\eta_{0},\bx_{0},q)$, with the observing time and location $(\eta_0,\bx_0)$ omitted for brevity hereafter, i.e.,  
\begin{equation}\label{eq:reduced-angular-power-spectrum-def}
    \langle\delta_{\uGW,\ell m}(q) \delta_{\uGW,\ell' m'}^\ast(q')\rangle
        = \delta_{\ell \ell'} \delta_{mm'} \widetilde{C}_\ell (q,q')\ ,
\end{equation}
where the tilde stands for a reduced quantity. 
In fact, this is cross-correlation at two frequency bands denoted by $q$ and $q'$. 
Further, the angular power spectrum is defined as the two-point correlator of $\delta\omega_{\uGW} (q)$, i.e., 
\begin{equation}\label{eq:angular-power-spectrum-def}
     \left\langle
            \delta\omega_{\uGW,\ell m} (q) \delta\omega_{\uGW,\ell' m'}^\ast (q')
        \right\rangle = \delta_{\ell \ell'} \delta_{mm'} C_\ell (q,q')  \ .
\end{equation}  
A relation between $\widetilde{C}_\ell$ and $C_\ell$ can be derived from \cref{eq:deltaGW-def}, namely, 
\begin{equation}\label{eq:Ct-C}
    \widetilde{C}_\ell (q,q') = \frac{(4\pi)^2 C_\ell (q,q')}{\bar{\Omega}_\uGW (q) \bar{\Omega}_\uGW (q')}\ ,
\end{equation}
where $\bar{\Omega}_\uGW (q)$ denotes the energy-density fraction spectrum in the observer frame with $(\eta_0,\bx_0)$. 
Here, besides correlations between the same frequency band (i.e., $q=q^\prime$), we also consider correlations between different frequency bands (i.e., $q\neq q^\prime$). 
Such a consideration would give rise to non-trivial theoretical results, as will be shown in \cref{sec:APS-Result}. 

\subsection{Boltzmann equation}\label{sec:Bm}


Following Refs.~\cite{Contaldi:2016koz,Bartolo:2019oiq,Bartolo:2019yeu}, we review the Boltzmann equation for gravitons in general. 
The energy density in \cref{eq:rho-def} is expressed in terms of the distribution function of gravitons $f(\eta,\bx, \bq)$, i.e.,  
\begin{equation}\label{eqn:rho-f}
    \rho_\uGW (\eta,\bx) 
        = \frac{1}{a^4} \int \ud^3 \bq\, q f(\eta,\bx,\bq)\ .
\end{equation} 
Combining it with \cref{eq:rho-Omega} and \cref{eq:omega-def}, we obtain a relation of the form
\begin{equation}\label{eq:f-omega}
    f(\eta,\bx,\bq) 
        = \rho_\uc \left(\frac{a}{q}\right)^4 \omega_\uGW (\eta,\bx,\bq) \ .
\end{equation}
Analogue to decomposition of $\omega_\uGW$ in \cref{sec:ed}, the distribution function can also be separated into a background $\bar{f}(\eta,q)$ and perturbations $\Gamma(\eta,\bx,\bq)$, i.e., 
\begin{equation}\label{eqn:Gamma-def}
    f(\eta,\bx,\bq)
        =\bar{f}(\eta,q)
        - q\frac{\partial \bar{f}}{\partial q}\Gamma(\eta,\bx,\bq)\ .
\end{equation}
The former is related with the energy-density fraction spectrum $\bar{\Omega}_\uGW (\eta,q)$ via \cref{eq:omegabar} and \cref{eq:f-omega}, i.e.,  
\begin{equation}\label{eq:fbar-Omegabar}
    \bar{f}(\eta,q)
        = \frac{\rho_\uc}{4\pi} \left(\frac{a}{q}\right)^4 \bar{\Omega}_\uGW (\eta,q) \ . 
\end{equation}
Therefore, the density contrast in \cref{eq:deltaGW-def} can be expressed in terms of $\Gamma(\eta,\bx,\bq)$ as follows 
\begin{equation}\label{eq:delta-Gamma}
    \delta_\uGW (\eta,\bx,\bq)
        = \left[
            4-\frac{\partial \ln \bar{\Omega}_\uGW (\eta,q)}{\partial\ln q}
        \right]
        \Gamma (\eta,\bx,\bq)
        = \left[4-n_\uGW (\eta,q)\right]
        \Gamma (\eta,\bx,\bq)\ ,
\end{equation}
where we define the tensor spectral index as  
\begin{equation}\label{eq:ngw-def}
    n_\uGW (\eta,q) = \frac{\partial \ln \bar{\Omega}_\uGW (\eta,q)}{\partial\ln q}\ .
\end{equation}

The evolution of distribution function follows the Boltzmann equation, i.e., 
${\ud f}/{\ud \eta}=\mathcal{I}(f) + \mathcal{C}(f)$,
where $\mathcal{I}$ denotes the emissivity term and $\mathcal{C}$ stands for the collision term. 
Due to absence of interaction of gravitons, the collision term is negligible, i.e., $\mathcal{C}=0$ \cite{Bartolo:2018igk,Flauger:2019cam}. 
The emissivity term for cosmological processes can be viewed as the initial condition, implying $\mathcal{I}=0$ \cite{Bartolo:2019oiq,Bartolo:2019yeu}. 
Therefore, the Boltzmann equation can be expressed as  
\begin{equation}\label{eq:f-total-derivation}
    \frac{\ud f}{\ud \eta}
        = \frac{\partial f}{\partial \eta}
        +\frac{\partial f}{\partial x^i}\frac{\ud x^i}{\ud \eta}
        +\frac{\partial f}{\partial q}\frac{\ud q}{\ud \eta} 
        +\frac{\partial f}{\partial n^i}\frac{\ud n^i}{\ud \eta}=0\ .
\end{equation}
For the Boltzmann equation up to first order, the massless condition and geodesic of gravitons lead to $\ud x^i / \ud \eta = n^i$, $\ud q / \ud \eta= \left(\partial_\eta\Psi - n^i \partial_i \Phi - n^i n^j\partial_\eta \chi_{ij} / 2\right) q$, and $\ud n^i / \ud \eta = 0$. 
The Boltzmann equation can be separated into 
\begin{eqnarray}
    \partial_\eta \bar{f} & = & 0\ ,\label{eq:Boltzmannfbar}\\
    \partial_\eta\Gamma + n^i \partial_i \Gamma 
    & = & \partial_\eta\Psi - n^i\partial_i \Phi - \frac{1}{2}n^i n^j\partial_\eta \chi_{ij}\ . \label{eq:Boltzmann-1st}
\end{eqnarray}
Eq.~(\ref{eq:Boltzmannfbar}) indicates that the background does not evolve with respect to time. 
Eq.~(\ref{eq:Boltzmann-1st}) can be transformed to Fourier space, i.e., 
\begin{equation}\label{eq:Boltzmann-k}
    \partial_\eta\Gamma + i k \mu \Gamma 
    = \partial_\eta\Psi - i k \mu \Phi - \frac{1}{2}n^i n^j\partial_\eta \chi_{ij}\ , 
\end{equation}
where we denote $k\mu = \bk\cdot\bn$ for simplicity.

Analogue to the Boltzmann equation for the anisotropies and polarization in \ac{CMB} \cite{Zaldarriaga:2003my}, \cref{eq:Boltzmann-k} also has the line-of-sight solution of the form 
\begin{eqnarray}\label{eq:Gamma-solution}
    \Gamma (\eta,\bk,\bq) 
    & = & e^{i k \mu (\eta_\uin - \eta)} 
        \left[
            \Gamma (\eta_\uin, \bk, \bq) + \Phi (\eta_\uin, \bk)
        \right]
        - \Phi (\eta, \bk) \nonumber\\
        && + \int_{\eta_\uin}^\eta \ud \eta'\,
            e^{i k \mu (\eta' - \eta)}
            \left\lbrace
                \partial_{\eta'} \left[\Psi (\eta',\bk) + \Phi (\eta',\bk)\right]
            - \frac{n^i n^j}{2}\partial_\eta \chi_{ij} (\eta',\bk) 
            \right\rbrace\ ,
\end{eqnarray}
where we use the suffix $_\uin$ to label quantities at initial time. 
We decompose the solution as
\begin{equation}
    \Gamma (\eta,\bk,\bq)
    = \Gamma_I (\eta,\bk,\bq) +\Gamma_S (\eta,\bk,\bq) +\Gamma_T (\eta,\bk,\bq) - \Phi (\eta, \bk)\ ,
\end{equation}
where $\Gamma_{I}$ stands for the initial term, and $\Gamma_{S}$ and $\Gamma_{T}$ denote the scalar and tensor sourced terms, respectively.
To be specific, we have 
\begin{eqnarray}
    \Gamma_I (\eta,\bk,\bq) 
        & = & e^{i k \mu (\eta_\uin - \eta)} \Gamma(\eta_\uin,\bk,\bq)\ ,\label{eq:Gamma-I}\\
    \Gamma_S (\eta,\bk,\bq) 
        & = & \int_{\eta_\uin}^\eta \ud \eta'\,
            e^{i k \mu (\eta' - \eta)}
            \left\{
                \Phi(\eta', \bk) \delta (\eta' - \eta_\uin)
                + \partial_{\eta'} \left[\Psi (\eta',\bk) + \Phi (\eta',\bk)\right]
        \right\}\ ,\label{eq:Gamma-S}\\
    \Gamma_T (\eta,\bk,\bq) 
        & = & - \frac{1}{2} n^i n^j \int_{\eta_\uin}^\eta \ud \eta'\,
            e^{i k \mu (\eta' - \eta)}
            \partial_{\eta'} \chi_{ij} (\eta',\bk) \ .\label{eq:Gamma-T}
\end{eqnarray}
By considering \cref{eq:delta-Gamma} and the relation of $\bx_0-\bx_{\uin} = (\eta_0-\eta_\uin)\bn_0$, we can relate $\Gamma(\eta,\bk,\bq)$ in Eq.~(\ref{eq:Gamma-solution}) with the inhomogeneities $\delta_\uGW(\eta,\bx,\bq)$.
Therefore, we obtain $\delta_\uGW (\bq)=\delta_\uGW (\eta_0,\bx_0,\bq)$ as follows 
\begin{eqnarray}\label{eq:deltaGW-CGW}
    \delta_\uGW (\bq) 
    & = & \left[4-n_\uGW (\eta_0,q)\right]
        \Biggl\lbrace
            \left[
                \frac{\delta_\uGW (\eta_\uin, \bx_\uin,\bq)}{4-n_\uGW (\eta_\uin,q)} 
                + \Phi (\eta_\uin, \bx_\uin)
            \right]
            - \Phi (\eta_0, \bx_0)\\
            && + \int \frac{\ud^3 \bk}{(2\pi)^{3/2}} e^{i\bk\cdot\bx_0}
            \int_{\eta_\uin}^{\eta_0} \ud \eta \,  
            e^{i k \mu (\eta - \eta_0)}
            \left[
                \partial_\eta \left[\Psi (\eta,\bk) + \Phi (\eta,\bk)\right]
                - \frac{n^i n^j}{2}\partial_\eta \chi_{ij} (\eta,\bk) 
            \right]
        \Biggr\rbrace\ ,\nonumber
\end{eqnarray}
where $\delta (\eta_\uin, \bx_\uin, \bq)$ represents the initial perturbations \footnote{In contrast, the initial perturbations for \ac{CMB} were completely erased by Compton scattering due to the tightly coupled limit before the free streaming of photons \cite{Dodelson:2003ft}.}, $\Phi (\eta_\uin, \bx_\uin)$ leads to the \ac{SW} effect \cite{Sachs:1967er}, $\Phi (\eta_0, \bx_0)$ is the monopole term that can be disregarded, and the integral refers to the \ac{ISW} effect \cite{Sachs:1967er}. 
By substituting Eq.~(\ref{eq:deltaGW-CGW}) into \cref{eq:delta-lm} and then into \cref{eq:reduced-angular-power-spectrum-def}, we can obtain an explicit formula of the angular power spectrum for the anisotropies in \ac{CGWB}.


\section{Scalar-induced gravitational waves}\label{sec:SIGW}

Following Refs.~\cite{Espinosa:2018eve,Kohri:2018awv}, we review the theory of \acp{SIGW} such as the equation of motion and its solution during radiation domination. 
Our theoretical formalism can be straightforwardly used for the study of \acp{SIGW} during other epochs, e.g., early-matter domination \cite{Domenech:2019quo}. 
In this section and the next section, we use $\eta$ to denote $\eta_{\uin}$ for simplicity.

\subsection{Equation of motion and its solution}\label{sec:motion}

We consider the case that the tensor perturbations are \acp{SIGW}, implying that the second-order tensor perturbations are considered.
We let $\chi_{ij} = h_{ij}/2$ in \cref{metric} and neglect the anisotropic stress, i.e., $\Psi = \Phi$. 
Therefore, the equation of motion of \acp{SIGW} is derived from the spatial components of Einstein's equation at second order, i.e., \cite{Ananda:2006af,Baumann:2007zm} 
\begin{equation}\label{eq:SIGW-motion} 
    \partial_\eta^2 h_{\lambda}(\eta, \bq)
    + 2\cH \partial_\eta h_{\lambda}(\eta,\bq) 
    +q^2 h_{\lambda}(\eta,\bq) 
    = 4 S_{\lambda}(\eta,\bq)\ , 
\end{equation} 
where $S_{\lambda}(\eta,\bq)$ is the source term quadratic in the scalar perturbations $\Phi$, i.e.,   
\begin{eqnarray}\label{eq:source} 
    \cS_\lambda(\eta,\bq) 
    & = & \int\frac{\ud^3 \bq_a}{(2\pi)^{3/2}}
            \epsilon_{ij}^{\lambda}(\bq) q_{a}^i q_{a}^j
            \Bigg\{
                2 \Phi (\eta,\bq - \bq_a)\Phi(\eta,\bq_a)
            \\
            && 
                +\frac{4}{3(1+w)\cH^2}
                \left[
                    \partial_\eta \Phi(\eta,\bq -\bq_a)
                    + \cH\Phi(\eta,\bq -\bq_a)
                \right]
                \left[
                    \partial_\eta \Phi(\eta,\bq_a) 
                    + \cH\Phi(\eta,\bq_a)
                \right]
            \Bigg\}\ .\nonumber
\end{eqnarray} 
Here, $w$ stands for the equation-of-state parameter of the universe. 
The above derivation can be finished via the \texttt{xpand} \cite{Pitrou:2013hga} package.

\cref{eq:SIGW-motion} can be solved with the Green's function method, as was demonstrated in Refs.~\cite{Espinosa:2018eve,Kohri:2018awv}. The solution can be expressed in the form of 
\begin{equation}\label{eq:h-G} 
    a(\eta) h_\lambda(\eta, \bq) 
    = 4 \int^{\eta}_{} \ud \eta' \,
        G_\bq(\eta,\eta') a(\eta') \cS_\lambda(\eta', \bq)\ , 
\end{equation}
where the Green's function $G_\bq(\eta,\eta')$ obeys  
\begin{equation}\label{eq:Green}
    \partial_{\eta}^2 G_\bq(\eta,\eta') 
        +  \left[q^2-\frac{\partial_\eta^2 a(\eta)}{a(\eta)}\right] G_\bq(\eta, \eta')
        = \delta(\eta - \eta')\ .
\end{equation}
As will be shown in \cref{sec:SIGW-RD}, we can solve \cref{eq:Green} once the evolution of $a(\eta)$ is known. 
To relate the linear perturbations $\Phi(\eta,\bq)$ with the initial value, we define the scalar transfer function $T(q\eta)$ as follows 
\begin{equation}\label{eq:T-zeta-def}
    \Phi(\eta, \bq)
        = \frac{3+3w}{5+3w} T(q \eta) \zeta(\bq)\ ,
\end{equation}
where $\zeta(\bq)$ denotes the primordial (comoving) curvature perturbations. 
Therefore, Eq.~(\ref{eq:source}) can be rewritten as 
\begin{equation}\label{eq:S}
    \cS_\lambda(\eta, \bq)
        = \int \frac{\ud^3 \bq_a}{(2\pi)^{3/2}} q^2 Q_{\lambda}(\bq,\bq_a)
            F(\abs{\bq-\bq_a}, q_a, \eta)
            \zeta(\bq_a)
            \zeta(\bq-\bq_a)\ ,
\end{equation}
where we introduce a functional $F(\abs{\bq-\bq_a}, q_a, \eta)$ and a projection factor $Q_{\lambda}(\bq, \bq_a)$. 
Denoting $p_a=\abs{\bq-\bq_a}$, we represent the functional $F(p_a, q_a, \eta)$ in terms of $T(\eta)$ and $\partial_\eta T(\eta)$, i.e.,  
\begin{eqnarray}\label{eq:F-def}
    F(p_a,q_a,\eta)
    & = & \frac{3 (1 + w)}{(5 + 3 w)^2}
        \biggl[
            2 (5 + 3 w) T(p_a \eta) T(q_a \eta)
            + \frac{4}{\cH^2} \partial_\eta T(p_a \eta) \partial_\eta T (q_a \eta)\nonumber\\ 
    &&\hphantom{ \quad \frac{3 (1 + w)}{(5 + 3 w)^2} \biggl[ }
            + \frac{4}{\cH} \left(
                T(p_a \eta) \partial_\eta T(q_a \eta) + \partial_\eta T(p_a \eta) T(q_a \eta)
            \right)
        \biggr]\ .
\end{eqnarray}
On the other hand, the projection factor $Q_{\lambda}(\bq, \bq_a)$ is defined as follow 
\begin{equation}\label{eq:Qsai}
    Q_{\lambda}(\bq, \bq_a) 
    = \epsilon_{ij}^{\lambda}(\bq) \frac{q_{a}^i q_{a}^j}{q^2}
    = \frac{\sin^2 \theta}{\sqrt{2}}
     \times
        \begin{cases}
            \cos(2\phi_a) &\lambda = + \\
            \sin(2\phi_a) &\lambda = \times 
        \end{cases}\ , 
\end{equation}
where $\theta$ is the separation angle between $\bq$ and $\bq_a$, while $\phi_a$ represents the azimuthal angle of $\bq_a$ when $\bq$ is along the $\mathbf{z}$ axis. 
In addition, the evolution of $\Phi(\eta,\bq)$ and then $T(q\eta)$ follows a master equation derived from the Einstein's equation at first order. 
In absence of entropy perturbations, the master equation is \cite{Maggiore:2018sht} 
\begin{equation}\label{eq:master}
    \partial_\eta^2 \Phi + 3\cH \left(1+c_s^2\right) \partial_\eta \Phi + 3\left(c_s^2 -w\right) \cH^2\Phi + c_s^2 q^2 \Phi  = 0\ .
\end{equation}
where $c_s^2$ denotes the speed of sound.  
We will obtain the analytic expression of $T(q\eta)$ during radiation domination in \cref{sec:SIGW-RD}.

Based on the above discussion, we can rewrite the \ac{SIGW} strain in \cref{eq:h-G} as follows 
\begin{equation}\label{eq:h} 
    h_\lambda(\eta, \bq) 
    = 4 \int \frac{\ud^3 \bq_a}{(2\pi)^{3/2}} 
        \zeta(\bq_a) \zeta(\bq-\bq_a) Q_{\lambda}(\bq,\bq_a) 
        \hat{I} (\abs{\bq - \bq_a},q,\eta)\ , 
\end{equation}
where the kernel function is given as    
\begin{equation}\label{eq:I-def}
    \hat{I}(\abs{\bq - \bq_a},q_a,\eta)
    =\int^{\eta}_{} \ud \eta' \,
        q^2 G_\bq(\eta, \eta')
        \frac{a(\eta')}{a(\eta)} 
        F(\abs{\bq - \bq_a}, q_a, \eta')\ .
\end{equation}
Note that \cref{eq:h} is the most important result in this subsection. 
The remaining work is to compute the kernel function, as will be done in \cref{sec:SIGW-RD}.

\subsection{Kernel function during radiation domination}\label{sec:SIGW-RD}

We focus on the \ac{RD} epoch in the following, implying that we have $w=c_s^2=1/3$, $a\propto \eta$, and $\cH = 1/\eta$. 
We will solve \cref{eq:Green} and \cref{eq:master}, and then get the analytic formula of kernel function via \cref{eq:I-def}, which eventually leads to \cref{eq:h}. 


Since we have $\partial_{\eta}^{2}a=0$, we rewrite \cref{eq:Green} as $\partial_{\eta}^{2}G_{\bq}+q^{2} G_{\bq}=\delta(\eta-\eta')$. 
Its solution gives the Green's function, i.e.,  
\begin{equation}\label{eq:G-RD}
    G_\bq (\eta,\eta')
        = \Theta (\eta -\eta') 
        \frac{\sin q(\eta - \eta')}{q}\ ,
\end{equation}
where $\Theta(x)$ is the Heaviside function with variable $x$. 
We rewrite \cref{eq:master} as $\partial_{\eta}^{2}\Phi+4\mathcal{H}\partial_{\eta}\Phi+q^{2}\Phi/3=0$. 
By using \cref{eq:T-zeta-def}, we further rewrite it as 
\begin{equation}
    \frac{\ud^2 T (x)}{\ud x^2} + \frac{4}{x}\frac{\ud T (x)}{\ud x} + \frac{T (x)}{3} = 0 \ ,
\end{equation}
where we denote $x= q\eta$ for simplicity. 
Therefore,  the scalar transfer function is given as 
\begin{equation}\label{eq:T-RD}
    T (x) = \frac{9}{x^2} 
            \left(
                \frac{\sin (x/\sqrt{3})}{x/\sqrt{3}}
                -\cos (x/\sqrt{3})
            \right) \ ,
\end{equation}
which satisfies the conditions of $T(q\eta_{\mathrm{out}})=1$ and $\partial_\eta T(q\eta_{\mathrm{out}})=0$ in the limit of superhorizon, i.e., $q\eta_{\mathrm{out}}\rightarrow 0$.

By substituting \cref{eq:T-RD} into Eq.~(\ref{eq:F-def}), we obtain an expression for the functional $F(\abs{\bq-\bq_a},q_a,\eta)$ during \ac{RD} epoch. 
Further combining it with \cref{eq:G-RD}, we get an expression for the kernel function $\hat{I} \left(\abs{\bq-\bq_a},q_a,\eta\right)$ in \cref{eq:I-def}. 
The kernel function can be further recast into $I_{\uRD}(u,v,x)$ of the form 
\begin{equation}\label{eq:irdsai}
    \hat{I} \left(\abs{\bq-\bq_a},q_a,\eta\right) 
        =  I_{\uRD} \left(\frac{\abs{\bq-\bq_a}}{q},\frac{q_a}{q},q\eta\right)
        = I_{\uRD} (u,v,x)\ .
\end{equation}
Here, we have introduced two new variables $u={\abs{\bq-\bq_a}}/{q}$ and $v={q_a}/{q}$ for simplicity. 
As was shown in Refs.~\cite{Kohri:2018awv,Espinosa:2018eve}, the analytic formula for $I(u,v,x)$ on subhorizon scales, i.e., $x \gg 1$, takes the form of 
\begin{subequations}\label{eq:I-RD} 
\begin{eqnarray}
I_{\uRD} (u,v,x \gg 1) &=& \frac{1}{x} I_A (u,v) 
            \left[I_B (u,v) \sin x - \pi I_C (u,v) \cos x\right]\ ,\\
I_A (u,v) &=& \frac{3\left(u^2 + v^2 - 3\right)}{4 u^3 v^3}\ , \\
I_B (u,v) &=& - 4 u v  
        + \left(u^2+v^2-3\right) \ln \abs{\frac{3-(u + v)^2}{3-(u - v)^2}}\ , \\
I_C (u,v) &=& \left(u^2 + v^2 - 3\right)
        \Theta\left(u + v - \sqrt{3}\right)\ .
\end{eqnarray} 
\end{subequations} 
As will be used in the next section, the oscillation average of two kernel functions with the same $x$ is given as \cite{Adshead:2021hnm}
\begin{eqnarray}\label{eq:I-ave-12}
    && \overbar{I_\uRD (u,v,x\rightarrow\infty) 
        I_\uRD (u',v',x\rightarrow\infty)} \nonumber\\
        & = & \frac{I_A (u,v) I_A (u',v')}{2 x^2} 
        \left[
                I_B (u,v) I_B (u',v')
                + \pi^2 I_C (u,v) I_C (u',v')
            \right]\ .
\end{eqnarray}
Such oscillation average can significantly simplify our computation in the following sections. 
In addition, \cref{eq:I-RD} should be substituted into \cref{eq:h} for a next step. 

\section{Monopole and degeneracies in model parameters}\label{sec:mono}

Following Ref.~\cite{Adshead:2021hnm}, we review the significant contributions of local-type primordial non-Gaussianity to the monopole in \acp{SIGW}. 
Further, we show serious degeneracies of the model parameters in the energy-density fraction spectrum, making it challenging to measure the \acl{PNG} with the monopole in \acp{SIGW} only.

Similar to \cref{eq:chi-cor}, in order to study the statistics of \acp{SIGW}, we define the power spectrum of \acp{SIGW} as follows 
\begin{equation}\label{eq:h-cor} 
    \langle h_\lambda (\eta,\bq) h_{\lambda'} (\eta,\bq')\rangle 
    = \delta_{\lambda\lambda'} \delta^{(3)} (\bq+\bq') P_{h_\lambda} (\eta,q) \ , 
\end{equation} 
By substituting Eq.~(\ref{eq:h}) into Eq.~(\ref{eq:h-cor}), we obtain 
\begin{eqnarray}\label{eq:Ph-zeta}
    \langle
        h_{\lambda}(\eta,\bq)
        h_{\lambda'}(\eta,\bq')
    \rangle
    & = & 16
        \int \frac{\ud^3 \bq_1}{(2\pi)^{3/2}} \frac{\ud^3 \bq_2}{(2\pi)^{3/2}} 
        \langle
            \zeta(\bq_1) \zeta(\bq - \bq_1) \zeta(\bq_2) \zeta(\bq' - \bq_2)
        \rangle \\
        & &\quad \times 
        Q_{\lambda}(\bq, \bq_1)
        \hat{I} (\abs{\bq - \bq_1}, q_1, \eta) 
        Q_{\lambda'}(\bq', \bq_2)
        \hat{I} (\abs{\bq' - \bq_2}, q_2, \eta) \ .\nonumber
\end{eqnarray}
Similar to \cref{eq:Omegabar}, the energy-density fraction spectrum for the monopole in \acp{SIGW} is defined as 
\begin{equation}\label{eq:Omegabar-h} 
    \bar{\Omega}_\uGW (\eta, q)
    = \frac{q^5}{96\pi^2 \cH^2} 
        \sum_{\lambda=+,\times} \overbar{P_{h_\lambda} (\eta,q)}\ .
\end{equation}
Utilizing \cref{eq:h-cor} and Eq.~(\ref{eq:Ph-zeta}), we can express the monopole as a four-point correlator of primordial curvature perturbations $\zeta$, namely, $\bar{\Omega}_\uGW \propto \langle\zeta^4\rangle$ schematically. 

\subsection{Primordial non-Gaussianity of local type}\label{sec:NG}

When $\zeta$ are Gaussian, the four-point correlator can be reduced to the two-point correlator, as have been done in the literature \cite{Ananda:2006af,Baumann:2007zm,Mollerach:2003nq,Assadullahi:2009jc,Espinosa:2018eve,Kohri:2018awv}. 
In contrast, we should carefully study contributions of non-Gaussianity to $\bar{\Omega}_{\uGW}$ when $\zeta$ is non-Gaussian. 
See Ref.~\cite{Adshead:2021hnm} for the complete analysis of local-type non-Gaussianity and Ref.~\cite{Garcia-Saenz:2022tzu} for a general analysis for any type of trispectrum shape, while other related studies can be found in Refs.~\cite{Nakama:2016gzw,Cai:2018dig,Unal:2018yaa,Yuan:2020iwf,Atal:2021jyo,Ragavendra:2021qdu}.


In this work, we focus on the local-type primordial non-Gaussianity of curvature perturbations. 
It can be expressed as \cite{Komatsu:2001rj} 
 \begin{equation}\label{eq:fnl-def}
     \zeta (\bx) = \zeta_g (\bx) + \frac{3}{5}\fnl \left[ \zeta_g^2(\bx) - \langle \zeta_g^{2}(\bx) \rangle \right]\ ,
 \end{equation}
where $\zeta_{g}$ stands for the Gaussian curvature perturbations, and $\fnl$ denotes the local-type non-Gaussian parameter. 
It can be transformed to Fourier space
\begin{equation}\label{eq:fnl-def-k}
    \zeta (\bq) = \zeta_g(\bq) + \frac{3}{5}\fnl \int \frac{\ud^3 \bk}{(2\pi)^{3/2}} \zeta_g(\bk) \zeta_g(\bq-\bk)\ ,
\end{equation}
where a delta-function term has been dropped. 
The statistics of $\zeta$ can be quantified by $\fnl$ and the power spectrum of $\zeta_{g}$.
In Fourier space, the latter is defined as  
\begin{equation}\label{eq:Pg-def}
    \langle \zeta_g (\bk) \zeta_g (\bk') \rangle 
    = \delta^{(3)} (\bk+\bk') P_g (k)\ ,
\end{equation}
for which the dimensionless power spectrum is given as $\Delta^2_g (k)=[k^3/(2 \pi^2)] P_g (k)$.

\subsection{Energy-density fraction spectrum}\label{sec:Omega}

\begin{figure}
    \centering
    \includegraphics[width =1. \columnwidth]{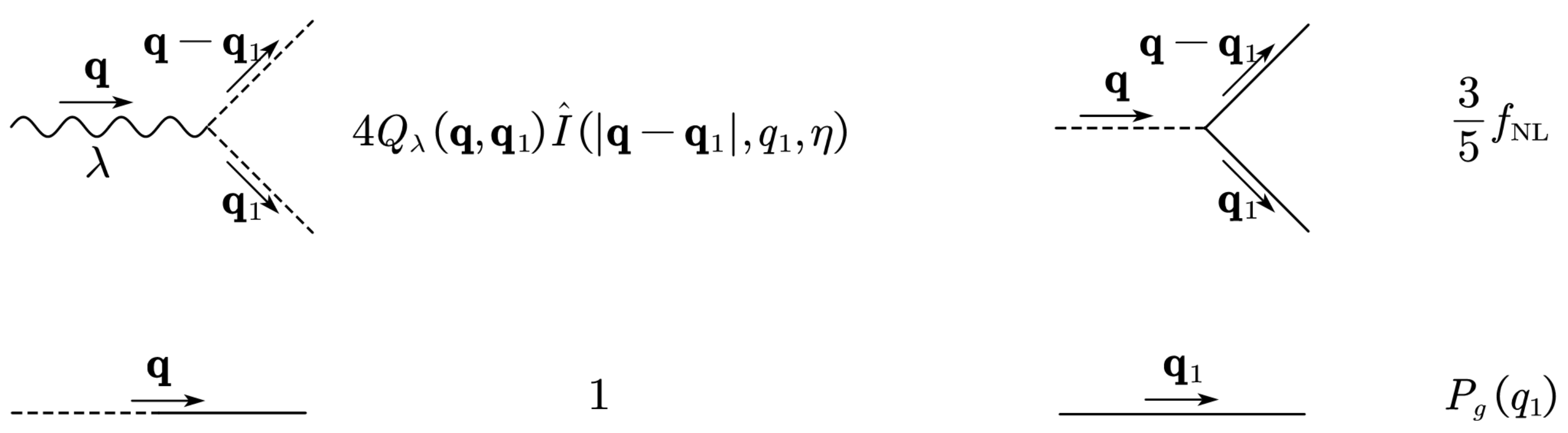}
    \caption{Feynman-like rules for the evaluation of \acp{SIGW}. Wavy lines denote \acp{GW}, dashes lines represent the transfer functions, and solid lines stand for the primordial curvature power spectra. Analogous to regular Feynman-like rules, the comoving 3-momenta flow along the directions of arrows. The 3-momentum is conserved at each vertex and the total 3-momentum is zero for each diagram. All loop 3-momenta should be integrated over. }
    \label{fig:F_Rules}
\end{figure}

In this subsection, we reproduce the theoretical results of Ref.~\cite{Adshead:2021hnm}. 
By substituting \cref{eq:fnl-def} into Eq.~(\ref{eq:Ph-zeta}), we decompose  $\langle\zeta^4\rangle$ into a four-point correlator $\langle\zeta_g^4\rangle$ at $\cO (\fnl^0)$ order, a six-point correlator $\langle\zeta_g^6\rangle$ at $\cO (\fnl^2)$ order, and an eight-point correlator $\langle\zeta_g^8\rangle$ at $\cO (\fnl^4)$ order. 
Based on the Wick's theorem, each correlator can be expressed in terms of two-point correlator $\langle\zeta_g^2\rangle$. 
The Feynman-like diagrams are useful to evaluation of these contractions. 
Therefore, the Feynman-like rules are explicitly shown in \cref{fig:F_Rules}.

\begin{figure}
    \includegraphics[width =0.5 \columnwidth]{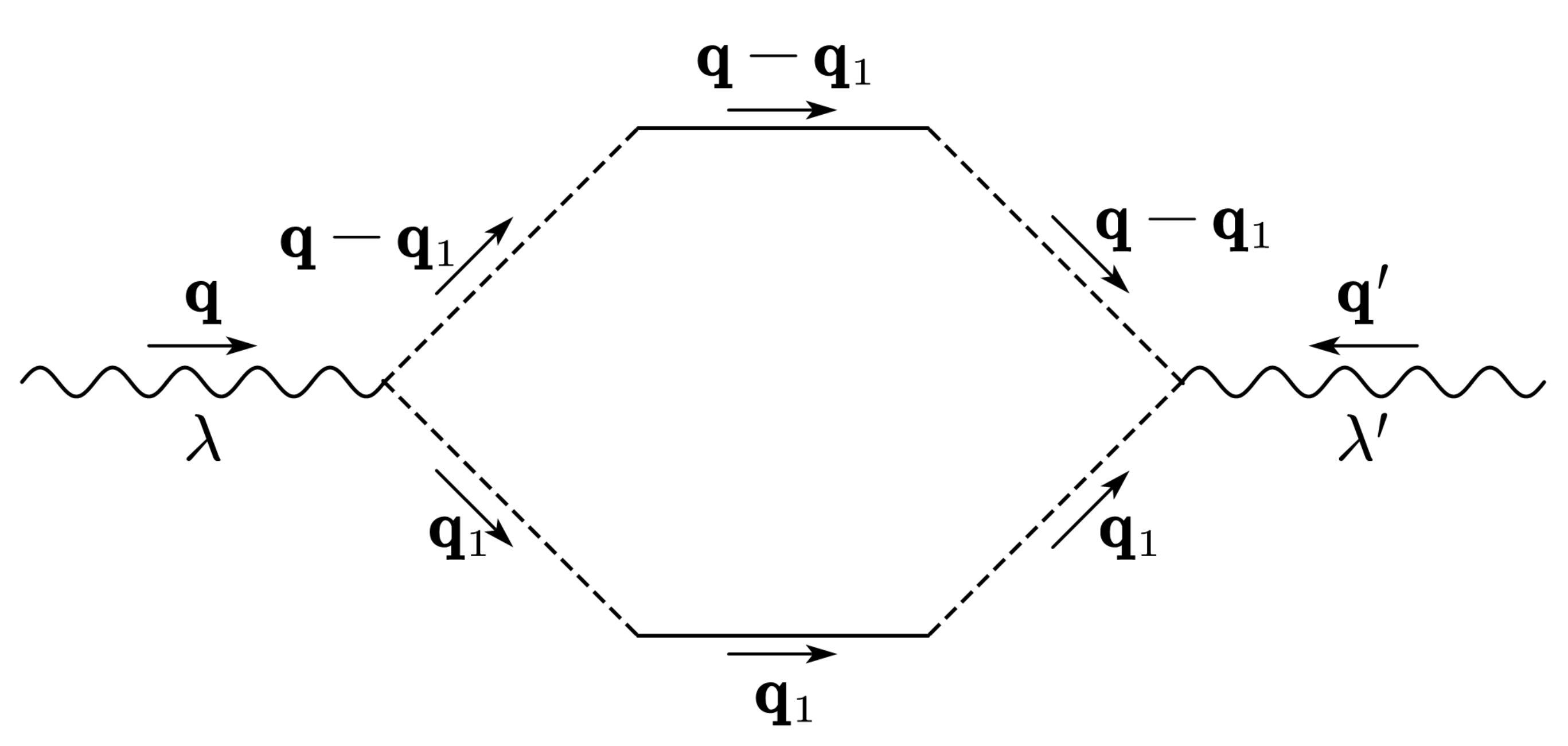}
    \includegraphics[width =0.5 \columnwidth]{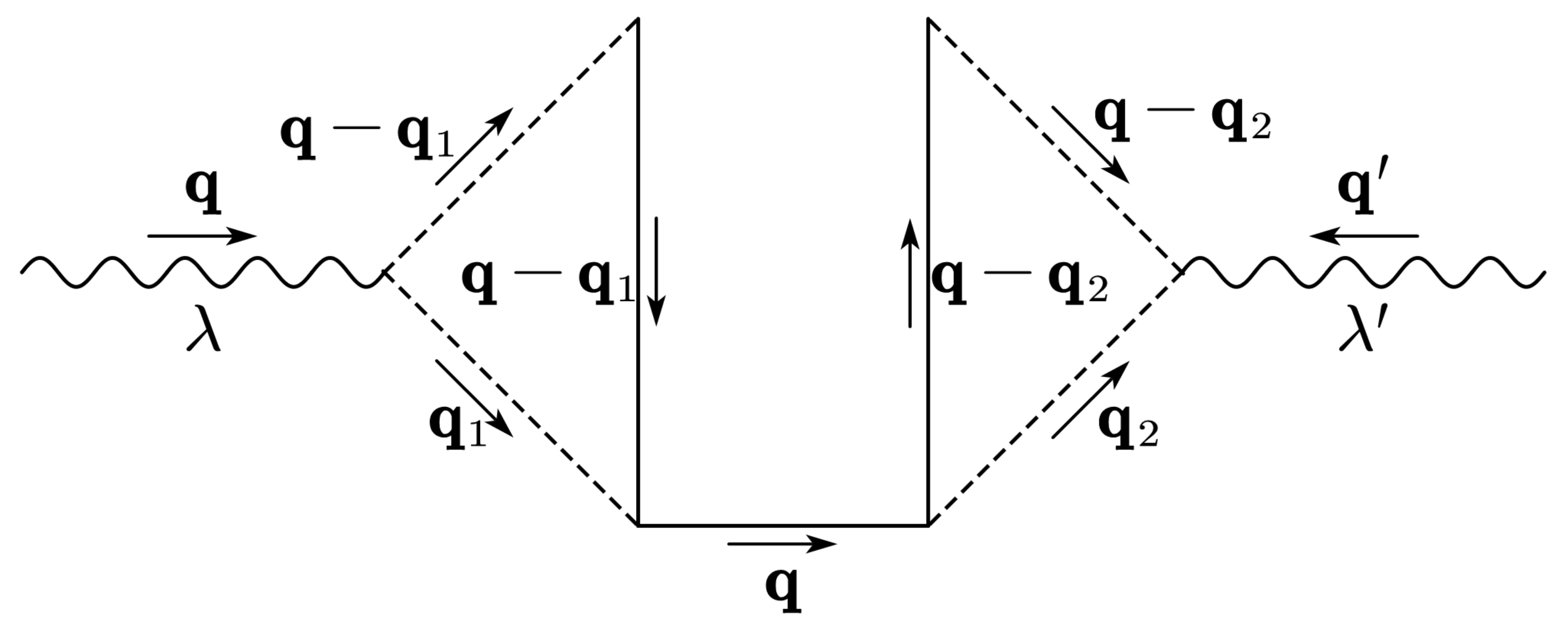}
    \caption{Left panel: The Feynman-like diagram at $\cO (\fnl^0)$ order. It is labeled as $G$.
    Right panel: The Feynman-like diagram that vanishes due to azimuthal angle in the integrand.}\label{fig:EDS-G}
\end{figure}

Based on the Feynman-like rules, we can represent the power spectrum in \cref{eq:h-cor} with the Feynman-like diagrams. 
However, disconnected diagrams that lead to the momenta of \acp{GW} $h_\lambda$ being zero violate the definition of the power spectrum in \cref{eq:h-cor}. 
They should be disregarded. 
Here, note that the meaning of ``disconnected diagram'' is different from that in Ref.~\cite{Adshead:2021hnm}. 
If diagrams have vertices in the top right panel of \cref{fig:F_Rules} with the solid lines being connected to form a loop, these diagrams will cease to exist, due to the definition in \cref{eq:fnl-def}. 
In addition, the contraction corresponded to the Feynman-like diagram in the right panel of \cref{fig:EDS-G} also vanishes, due to an azimuthal angle in the integrand.

\begin{figure}
    \includegraphics[width =0.5 \columnwidth]{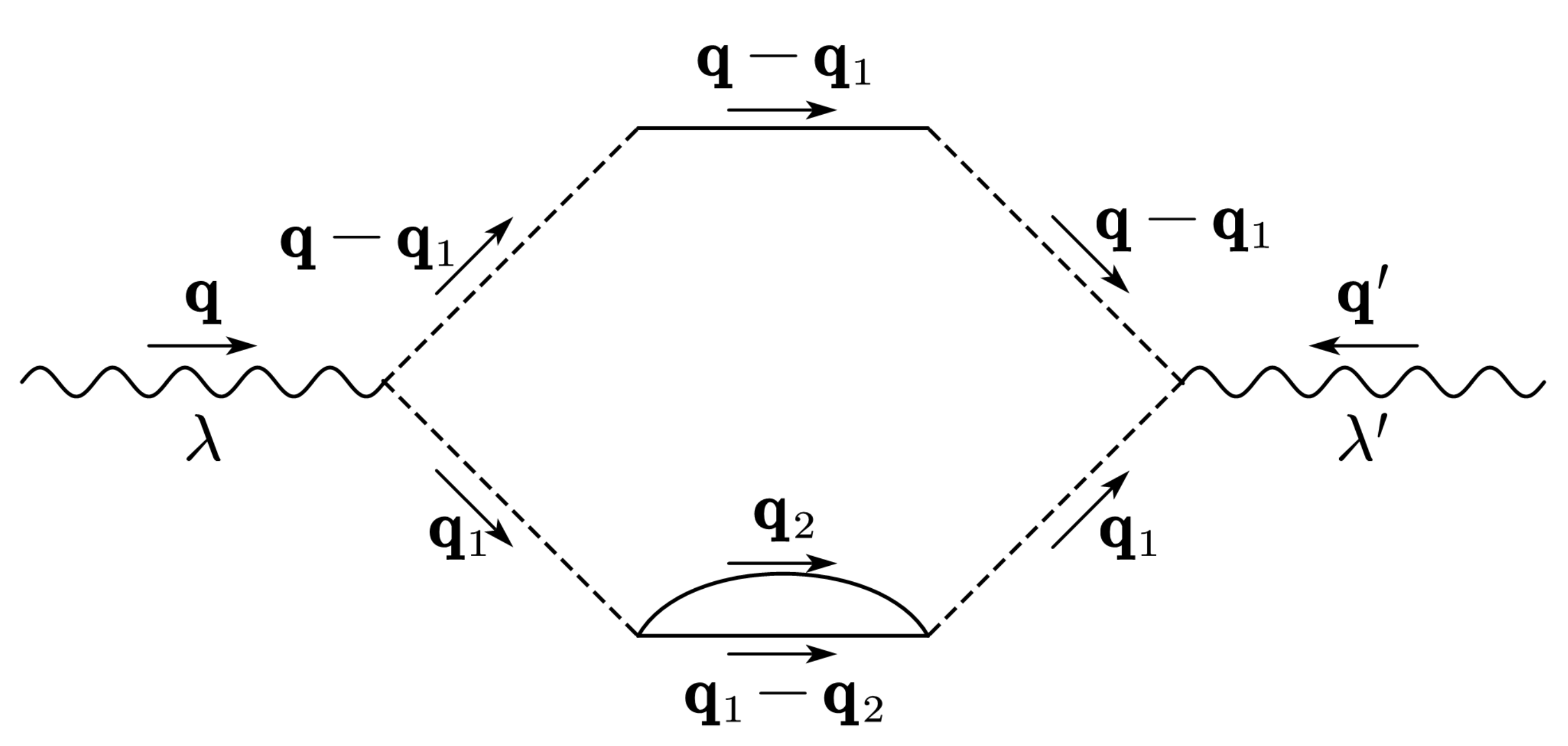}
    \includegraphics[width =0.5 \columnwidth]{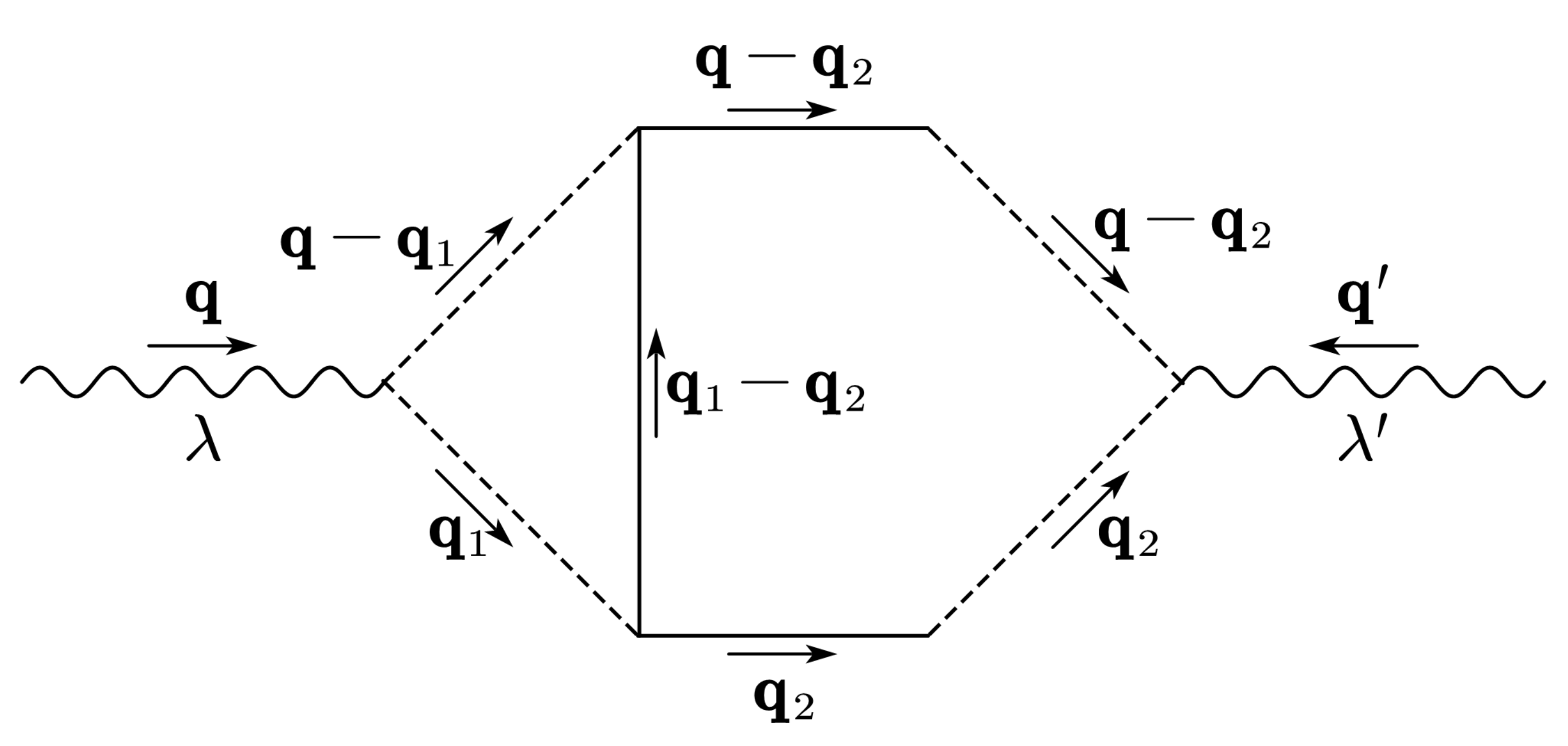}\\\\\\
    \includegraphics[width =0.5 \columnwidth]{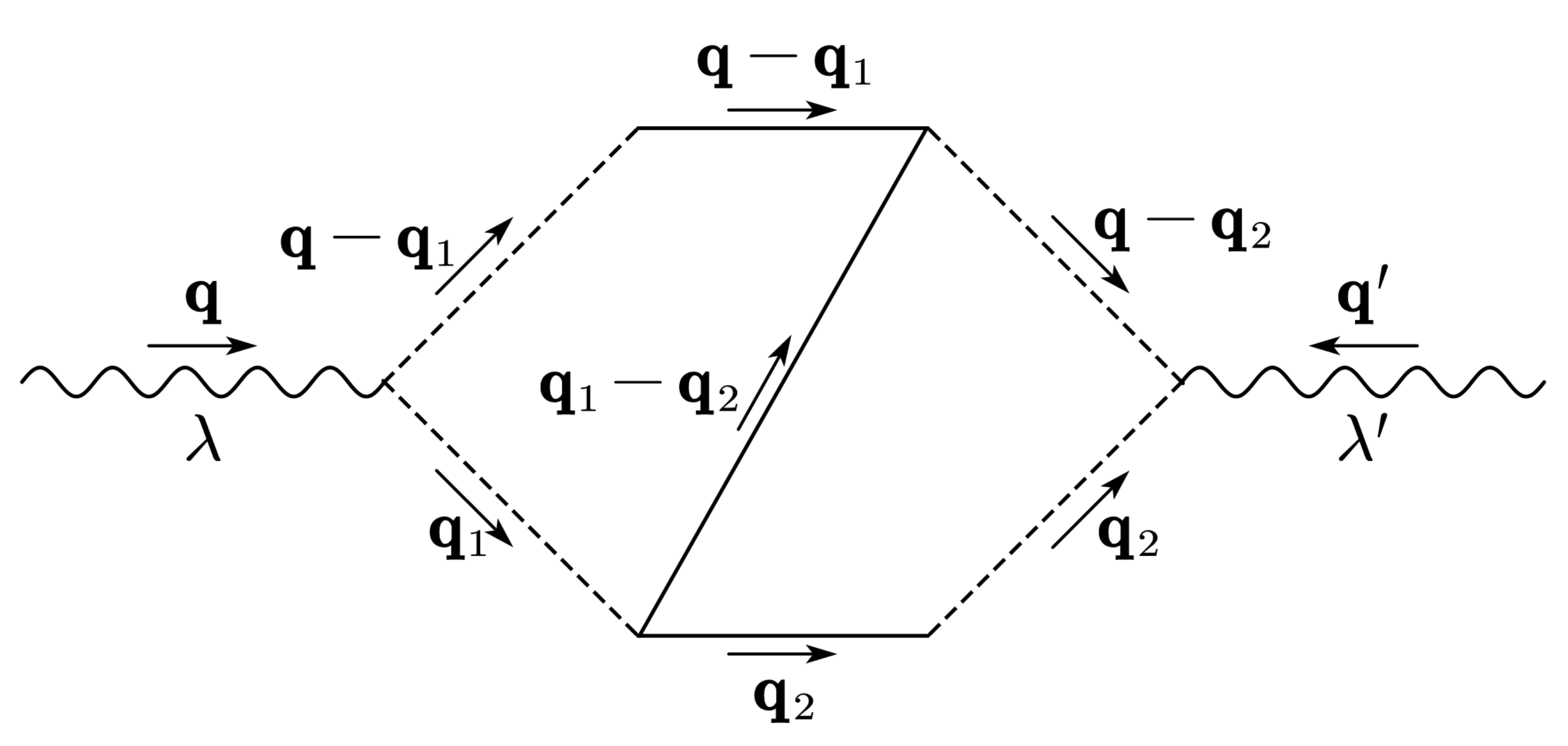}
    \includegraphics[width =0.5 \columnwidth]{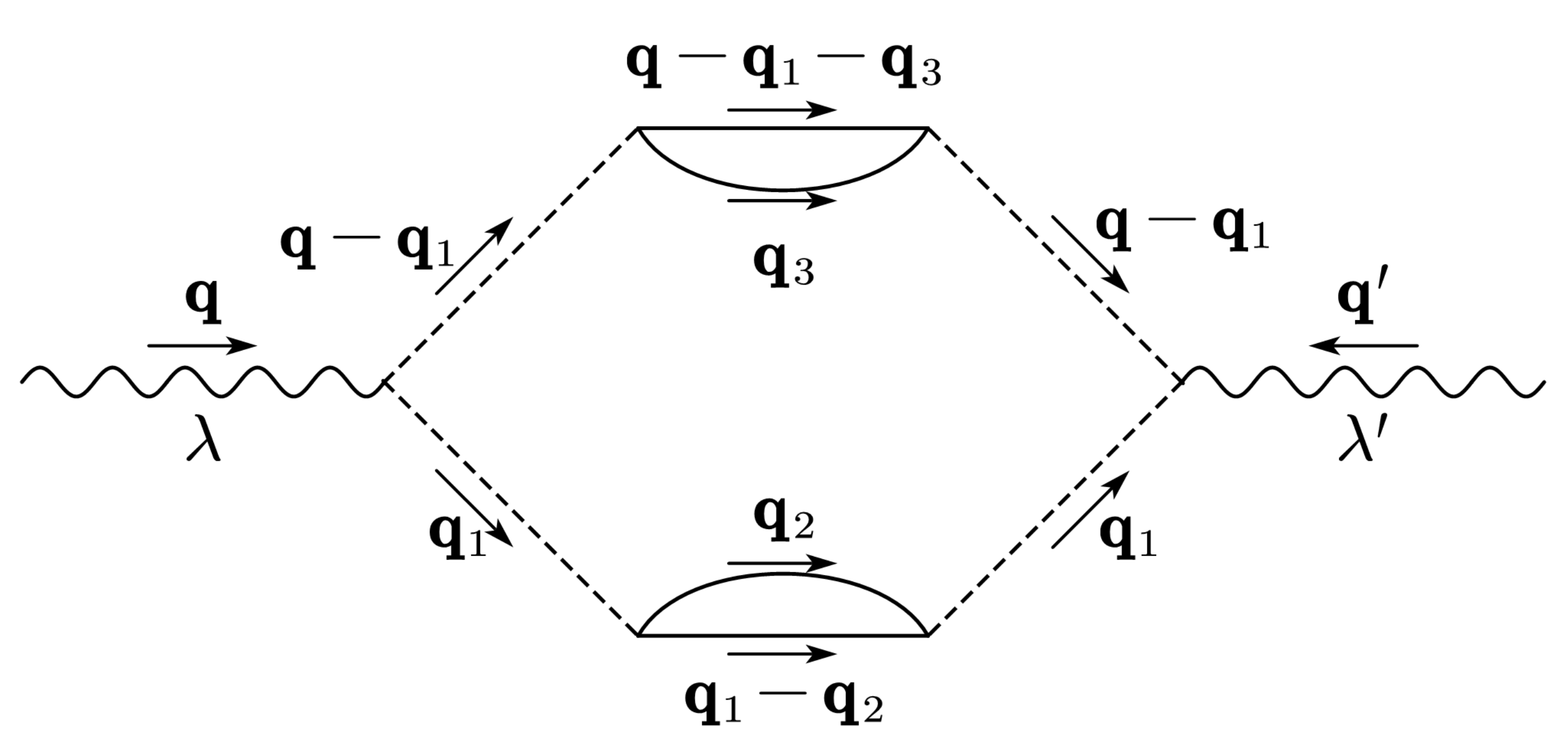}\\\\\\
    \includegraphics[width =0.5 \columnwidth]{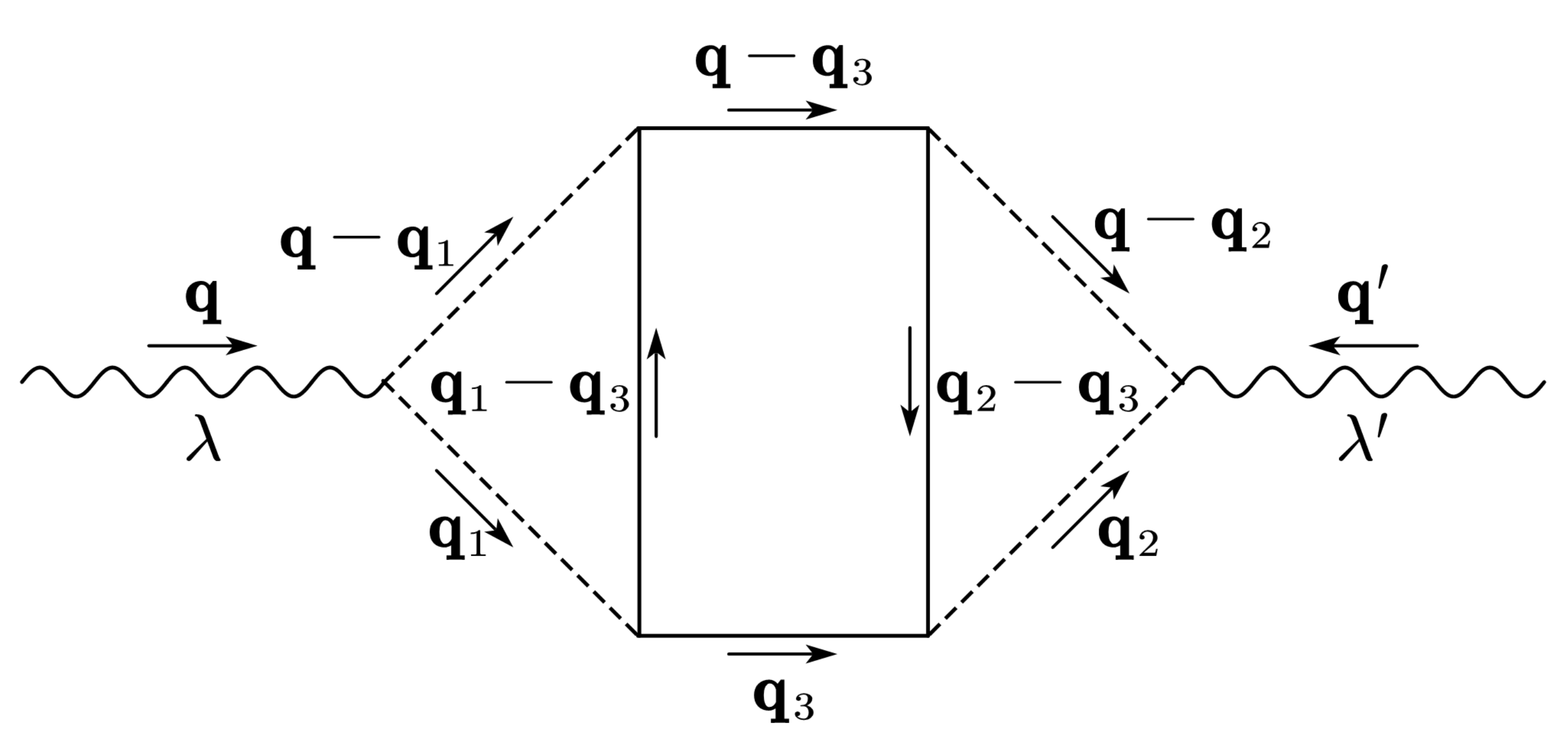}
    \includegraphics[width =0.5 \columnwidth]{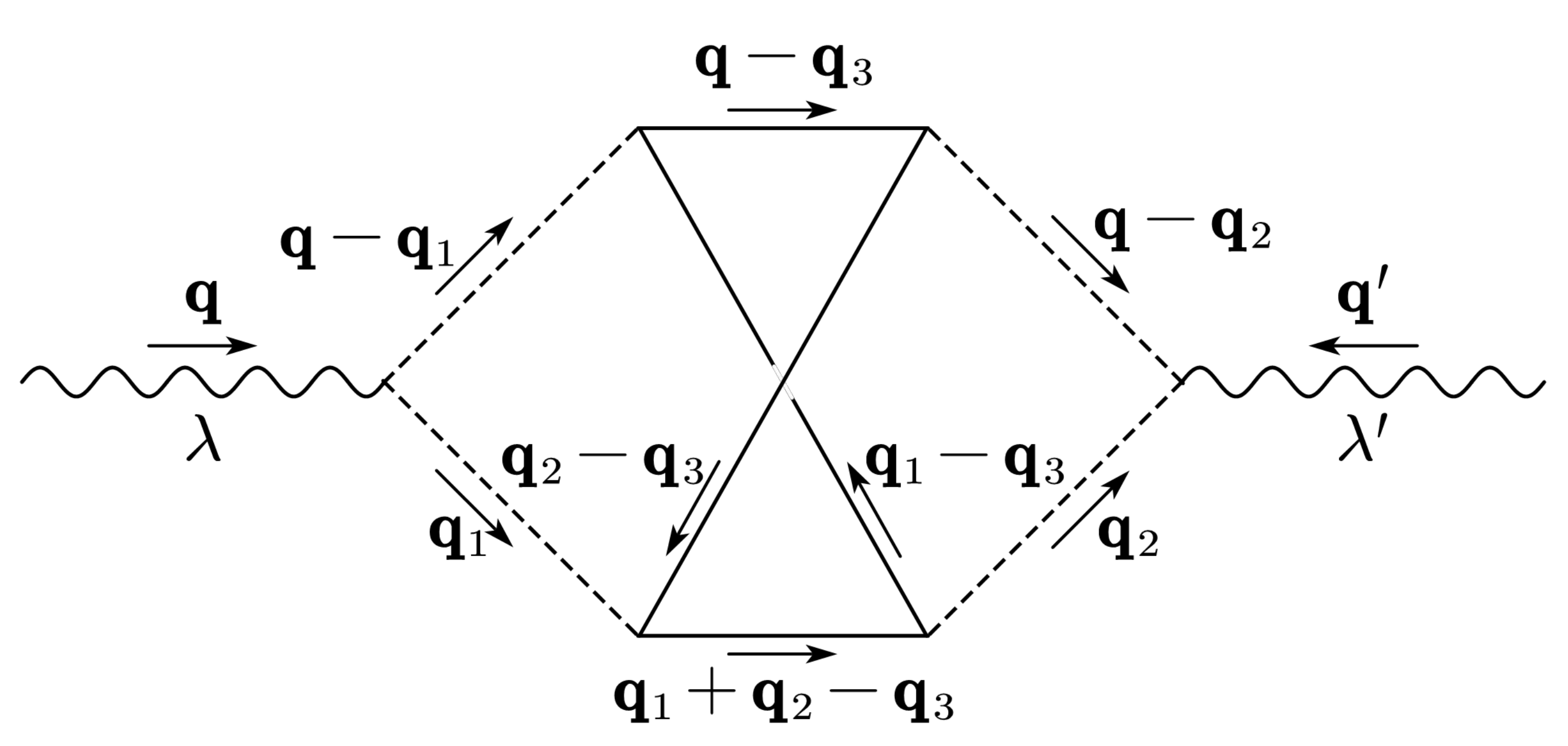}
    \caption{Feynman-like diagrams at $\cO(\fnl^{2})$ and $\cO(\fnl^{4})$ orders. They are specified as $H$ (Hybrid, top left), $C$ (top right), $Z$ (middle left), $R$ (Reducible, middle right), $P$ (Planar, bottom left), and $N$ (Non-Planar, bottom right) diagrams.}\label{fig:EDS-other}
\end{figure}

Therefore, we obtain seven Feynman-like diagrams that are related to nonvanishing contractions. 
They are depicted in the left panel of \cref{fig:EDS-G} and 
the panels of \cref{fig:EDS-other}. 
Their contributions to the power spectrum in \cref{eq:h-cor} can be obtained straightforwardly from the corresponding Feynman-like diagrams, as will be done in the following. 
Firstly, at $\cO(\fnl^0)$ order, the contribution labeled by $G$ is corresponded to the left panel of \cref{fig:EDS-G}, namely,  
\begin{equation}\label{eq:powerspectrum-zeta-0th} 
    P_{h_\lambda}^{G} (\eta,q) 
    = 2^5 \int \frac{\ud^3 \bq_1}{(2\pi)^{3}} Q_{\lambda}^2 (\bq, \bq_1) 
        \hat{I}^2 (\abs{\bq - \bq_1}, q_1, \eta) 
        P_g(q_1) P_g(\abs{\bq-\bq_1})\ .  
\end{equation}
It is exactly the result when $\zeta$ is Gaussian, as was studied in the literature \cite{Ananda:2006af,Baumann:2007zm,Mollerach:2003nq,Assadullahi:2009jc,Espinosa:2018eve,Kohri:2018awv}. 
In contrast, all of the non-Gaussian contributions are plotted in \cref{fig:EDS-other}. 
Secondly, at $\cO(\fnl^2)$ order \footnote{More exactly, it should be a combination of the form $\mathcal{O}(A \fnl^{2})$, where $A$ denotes the spectral amplitude in \cref{eq:Lognormal}. }, they are labeled by $H$, $C$, and $Z$, and can be expressed as follows 
\begin{eqnarray}\label{eqs:powerspectrum-zeta-2nd} 
    P_{h_\lambda}^{H} (\eta,q) 
    &=& 2^7 \left(\frac{3}{5}\fnl\right)^2 \int \frac{\ud^3 \bq_1}{(2\pi)^{3}} 
        \frac{\ud^3 \bq_2}{(2\pi)^{3}} 
        P_g(q_2) P_g(\abs{\bq-\bq_1})P_g(\abs{\bq_1-\bq_2}) \nonumber\\
        &&\hphantom{\  2^8 \left(\frac{3}{5}\fnl\right)^2 } 
        \times Q_{\lambda}^2(\bq, \bq_1) \hat{I}^2 (\abs{\bq - \bq_1}, q_1, \eta) \ , \label{eq:powerspectrum-zeta-H}\\  
    P_{h_\lambda}^{C} (\eta,q) 
    &=& 2^8 \left(\frac{3}{5}\fnl\right)^2 
        \int \frac{\ud^3 \bq_1}{(2\pi)^{3}} \frac{\ud^3 \bq_2}{(2\pi)^{3}}
        P_g(q_2) P_g(\abs{\bq-\bq_2}) P_g(\abs{\bq_1-\bq_2}) \nonumber\\
        &&\hphantom{\  2^8 \left(\frac{3}{5}\fnl\right)^2 } 
        \times Q_{\lambda}(\bq, \bq_1) \hat{I} (\abs{\bq - \bq_1}, q_1, \eta) 
        Q_{\lambda}(\bq, \bq_2) \hat{I} (\abs{\bq - \bq_2}, q_2, \eta) \ , \label{eq:powerspectrum-zeta-C}\\ 
    P_{h_\lambda}^{Z} (\eta,q) 
    &=& 2^8 \left(\frac{3}{5}\fnl\right)^2 
        \int \frac{\ud^3 \bq_1}{(2\pi)^{3}} \frac{\ud^3 \bq_2}{(2\pi)^{3}}
        P_g(q_2) P_g(\abs{\bq-\bq_1}) P_g(\abs{\bq_1-\bq_2}) \nonumber\\
        &&\hphantom{\  2^8 \left(\frac{3}{5}\fnl\right)^2 } 
        \times Q_{\lambda}(\bq, \bq_1) \hat{I} (\abs{\bq - \bq_1}, q_1, \eta) 
        Q_{\lambda}(\bq, \bq_2) \hat{I} (\abs{\bq - \bq_2}, q_2, \eta) \ .\label{eq:powerspectrum-zeta-Z} 
\end{eqnarray}
Thirdly, at $\cO(\fnl^4)$ order, they are labeled by $R$, $P$, and $N$, and can be expressed as follows 
\begin{eqnarray}\label{eqs:powerspectrum-zeta-4th} 
    P_{h_\lambda}^{R} (\eta,q) 
    &=& 2^7 \left(\frac{3}{5}\fnl\right)^4 
        \int \frac{\ud^3 \bq_1}{(2\pi)^{3}} \frac{\ud^3 \bq_2}{(2\pi)^{3}} \frac{\ud^3 \bq_3}{(2\pi)^{3}} 
        P_g(q_2) P_g(q_3) P_g(\abs{\bq_1-\bq_2}) \\
        &&\hphantom{\ 2^7 \left(\frac{3}{5}\fnl\right)^4  }
        \times P_g(\abs{\bq-\bq_1-\bq_3}) Q_{\lambda}^2 (\bq, \bq_1) \hat{I}^2 (\abs{\bq - \bq_1}, q_1, \eta) \ , \label{eq:powerspectrum-zeta-R}\nonumber\\ 
    P_{h_\lambda}^{P} (\eta,q) 
    &=& 2^9 \left(\frac{3}{5}\fnl\right)^4 
        \int \frac{\ud^3 \bq_1}{(2\pi)^{3}} \frac{\ud^3 \bq_2}{(2\pi)^{3}} \frac{\ud^3 \bq_3}{(2\pi)^{3}}
        P_g(q_3) P_g(\abs{\bq-\bq_3}) P_g(\abs{\bq_1-\bq_3}) \nonumber\\
        &&\hphantom{\ 2^7 \left(\frac{3}{5}\fnl\right)^4  }
        \times P_g(\abs{\bq_2-\bq_3}) 
        Q_{\lambda}(\bq, \bq_1) \hat{I} (\abs{\bq - \bq_1}, q_1, \eta) \label{eq:powerspectrum-zeta-P}\\
        &&\hphantom{\ 2^7 \left(\frac{3}{5}\fnl\right)^4  }
        \times Q_{\lambda}(\bq, \bq_2) \hat{I} (\abs{\bq - \bq_2}, q_2, \eta) \ , \nonumber\\ 
    P_{h_\lambda}^{N} (\eta,q) 
    &=& 2^8 \left(\frac{3}{5}\fnl\right)^4 
        \int \frac{\ud^3 \bq_1}{(2\pi)^{3}} \frac{\ud^3 \bq_2}{(2\pi)^{3}} \frac{\ud^3 \bq_3}{(2\pi)^{3}} 
        P_g(\abs{\bq-\bq_3}) P_g(\abs{\bq_1-\bq_3}) \nonumber\\
        &&\hphantom{\ 2^7 \left(\frac{3}{5}\fnl\right)^4  }
        \times P_g(\abs{\bq_1+\bq_2-\bq_3}) P_g(\abs{\bq_2-\bq_3}) Q_{\lambda}(\bq, \bq_1) \label{eq:powerspectrum-zeta-N} \\
        &&\hphantom{\ 2^7 \left(\frac{3}{5}\fnl\right)^4  } 
        \times \hat{I} (\abs{\bq - \bq_1}, q_1, \eta) 
        Q_{\lambda}(\bq, \bq_2) \hat{I} (\abs{\bq - \bq_2}, q_2, \eta) \ .\nonumber  
\end{eqnarray}
Here, we have taken into account the symmetry factor for each Feynman-like diagram. 
Following \cref{eq:Omegabar-h}, for each Feynman-like diagram, we can determine its contribution to $\bar{\Omega}_\uGW(\eta,q)$, which is labeled by the same superscript as the one labelling the Feynman-like diagram. 

\begin{figure}
    \includegraphics[width =.9 \columnwidth]{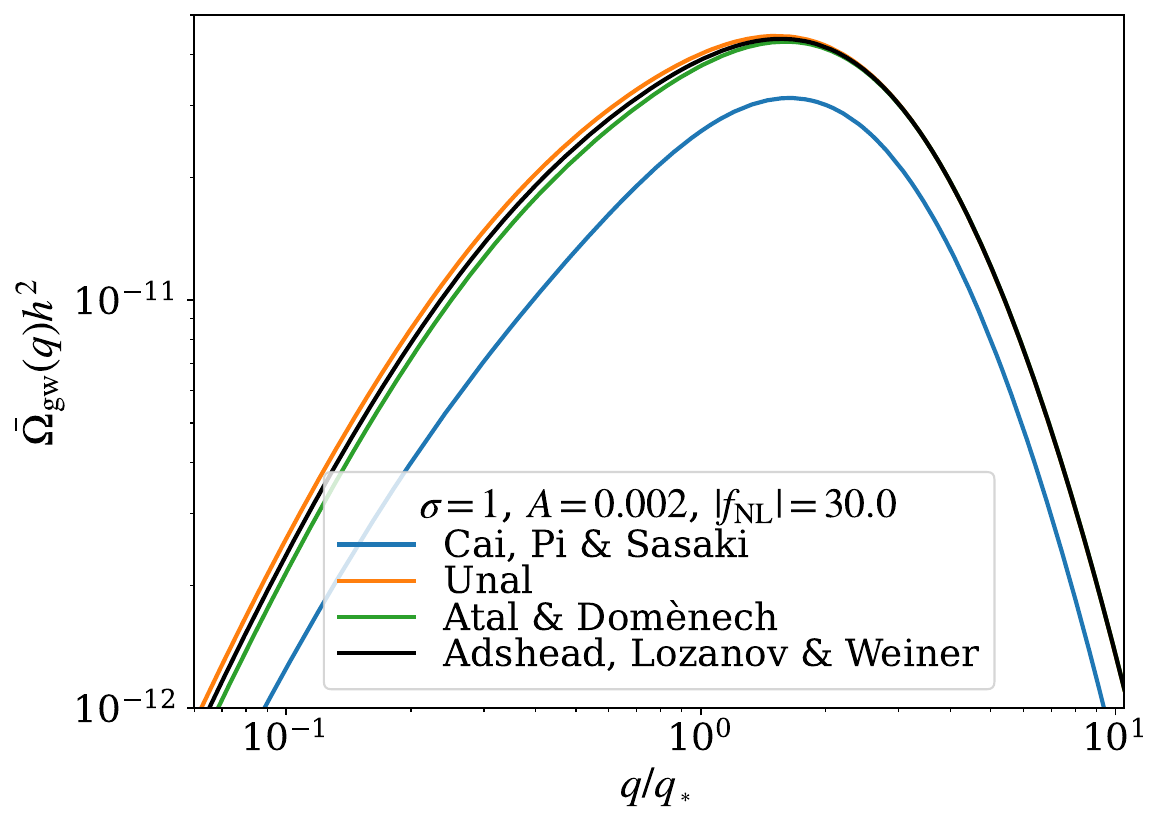}
    \caption{Comparison of the energy-density fraction spectra of \ac{SIGW} provided by Ref.~\cite{Cai:2018dig} (blue curve), Ref.~\cite{Unal:2018yaa} (orange curve), Ref.~\cite{Atal:2021jyo} (green curve), and Ref.~\cite{Adshead:2021hnm} (black curve). Here, we reproduce the complete results analyzed by Ref.~\cite{Adshead:2021hnm}. }
    \label{fig:kk}
\end{figure}

The study of \acp{GW} induced by the primordial scalar perturbations with local-type non-Gaussianity was first conducted in Ref.~\cite{Cai:2018dig}. The authors considered only the contributions labeled by $G$, $H$, and $R$. 
Subsequently, other contributions, except the one labeled by $Z$, were investigated in Ref.~\cite{Unal:2018yaa}. 
The contribution labeled by $Z$ was first computed in Ref.~\cite{Atal:2021jyo}. 
It is worth noting that the contribution labeled by $C$ is referred to as ``walnut'' in Ref.~\cite{Unal:2018yaa}, while ``walnut'' is used to denote the one labeled by $Z$ in Ref.~\cite{Atal:2021jyo}. 
The first complete analysis and the Feynman-like rules and diagrams have been provided by Ref.~\cite{Adshead:2021hnm}, which is followed by our current work. 
Corresponding to the above papers, we compare their results in \cref{fig:kk}, which will be numerically reproduced in the next subsection. 
In addition, the scale-dependent non-Gaussianity was studied in Ref.~\cite{Ragavendra:2021qdu}.

Based on \cref{eq:Omegabar-h}, we straightforwardly obtain contributions from the above seven integrals to the energy-density fraction spectrum. 
Therefore, the total spectrum is determined by a sum of them, i.e.,  
\begin{equation}\label{eq:Omegabar-total}
    \bar{\Omega}_\uGW (\eta,q) 
    = \bar{\Omega}_\uGW^{G}  + \bar{\Omega}_\uGW^{H}  + \bar{\Omega}_\uGW^{C}  + \bar{\Omega}_\uGW^{Z}  + \bar{\Omega}_\uGW^{R}  + \bar{\Omega}_\uGW^{P}  + \bar{\Omega}_\uGW^{N} \ .
\end{equation}
Based on the above derivations, it is obvious that each contribution to the total spectrum $\bar{\Omega}_{\uGW}(\eta,q)$ does not explicitly contain $\eta$ in the limit of $x\gg 1$. 
Therefore, $n_\uGW(\eta,q)$ defined in \cref{eq:ngw-def} is independent of $\eta$. 
In light of the tensor transfer function, the spectrum at current time $\eta_0$ had been presented, e.g., in Ref.~\cite{Wang:2019kaf}.  
It is given as 
\begin{eqnarray}\label{eq:ogwetaa0qsai}
    \bar{\Omega}_{\uGW} (q) 
    & = & \Omega_{\mathrm{rad}, 0} 
        \left(\frac{g_{*,\rho, \mathrm{e}}}{g_{*,\rho, 0}} \right)
        \left(\frac{g_{*,s, 0}}{g_{*,s, \mathrm{e}}} \right)^{4/3} \bar{\Omega}_\uGW (\eta, q) \ ,
\end{eqnarray}
where the subscripts $_0$ and $_\mathrm{e}$ label the present time and the emission time, respectively. 
The energy-density fraction of radiations today is  $\Omega_{\mathrm{rad}, 0} = 4.2 \times 10^{-5} h^{-2}$, where $h=0.6736$ is the dimensionless Hubble constant \cite{Planck:2018vyg}. 
The effective numbers of relativistic species, i.e., $g_{\ast,\rho}$ and $g_{\ast,s}$, can be obtained from the tabulated data shown in Ref.~\cite{Saikawa:2018rcs}. 

\subsection{Numerical results}\label{sec:EDS-Result}

In order to calculate the above seven integrals numerically, it is convenient to classify them into two categories. 
The first category contains the integrals labeled by $G$, $H$, and $R$, while the second one contains those labeled by $C$, $Z$, $P$, and $N$. 

For the first category, we introduce three sets of new variables $(u_i,v_i)$, namely, 
\begin{subequations}\label{eq:uv1-def}
\begin{eqnarray}
    v_1 &=& \frac{q_1}{q}\ ,\qquad\qquad\qquad
    u_1 = \frac{\abs{\bq-\bq_1}}{q}\ ,\label{eq:uv1-def1}\\
    v_2 &=& \frac{q_2}{q_1}\ , \qquad\qquad\qquad
    u_2 = \frac{\abs{\bq_1-\bq_2}}{q_1}\ , \\
    v_3 &=& \frac{q_3}{\abs{\bq-\bq_1}}\ ,\qquad\quad\ \,
    u_3 = \frac{\abs{\bq-\bq_1-\bq_3}}{\abs{\bq-\bq_1}}\ , 
\end{eqnarray}
\end{subequations}
During \ac{RD} epoch, we have $\hat{I}(\abs{\bq - \bq_1}, q_1, \eta) = I_{\uRD} (u_1,v_1,x)$ with $x=q\eta$, as was shown in \cref{eq:irdsai}. 
For simplification, we introduce a new quantity 
\begin{equation}\label{eq:J-def}
    J (u_1,v_1,x) 
    = \frac{x}{8}\bigl[(v_1+u_1)^2-1\bigr] \bigl[1-(v_1-u_1)^2\bigr] I_{\uRD} (u_1,v_1,x)\ , 
\end{equation}
where the explicit expression of $I_{\uRD}(u_1,v_1,x)$ when $x\rightarrow\infty$ was shown in \cref{eq:I-RD}. 
After explicit computation, we get  
\begin{equation}
J^2 (u_1,v_1,x)=x^2 \sum_\lambda Q_{\lambda}^2 (\bq, \bq_1) \hat{I}^2 (\abs{\bq - \bq_1}, q_1, \eta)\ .
\end{equation}
Due to the oscillation average, we have 
\begin{equation}\label{eq:J-ave-11}
    \overbar{J^2 (u_1,v_1,x\rightarrow\infty)} 
    = \frac{x^2}{64}\bigl[(v_1+u_1)^2-1\bigr]^2 \bigl[1-(v_1-u_1)^2\bigr]^2 
        \overbar{I_\uRD^2 (u_1,v_1,x\rightarrow\infty)}\ , 
\end{equation}
which can be obtained from Eq.~(\ref{eq:I-ave-12}). 
In fact, \cref{eq:J-ave-11} is independent of $x$. 
To transform the integration region into a rectangle, we define the transformation of variables as follows 
\begin{equation}\label{eq:transvarsai}
    s_i = u_i - v_i\ ,\qquad\qquad\quad
    t_i = u_i + v_i -1\ .
\end{equation}
After considering the Jacobian, we obtain the contributions to $\bar{\Omega}_\uGW(\eta,q)$ from the integrals labeled as $G$, $H$ and $R$, respectively. They are given as   
\begin{eqnarray}
    \bar{\Omega}_\uGW^G (\eta, q) 
    &=& \frac{1}{3} \int_0^\infty \ud t_1 \int_{-1}^1 \ud s_1 
        \overbar{J^2 (u_1,v_1,x\rightarrow\infty)} \frac{1}{(u_1 v_1)^2} 
        \Delta^2_g (v_1 q) \Delta^2_g (u_1 q) \ ,\label{eq:Omega-G}\\ 
    \bar{\Omega}_\uGW^H (\eta, q) 
    &=& \frac{1}{3} \left(\frac{3\fnl}{5}\right)^2 
        \prod_{i=1}^2 \biggl[\int_0^\infty \ud t_i \int_{-1}^1 \ud s_i\biggr] 
        \overbar{J^2 (u_1,v_1,x\rightarrow\infty)}\frac{1}{(u_1 v_1 u_2 v_2)^2}\nonumber\\ 
    &&\hphantom{\left(\frac{3\fnl}{5}\right)^2 \frac{1}{12} 
        \biggl[\int_0^\infty \biggr]}
        \times \Delta^2_g (v_1 v_2 q) \Delta^2_g (u_1 q) \Delta^2_g (v_1 u_2 q) \ ,\label{eq:Omega-H}\\ 
    \bar{\Omega}_\uGW^R (\eta, q) 
    &=& \frac{1}{12} \left(\frac{3\fnl}{5}\right)^4 
        \prod_{i=1}^3 \biggl[\int_0^\infty \ud t_i \int_{-1}^1 \ud s_i\biggr] 
        \overbar{J^2 (u_1,v_1,x\rightarrow\infty)}\frac{1}{(u_1 v_1 u_2 v_2 u_3 v_3)^2}\nonumber\\ 
    &&\hphantom{\left(\frac{3\fnl}{5}\right)^2 \frac{1}{12} 
        \biggl[\int_0^\infty \biggr]}
        \times \Delta^2_g (v_1 v_2 q) \Delta^2_g (v_1 u_2 q) \Delta^2_g (u_1 v_3 q) \Delta^2_g (u_1 u_3 q) \ .\label{eq:Omega-R}
\end{eqnarray}
The above three integrals can be numerically computed by the \texttt{vegas} \cite{Lepage:2020tgj} package. 

For the second category, we also introduce three sets of new variables, still labeled by $(u_i,v_i)$ with $i=1,2,3$, namely, 
\begin{equation}\label{eq:uv2-def}
    v_i = \frac{q_i}{q}\ ,\qquad\qquad\quad
    u_i = \frac{\abs{\bq-\bq_i}}{q}\ ,
\end{equation}
which are different from those in Eq.~(\ref{eq:uv1-def}). 
Since the definition of $(v_1,u_1)$ in \cref{eq:uv2-def} is the same as that in \cref{eq:uv1-def1}, the definition of $J(u_1,v_1,x)$ in \cref{eq:J-def} is still applicable for $(u_1,v_1)$ in \cref{eq:uv2-def}. 
Here, we further redefine it in a more general way, i.e., 
\begin{equation}\label{eq:J-def-general}
    J (u_i,v_i,x) 
    = \frac{x}{8}\bigl[(v_i+u_i)^2-1\bigr] \bigl[1-(v_i-u_i)^2\bigr] I_{\uRD} (u_i,v_i,x)\ , 
\end{equation}
where the explicit expression of $I_{\uRD}(u_i,v_i,x)$ when $x\rightarrow\infty$ was still shown in \cref{eq:I-RD}. 
After explicit computation, we get  
\begin{equation}
    J(u_1,v_1,x)J(u_2,v_2,x) \cos 2\varphi_{12} 
    = x^2 \sum_\lambda Q_{\lambda}(\bq, \bq_1) \hat{I} (\abs{\bq - \bq_1}, q_1, \eta) Q_{\lambda}(\bq, \bq_2) \hat{I} (\abs{\bq - \bq_2}, q_2, \eta)\ ,
\end{equation} 
where we denote $\varphi_{ij}=\phi_i-\phi_j$ for the sake of brevity, and the azimuthal angle $\phi_i$ has been defined in \cref{sec:motion}. 
The oscillation average of $J (u_1,v_1,x)J (u_2,v_2,x)$ can be obtained via Eq.~(\ref{eq:I-ave-12}), i.e., 
\begin{eqnarray}\label{eq:J-ave-12}
    \overbar{J (u_1,v_1,x\rightarrow\infty)J (u_2,v_2,x\rightarrow\infty)} 
    &=& \frac{x^2}{64} \bigl[(v_1+u_1)^2-1\bigr] \bigl[1-(v_1-u_1)^2\bigr]\nonumber\\
    && \times    \bigl[(v_2+u_2)^2-1\bigr] \bigl[1-(v_2-u_2)^2\bigr]\\ 
    && \times \overbar{I_\uRD (u_1,v_1,x\rightarrow\infty) I_\uRD (u_2,v_2,x\rightarrow\infty)} \ ,\nonumber 
\end{eqnarray}
which is also independent of $x$. 
The transformation from $(u_i,v_i)$ to $(s_i,t_i)$ remains the same as that in Eq.~(\ref{eq:transvarsai}). 
For simplification, we introduce a new quantity  
\begin{eqnarray}
    y_{ij} = \frac{\bq_i \cdot \bq_j}{q^2} 
        &=& \frac{\cos\varphi_{ij}}{4}\sqrt{t_i (t_i + 2) (1 - s_i^2) t_j (t_j + 2) (1 - s_j^2)} 
            \nonumber\\
        &&  + \frac{1}{4}[1 - s_i (t_i + 1)][1 - s_j (t_j + 1)]\ ,
\end{eqnarray}
and further define two new quantities as follows  
\begin{eqnarray}
    w_{ij} &=& \frac{\abs{\bq_i-\bq_j}}{q} =  \sqrt{v_i^2 + v_j^2 - y_{ij}}\ ,\\
    w_{123} &=& \frac{\abs{\bq_1+\bq_2-\bq_3}}{q} = \sqrt{v_1^2 + v_2^2 + v_3^2 + y_{12} - y_{13} - y_{23}}\ .
\end{eqnarray}
After considering the Jacobian, we obtain the contributions to $\bar{\Omega}_\uGW(\eta,q)$ from the integrals labeled as $C$, $Z$, $P$, and $N$, respectively. They are given as   
\begin{eqnarray}
    \bar{\Omega}_\uGW^{C} (\eta,q) 
    &=& \frac{1}{3\pi} \left(\frac{3\fnl}{5}\right)^2 
        \prod_{i=1}^2 \biggl[\int_0^\infty \ud t_i \int_{-1}^1 \ud s_i\, v_i u_i\biggr] 
        \int_0^{2\pi} \ud \varphi_{12}\, \cos 2\varphi_{12}  \nonumber\\
        &&\hphantom{\left(\frac{3\fnl}{5}\right)^2 \frac{1}{12\pi} \prod_{i=1}^2 }
        \times \overbar{J (u_1,v_1,x\rightarrow\infty)J (u_2,v_2,x\rightarrow\infty)} \label{eq:Omega-C}\\ 
        &&\hphantom{\left(\frac{3\fnl}{5}\right)^2 \frac{1}{12\pi} \prod_{i=1}^2 }
        \times \frac{\Delta^2_g (v_2 q)}{v_2^3} \frac{\Delta^2_g (u_2 q)}{u_2^3} \frac{\Delta^2_g (w_{12} q)}{w_{12}^3} 
            \ ,\nonumber\\
    \bar{\Omega}_\uGW^{Z} (\eta,q) 
    &=& \frac{1}{3\pi} \left(\frac{3\fnl}{5}\right)^2 
        \prod_{i=1}^2 \biggl[\int_0^\infty \ud t_i \int_{-1}^1 \ud s_i\, v_i u_i\biggr] 
        \int_0^{2\pi} \ud \varphi_{12}\, \cos 2\varphi_{12}  \nonumber\\
        &&\hphantom{\left(\frac{3\fnl}{5}\right)^2 \frac{1}{12\pi} \prod_{i=1}^2 }
        \times \overbar{J (u_1,v_1,x\rightarrow\infty)J (u_2,v_2,x\rightarrow\infty)} \label{eq:Omega-Z}\\
        &&\hphantom{\left(\frac{3\fnl}{5}\right)^2 \frac{1}{12\pi} \prod_{i=1}^2 }
        \times \frac{\Delta^2_g (v_2 q)}{v_2^3} \frac{\Delta^2_g (u_1 q)}{u_1^3} \frac{\Delta^2_g (w_{12} q)}{w_{12}^3} \ , \nonumber\\
    \bar{\Omega}_\uGW^{P} (\eta,q) 
    &=& \frac{1}{24\pi^2} \left(\frac{3\fnl}{5}\right)^4 
        \prod_{i=1}^3 \biggl[\int_0^\infty \ud t_i \int_{-1}^1 \ud s_i\, v_i u_i\biggr] 
        \int_0^{2\pi} \ud \varphi_{12}\ud \varphi_{23}\, \cos 2\varphi_{12} \nonumber\\ 
        &&\hphantom{\left(\frac{3\fnl}{5}\right)^4 \frac{1}{96\pi^2} \prod_{i=1}^3 }
        \times \overbar{J (u_1,v_1,x\rightarrow\infty) J (u_2,v_2,x\rightarrow\infty)} \label{eq:Omega-P}\\ 
        &&\hphantom{\left(\frac{3\fnl}{5}\right)^4 \frac{1}{96\pi^2} \prod_{i=1}^3 } 
        \times \frac{\Delta^2_g (v_3 q)}{v_3^3} \frac{\Delta^2_g (u_3 q)}{u_3^3} \frac{\Delta^2_g (w_{13} q)}{w_{13}^3} \frac{\Delta^2_g (w_{23} q)}{w_{23}^3} \ ,  \nonumber \\
    \bar{\Omega}_\uGW^{N} (\eta,q) 
    &=& \frac{1}{24\pi^2} \left(\frac{3\fnl}{5}\right)^4 
        \prod_{i=1}^3 \biggl[\int_0^\infty \ud t_i \int_{-1}^1 \ud s_i\, v_i u_i\biggr] 
        \int_0^{2\pi} \ud \varphi_{12}\ud \varphi_{23}\, \cos 2\varphi_{12} \nonumber\\ 
        &&\hphantom{\left(\frac{3\fnl}{5}\right)^4 \frac{1}{96\pi^2} \prod_{i=1}^3 }
        \times \overbar{J (u_1,v_1,x\rightarrow\infty) J (u_2,v_2,x\rightarrow\infty)} \label{eq:Omega-N}\\ 
        &&\hphantom{\left(\frac{3\fnl}{5}\right)^4 \frac{1}{96\pi^2} \prod_{i=1}^3 } 
        \times \frac{\Delta^2_g (u_3 q)}{u_3^3} \frac{\Delta^2_g (w_{13} q)}{w_{13}^3} 
            \frac{\Delta^2_g (w_{23} q)}{w_{23}^3}
            \frac{\Delta^2_g (w_{123} q)}{w_{123}^3} \ . \nonumber 
\end{eqnarray}
The above four integrals can also be numerically computed by the \texttt{vegas} \cite{Lepage:2020tgj} package.
We postulate that the dimensionless power spectrum of the Gaussian primordial curvature perturbations $\zeta_g$ is given by a normal function with respect to $\ln q$, i.e.,  
\begin{equation}\label{eq:Lognormal}
    \Delta^2_g (q) = \frac{A}{\sqrt{2\pi\sigma^2}}\exp\left(-\frac{\ln^2 (q/q_\ast)}{2 \sigma^2}\right)\ ,
\end{equation}
where $q_\ast$ denotes the spectral peak, $\sigma$ stands for the standard deviation, and $A$ is the the spectral amplitude at $q_\ast$. 
This power spectrum has been broadly used in the literature, e.g., Refs.~\cite{Pi:2020otn,Adshead:2021hnm,Zhao:2022kvz,Dimastrogiovanni:2022eir}. 

\begin{figure}
    \includegraphics[width =1. \columnwidth]{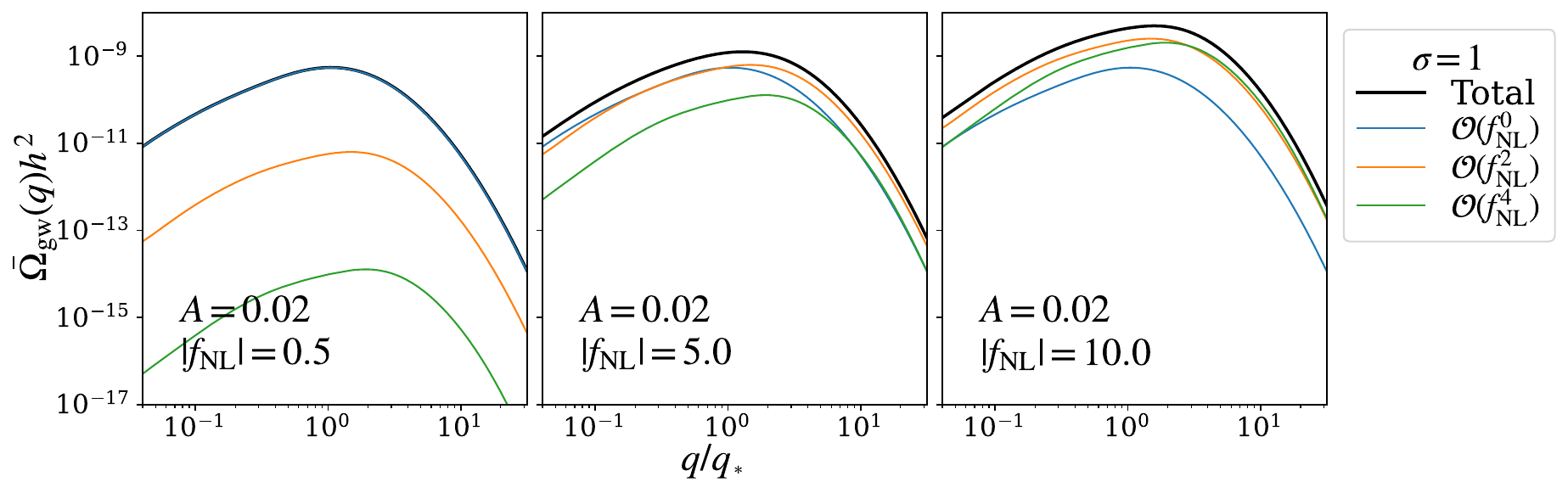}
    \caption{Energy-density fraction spectrum of \acp{SIGW} in the current universe. We let $A=0.02$, $\sigma = 1$, and $\fnl=0.5,~5.0,~10.0$ from left to right panels. The Gaussian contribution, which is of $\cO(\fnl^{0})$ order, is denoted by a blue line in each panel, while the non-Gaussian contributions of $\cO (\fnl^2)$ and $\cO(\fnl^4)$ orders are denoted by orange and green curves, respectively. The total spectra are shown as black curves. }\label{fig:Omega_order}
\end{figure}

In \cref{fig:Omega_order}, we show the contributions of primordial non-Gaussianity, which depend on powers of $\fnl^{2}$ (exactly speaking, powers of $A\fnl^2$), to the energy-density fraction spectrum $\bar{\Omega}_{\uGW,0}(q)$ in Eq.~(\ref{eq:ogwetaa0qsai}). 
Here, we let $A=0.02$, $\sigma=1$, but vary $|\fnl|$ via letting it to be $0.5$, $5.0$, and $10.0$ from the left to right panels. 
Throughout this paper, we manipulate $\fnl$ to insure that the non-Gaussian contribution to $\zeta$ in \cref{eq:fnl-def} lies in perturbative regime, i.e., $(3\fnl/5)^2 A < 1$. 
However, we would not require the constraints on $\fnl$ from \ac{CMB}, since we are discussing couplings between long-wavelength modes, that could be related to \ac{CMB}, and extremely-short-wavelength modes that are beyond the scope of \ac{CMB} observations. 
Therefore, the \ac{CMB} bounds are irrelevant to our current work. 
In addition, we depict the Gaussian contribution that is of $\cO(\fnl^{0})$ order, following Refs.~\cite{Espinosa:2018eve,Kohri:2018awv,Adshead:2021hnm}. 
It is denoted by blue curves in the panels. 
As were shown in Refs.~\cite{Garcia-Bellido:2017aan,Domenech:2017ems,Cai:2018dig,Unal:2018yaa,Yuan:2020iwf,Atal:2021jyo,Adshead:2021hnm}, the non-Gaussian contributions become more significant with increase of $|\fnl|$, and could be dominant for $|\fnl|\sim\cO(10)$. 
Compared with the Gaussian contribution, they are negligible for $|\fnl|\sim\cO(0.1)$, comparable for $|\fnl|\sim\cO(1)$, and one order of magnitude larger for $|\fnl|\sim\cO(10)$. 
Further, the contribution of $\cO(\fnl^{4})$ order also becomes more significant with increase of $|\fnl|$, and could be comparable to that of $\cO(\fnl^{2})$ order for $|\fnl|\sim\cO(10)$. 
In fact, the above results are available for $A\sim\cO(10^{-3}-10^{-1})$, i.e., the value of spectral amplitude commonly used in scenarios of \ac{PBH} production.


\begin{figure}
    \includegraphics[width =1. \columnwidth]{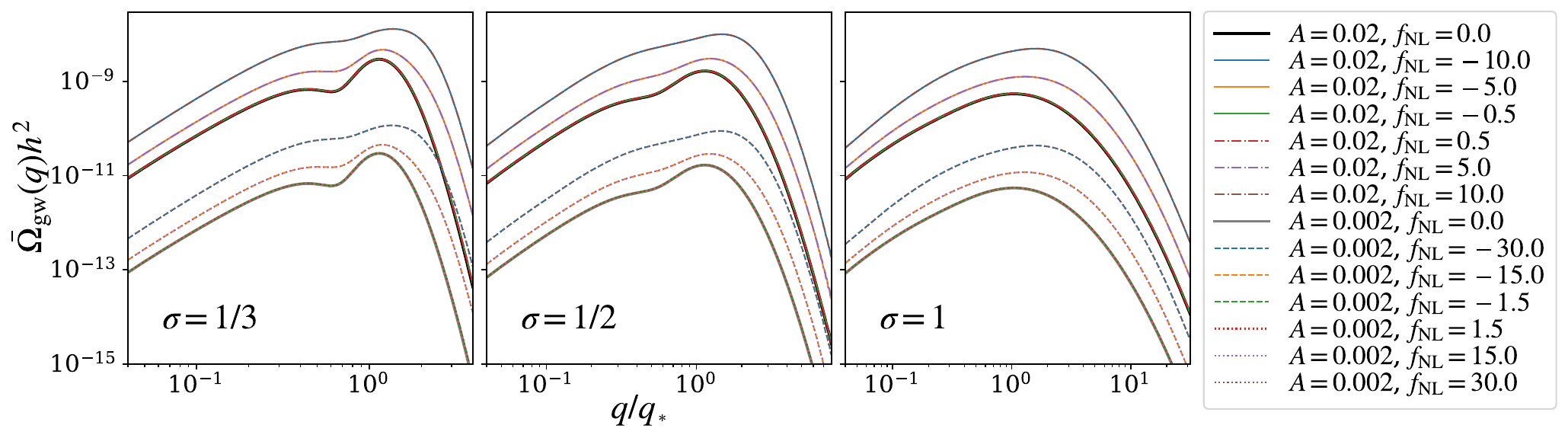}
    \caption{Dependence of the energy-density fraction spectrum of \acp{SIGW} in the current universe on the parameters $A$ and $\sigma$, as well as the sign degeneracy of $\fnl$. For $A=0.02$ ($A=0.002$), positive values of $\fnl$ are denoted by solid (dashed) curves, while negative ones are denoted by dot-dashed (dotted) curves.  For the Gaussian perturbations, i.e., $\fnl=0$, we denote the results of $A=0.02$ and $A=0.002$ with black and gray curves, respectively. }\label{fig:Omega_all}
\end{figure}

In \cref{fig:Omega_all}, we also show the dependence of $\bar{\Omega}_{\uGW}(q)$ on the parameters $A$ and $\sigma$, as well as the sign degeneracy of $\fnl$. 
The spectral magnitude strongly depends on $A$ (as well as $\fnl$), as is shown in Refs.~\cite{Ananda:2006af,Adshead:2021hnm}. 
Larger value of $A$ leads to a larger spectral magnitude, roughly following $\bar{\Omega}_{\uGW}\propto A^{2}$, and vice versa. 
In contrast, the variation of $\sigma$ mainly alters the shape of spectral profile, as is demonstrated in \cref{fig:Omega_all} from the left to right panels. 
Furthermore, there is a sign degeneracy of $\fnl$, because $\bar{\Omega}_{\uGW}(q)$ depends on powers of $\fnl^{2}$ in Eq.~(\ref{eq:Omega-G})--Eq.~(\ref{eq:Omega-R}) and Eq.~(\ref{eq:Omega-C})--Eq.~(\ref{eq:Omega-N}). 
Therefore, we can obtain at most the value of $|\fnl|$, rather than $\fnl$, via measuring the monopole in \acp{SIGW}. 

\begin{figure}
    \includegraphics[width =1. \columnwidth]{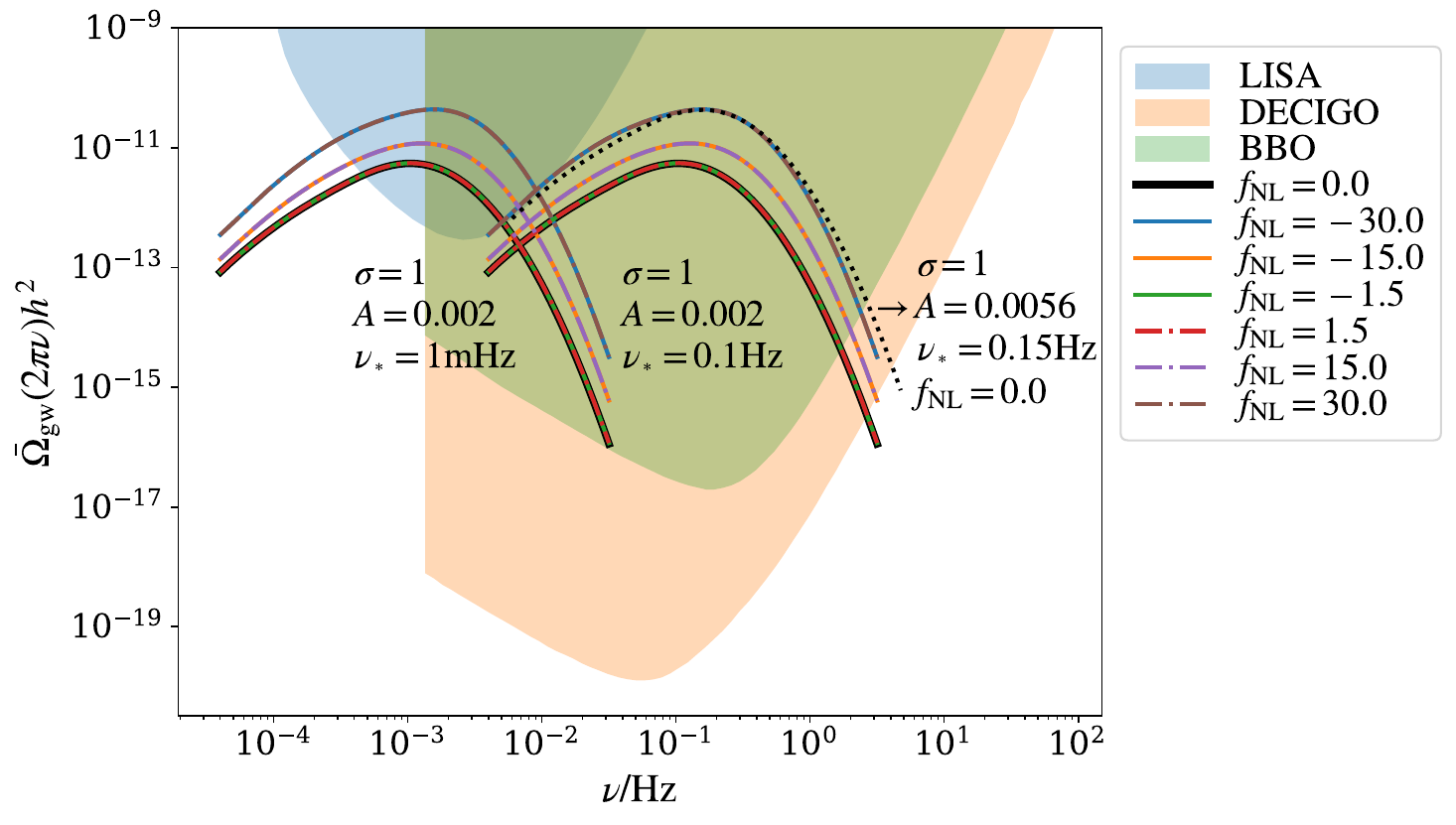}
    \caption{Energy-density fraction spectra of \ac{SIGW} in the current universe against the sensitivity curves of \acp{LISA} (blue shaded region) \cite{Baker:2019nia,Smith:2019wny}, \acp{DECIGO} (orange shaded region) \cite{Seto:2001qf,Kawamura:2020pcg}, and \acp{BBO} (green shaded region) \cite{Crowder:2005nr,Smith:2016jqs}. Positive and vanishing values of $\fnl$ are denoted by solid curves, while negative ones are denoted by dot-dashed curves. In addition, the dotted curve denotes the spectrum with $A = 0.0056$, $\sigma=1$, $\fnl=0.0$, and $\nu_\ast=0.15$ Hz. }\label{fig:Omega_sensityvity}
\end{figure}

As shown in \cref{fig:Omega_sensityvity}, the anticipated spectrum $\bar{\Omega}_{\uGW}$ is potentially measurable for future space-borne \ac{GW} detectors, e.g., \ac{LISA} \cite{Baker:2019nia,Smith:2019wny}, \ac{DECIGO} \cite{Seto:2001qf,Kawamura:2020pcg}, and \ac{BBO} \cite{Crowder:2005nr,Smith:2016jqs}. 
Here, the \ac{GW} frequency is $\nu=q/(2\pi)$ and the pivot frequency is $\nu_\ast=q_\ast/(2\pi)$. 
Letting $A=0.002$ and $\sigma=1$, but varying the value of $\fnl$, we depict the spectra $\bar{\Omega}_{\uGW}(2\pi\nu)$ for $\nu_\ast=1$ mHz and $\nu_\ast=0.1$ Hz, which are corresponded to the LISA band and the DECIGO/BBO band, respectively. 
Based on \cref{fig:Omega_sensityvity}, we expect these detectors to probe \acp{SIGW} related to the parameter region (particularly, the intervals of $A$ and $\fnl$) that exerts a significant impact on the formation of \acp{PBH}, in particular, the abundance \cite{Byrnes:2012yx,Young:2013oia,Franciolini:2018vbk}.

Besides the sign degeneracy of $\fnl$, there are other degeneracies in the model parameters including $A$, $\sigma$, $\fnl$, and $\nu_\ast$. 
As an example, we depict the spectrum $\bar{\Omega}_{\uGW}(\nu)$ for $A=0.0056$,  $\sigma=1$, $\fnl=0.0$, and $\nu_\ast=0.15$ Hz, as is denoted by the dotted curve in \cref{fig:Omega_sensityvity}. 
We find that it almost coincides with the spectra with $A = 0.002$, $\sigma=1$, $\fnl=\pm30.0$, and $\nu_\ast=0.1$ Hz, indicating that it is very challenging to determine the value of $|\fnl|$ with measurements of the monopole in \acp{SIGW} only. 
In fact, only a combination of the form $A\fnl^2$, rather than $\fnl$ itself, contributes to the energy-density fraction spectrum.


In summary, it is imperative to develop new probes of the primordial non-Gaussianity through potential measurements of \acp{SIGW}. 


\section{Multipoles and primordial non-Gaussianity}\label{sec:multi}

In this section, we study the anisotropies in \acp{SIGW} contributed by the local-type primordial non-Gaussianity in curvature perturbations, and then show the first complete analysis to the angular power spectrum of \acp{SIGW}. 
The method and analytic formulae developed in this section could be generalized straightforwardly, e.g., to study the anisotropies in \acp{SIGW} produced during matter domination.

\subsection{Angular power spectrum}\label{sec:Cl}

The correlation of initial perturbations, i.e., $\delta (\eta_\uin, \bx_\uin, \bq)$ in Eq.~(\ref{eq:deltaGW-CGW}), at two different locations separated by a large angle (i.e., low multipoles) can only arise from the \acl{PNG}.
The local-type \acl{PNG} leads to the coupling between modes of short-wavelength and long-wavelength \cite{Tada:2015noa}.
In this subsection, we will adopt the Feynman-like diagrams to compute two-point correlations of the initial inhomogeneities.

In Fourier space, we can decompose the Gaussian component of curvature perturbation $\zeta$ in \cref{eq:fnl-def} as follows
\begin{equation}\label{eq:zeta-g-SL} 
    \zeta_g (\bk) = \zeta_S (\bk) + \zeta_L (\bk)\ , 
\end{equation} 
where the suffixes $_S$ and $_L$ denote the short-wavelength and long-wavelength modes, respectively. 
We define the power spectra of these modes as  
\begin{subequations} 
\begin{eqnarray}\label{eqn:PSL-def} 
    \langle\zeta_S (\bq)\zeta_S (\bq')\rangle & = & \delta^{(3)}(\bq+\bq') P_S (q)\ , \\ 
    \langle\zeta_L (\bk)\zeta_L (\bk')\rangle & = & \delta^{(3)}(\bk+\bk') P_L (k)\ , \\
    \langle\zeta_S (\bq) \zeta_L (\bk)\rangle & = & 0\ . 
\end{eqnarray} 
\end{subequations} 
For the long-wavelength modes, the dimensionless power spectrum $\Delta^2_L$ is nearly scale-invariant, with the spectral amplitude $A_L\simeq2.1\times 10^{-9}$ \cite{Planck:2018vyg}. 
In contrast, for the short-wavelength modes, the spectral amplitude $A_S$ is nearly unconstrained by current observations. 
In this work, assuming the dimensionless power spectrum in \cref{eq:Lognormal}, we consider the spectral amplitude $A_S=A\sim\cO(10^{-3}-10^{-1})$, which is related to the formation scenarios of \acp{PBH} (e.g., see Refs.~\cite{Green:2020jor,Carr:2020gox}).

To simplify computation of the angular power spectrum in \cref{eq:reduced-angular-power-spectrum-def}, we make several approximations to the density contrast in \acp{SIGW} in Eq.~(\ref{eq:deltaGW-CGW}). 
Firstly, besides $\Phi(\eta_0,\bx_0)$, we disregard the tensor sourced term which is contributed by the primordial \acp{GW} and \acp{SIGW} smaller than the linear scalar perturbations. 
Secondly, the \ac{ISW} effect is subdominant and thus can be neglected, as was shown in Ref.~\cite{Bartolo:2019zvb}. 
Thirdly, we have demonstrated  in \cref{sec:Omega} that $n_\uGW(\eta,q)$ is independent of $\eta$. 
Therefore, we approximate Eq.~(\ref{eq:deltaGW-CGW}) as  
\begin{eqnarray}\label{eq:saidelta}
        \delta_\uGW (\bq) 
     =  \delta_\uGW (\eta_\uin, \bx_\uin,\bq)
       + \left[4-n_\uGW (q)\right] \Phi (\eta_\uin, \bx_\uin) \ ,
\end{eqnarray}
where we denote $n_\uGW(q)=n_\uGW(\eta_0,q)$ for simplicity. 
On the right hand side of Eq.~(\ref{eq:saidelta}), the first term denotes the initial inhomogeneities, while the second one leads to the \ac{SW} effect. 
Since the angular resolution is finite for a \ac{GW} detector, the signal along a line-of-sight is actually an ensemble average of the energy density of \acp{SIGW} over a large quantity of Hubble horizons. 
In this sense, the initial inhomogeneities in a neighborhood of $\bx_\uin$ can be viewed to be isotropic. 
However, the initial inhomogeneities around $\bx_\uin$ and $\bx_\uin '$ separated by a long distance could be correlated due to the \acl{PNG}, as will be computed with the Feynman-like diagrams in the following. 

In addition, the \ac{SW} effect is produced by the long-wavelength scalar modes that reentered into the Hubble horizon during matter domination, indicating $w=c_s^{2}=0$. 
Based on Eq.~(\ref{eq:master}) and Eq.~(\ref{eq:T-zeta-def}), we get the scalar transfer function to be $T(k\eta)=1$ during matter domination. 
Therefore, we have 
\begin{equation}\label{eq:phisail}
\Phi(\eta_\uin,\bx_\uin)=\frac{3}{5} \int \frac{\ud^3 \bk}{(2\pi)^{3/2}} e^{i\bk\cdot\bx_\uin} \zeta_{L}(\bk)\ ,
\end{equation}
which should be substituted back into Eq.~(\ref{eq:saidelta}). 
Here, we consider the linear term in $\zeta_{L}$ only, since higher-order terms are much smaller than this term due to $A_L\sim10^{-9}$.

\subsubsection{Feynman-like rules}

To get the initial inhomogeneities $\delta_\uGW(\eta_\uin,\bx_\uin,\bq)$, it is necessary to compute the initial energy-density full spectrum $\omega_\uGW(\eta_\uin,\bx_\uin,\bq)$ first of all, based on \cref{eq:delta-omega-def} and \cref{eq:deltaGW-def}. 
Following \cref{eq:omega-chi}, we obtain the latter to be 
\begin{equation}\label{eq:omega-h}
     \omega_\uGW(\eta_\uin,\bx_\uin,\bq) 
     = - \frac{q^3}{48 \cH^2} \int \frac{\ud^3 \bk}{(2\pi)^{3}} e^{i\bk\cdot\bx_\uin} 
        \left(\bk-\bq\right) \cdot \bq  
        \sum_{\lambda,\lambda'} \epsilon_{ij}^{\lambda}(\bk-\bq) \epsilon_{ij}^{\lambda'}(\bq) 
        \overbar{h_\lambda(\eta, \bk-\bq) h_{\lambda'}(\eta, \bq)}\ . 
\end{equation} 
Here, $\bq$ denotes a comoving momentum of \acp{GW} that is corresponded to the short-wavelength, while $\bk$ is associated with a Fourier mode of the inhomogeneities in \acp{SIGW}, that is corresponded to the long-wavelength. 
We will take $q \gtrsim \cH^{-1} \gg k$ in the following. 

\begin{figure}
    \includegraphics[width =1. \columnwidth]{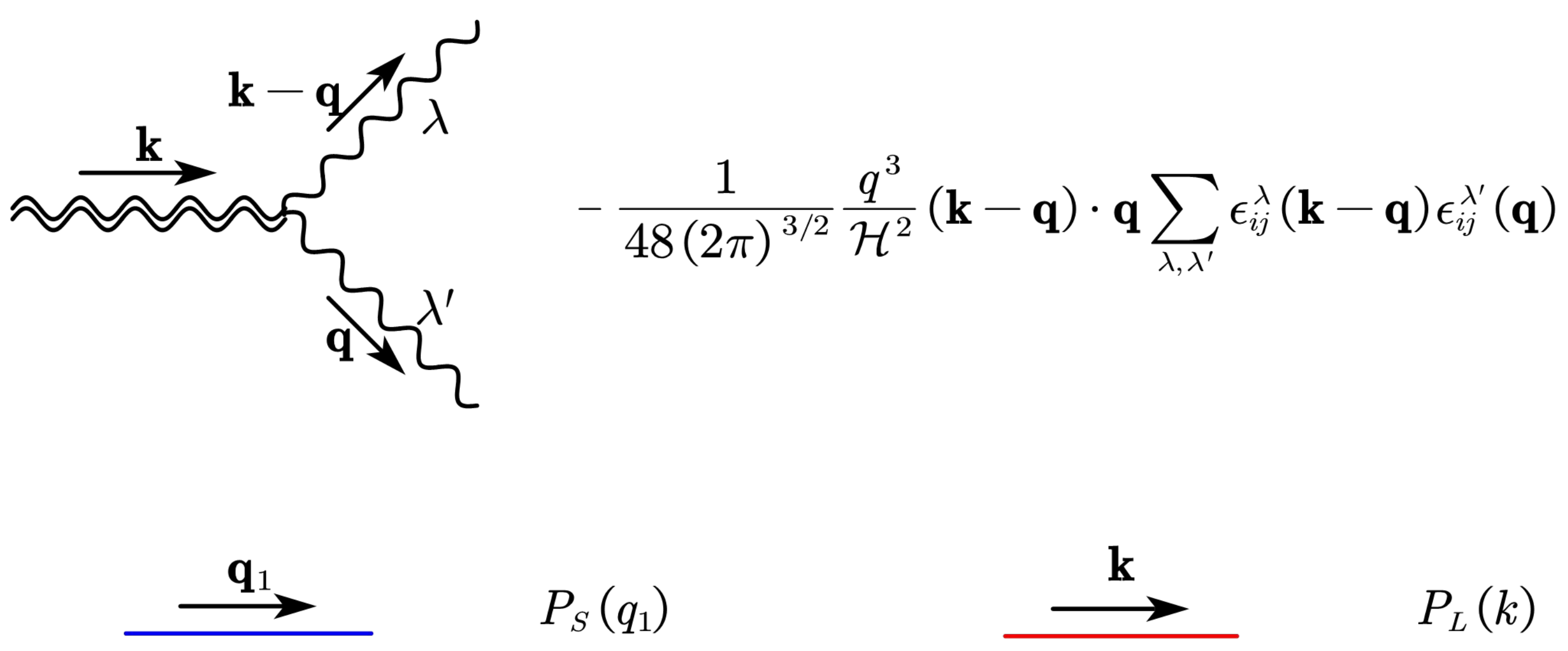}
    \caption{Feynman-like rules supplemented for the evaluation of \acp{SIGW}. The double wavy line denotes the energy-density full spectrum $\omega_\uGW (\eta,\bk,\bq)$, the blue solid line represents $P_S$, and the red solid line stands for $P_L$. Note that the Feynman-like rule in the top panel includes an operator. }
    \label{fig:F_Rules_add}
\end{figure}

The statistics of the inhomogeneities in \acp{SIGW} is expressed as a two-point correlator $\langle\omega_\uGW(\eta_\uin,\bx_\uin,\bq) \omega_\uGW(\eta_\uin,\bx_\uin',\bq')\rangle$. 
By substituting \cref{eq:h} into \cref{eq:omega-h}, we can rewrite the latter in terms of an eight-point correlator of $\zeta$. 
Utilizing \cref{eq:fnl-def}, \cref{eq:zeta-g-SL}, and \cref{eqn:PSL-def}, we further rewrite it in terms of two-point correlators of the Gaussian components $\zeta_S$ and $\zeta_L$, based on the Wick's theorem. 
The above derivation is straightforward but tedious. 
However, as was done in \cref{sec:Omega}, the method of Feynman-like diagrams is still applicable to simplify it. 
Therefore, besides the Feynman-like rules in \cref{fig:F_Rules}, we augment three Feynman-like rules shown in \cref{fig:F_Rules_add}. 
To be specific, besides the double wavy curves represent $\omega_\uGW(\eta,\bk,\bq)$, the solid black line in \cref{fig:F_Rules} is now replaced with two colored lines, with the blue one denoting $P_S$ and the red one denoting $P_L$. 
Note that the Feynman-like rule for vertex in \cref{fig:F_Rules} remains the same irrespective of the colors of solid lines.

\subsubsection{Feynman-like diagrams}

\begin{figure}
    \includegraphics[width =0.5 \columnwidth]{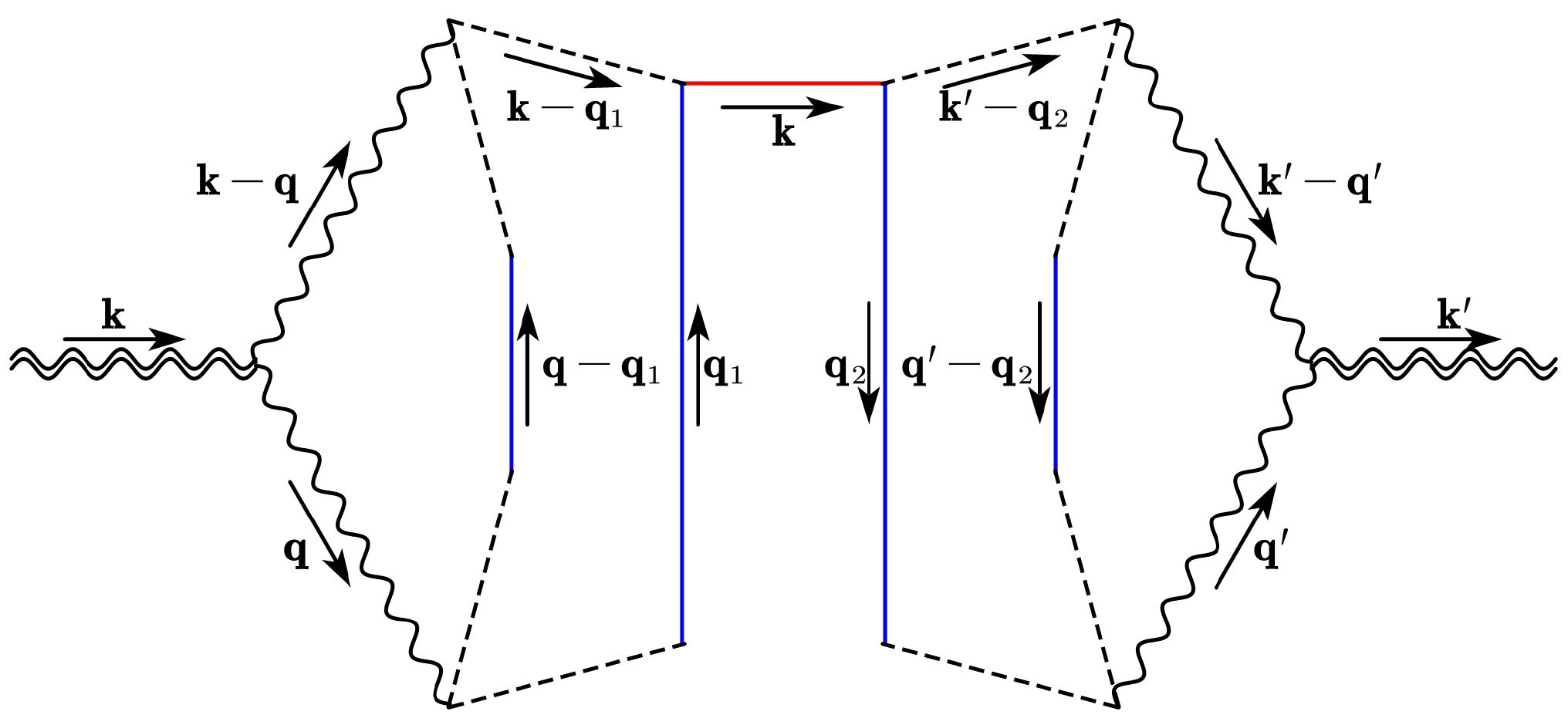}
    \includegraphics[width =0.5 \columnwidth]{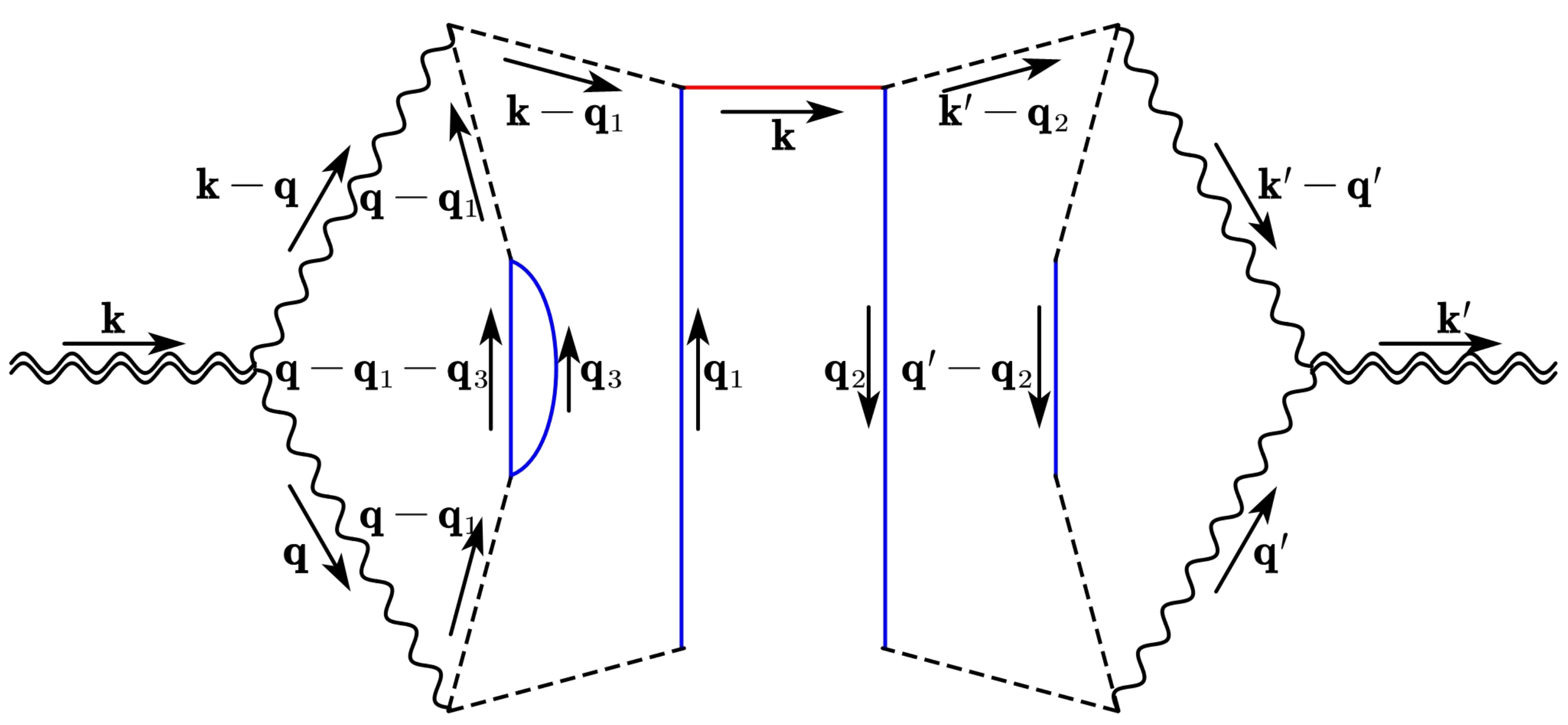}\\\\\\
    \includegraphics[width =0.5 \columnwidth]{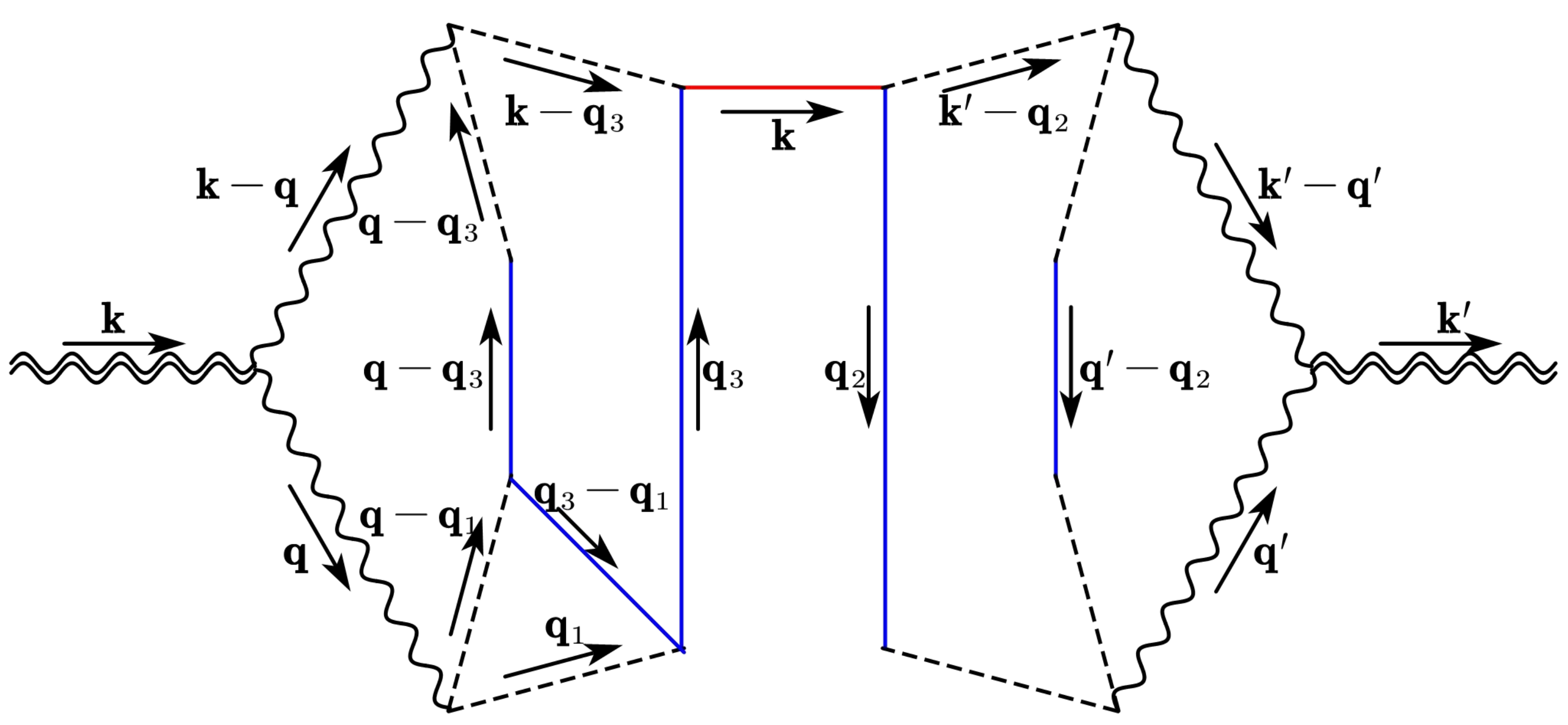}
    \includegraphics[width =0.5 \columnwidth]{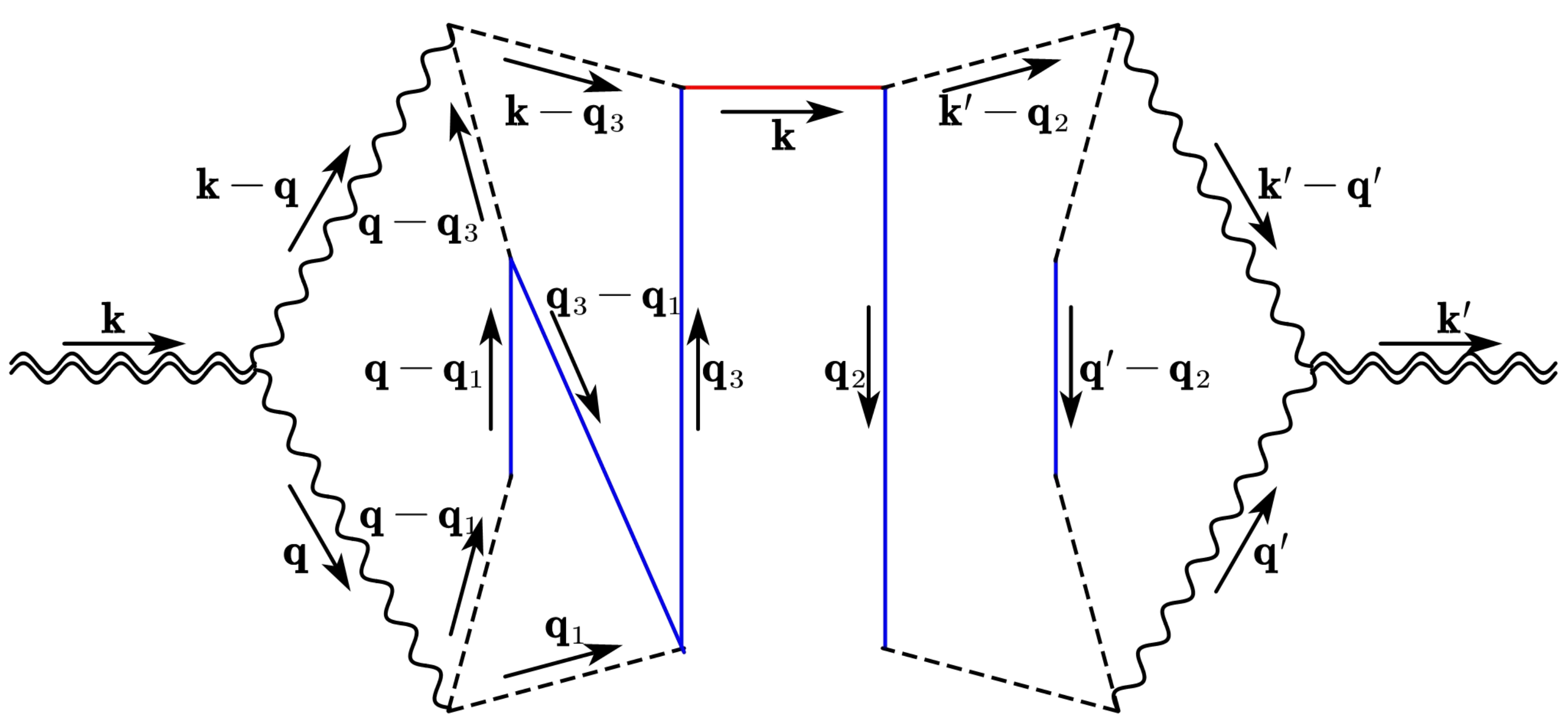}\\\\\\
    \includegraphics[width =0.5 \columnwidth]{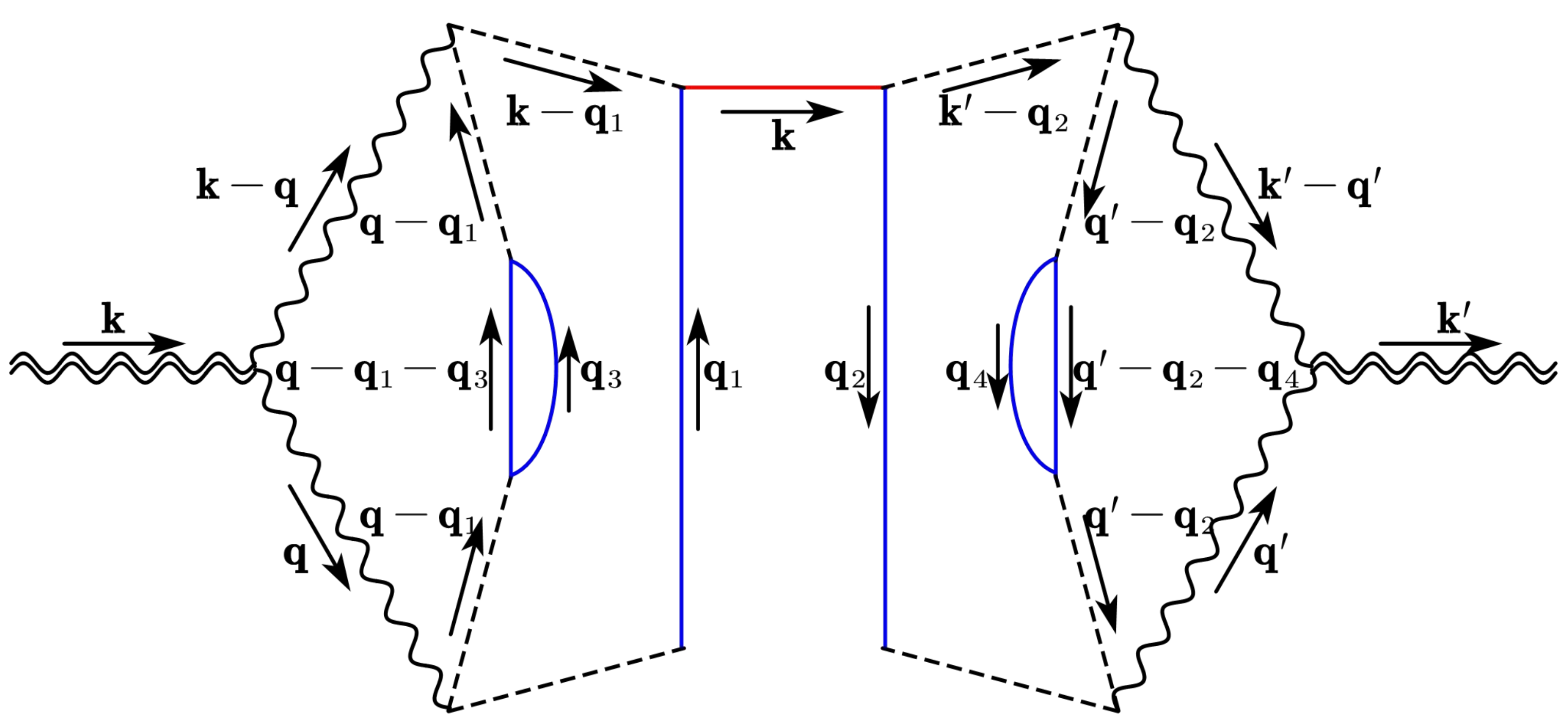}
    \includegraphics[width =0.5 \columnwidth]{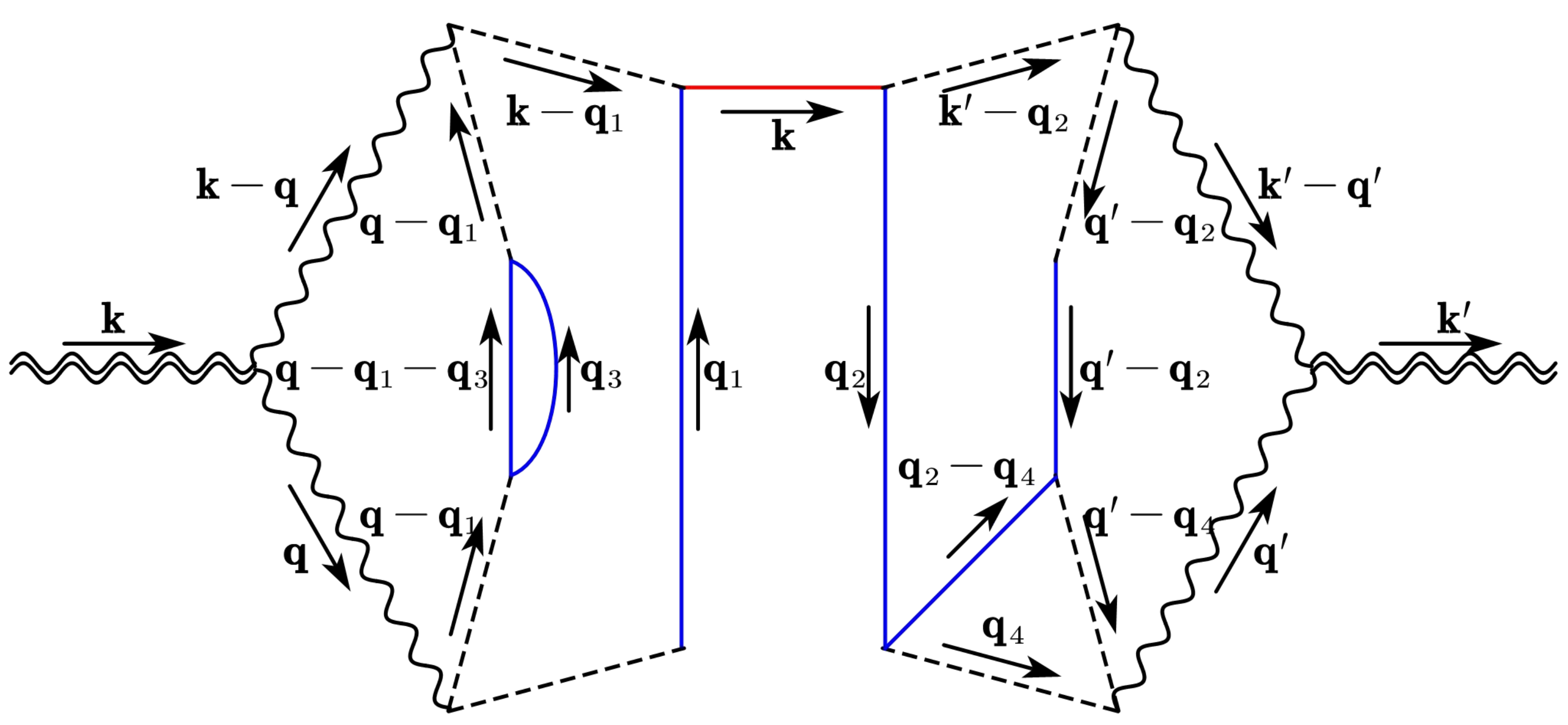}\\\\\\
    \includegraphics[width =0.5 \columnwidth]{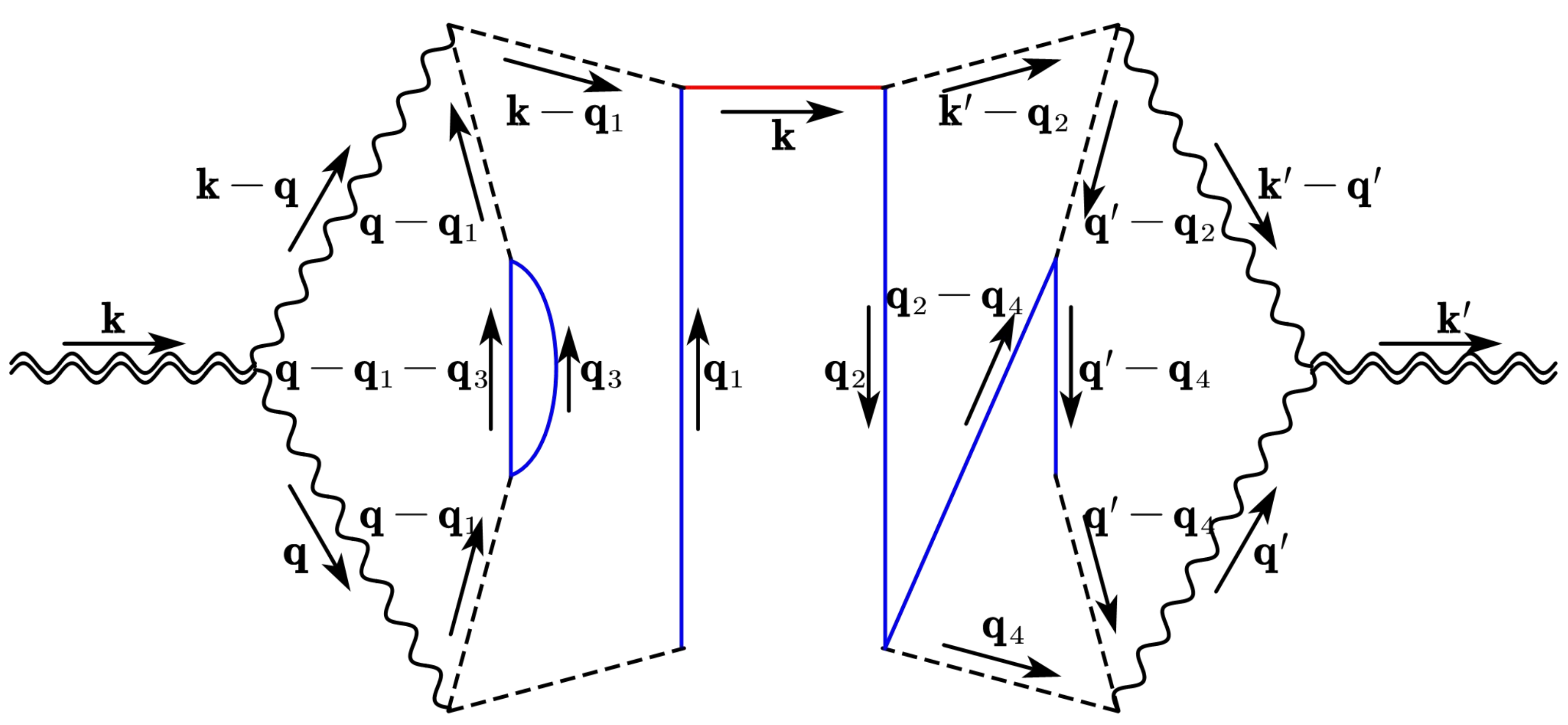}
    \includegraphics[width =0.5 \columnwidth]{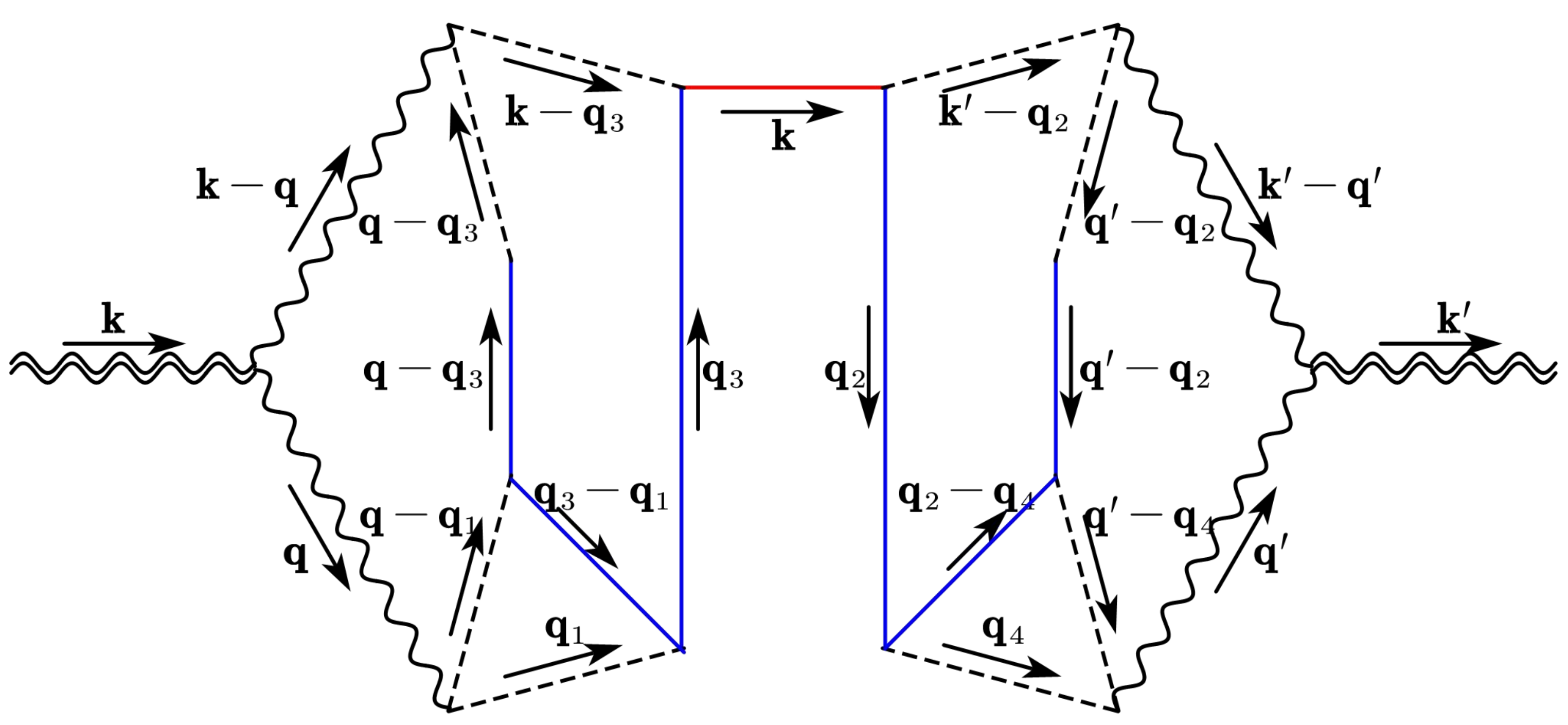}\\\\\\
    \includegraphics[width =0.5 \columnwidth]{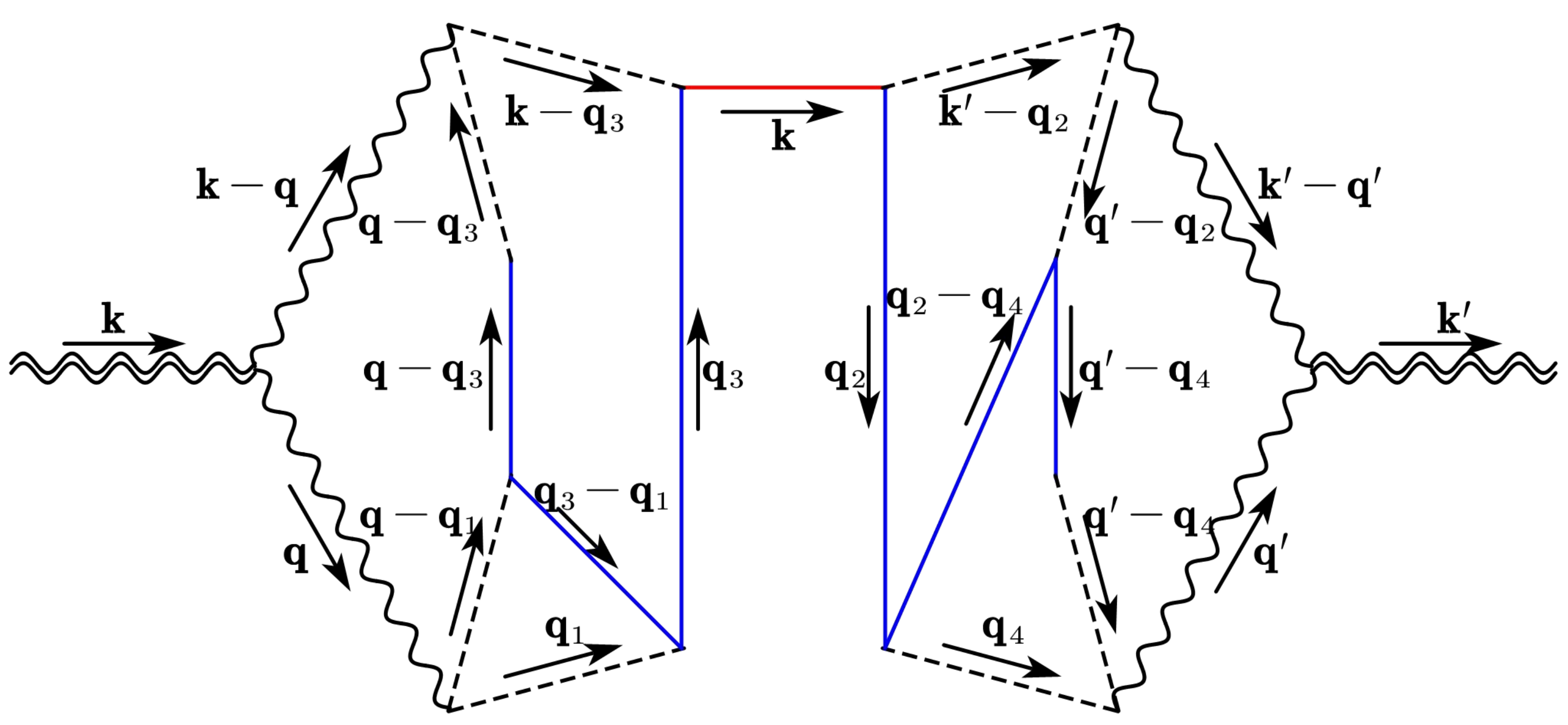}
    \includegraphics[width =0.5 \columnwidth]{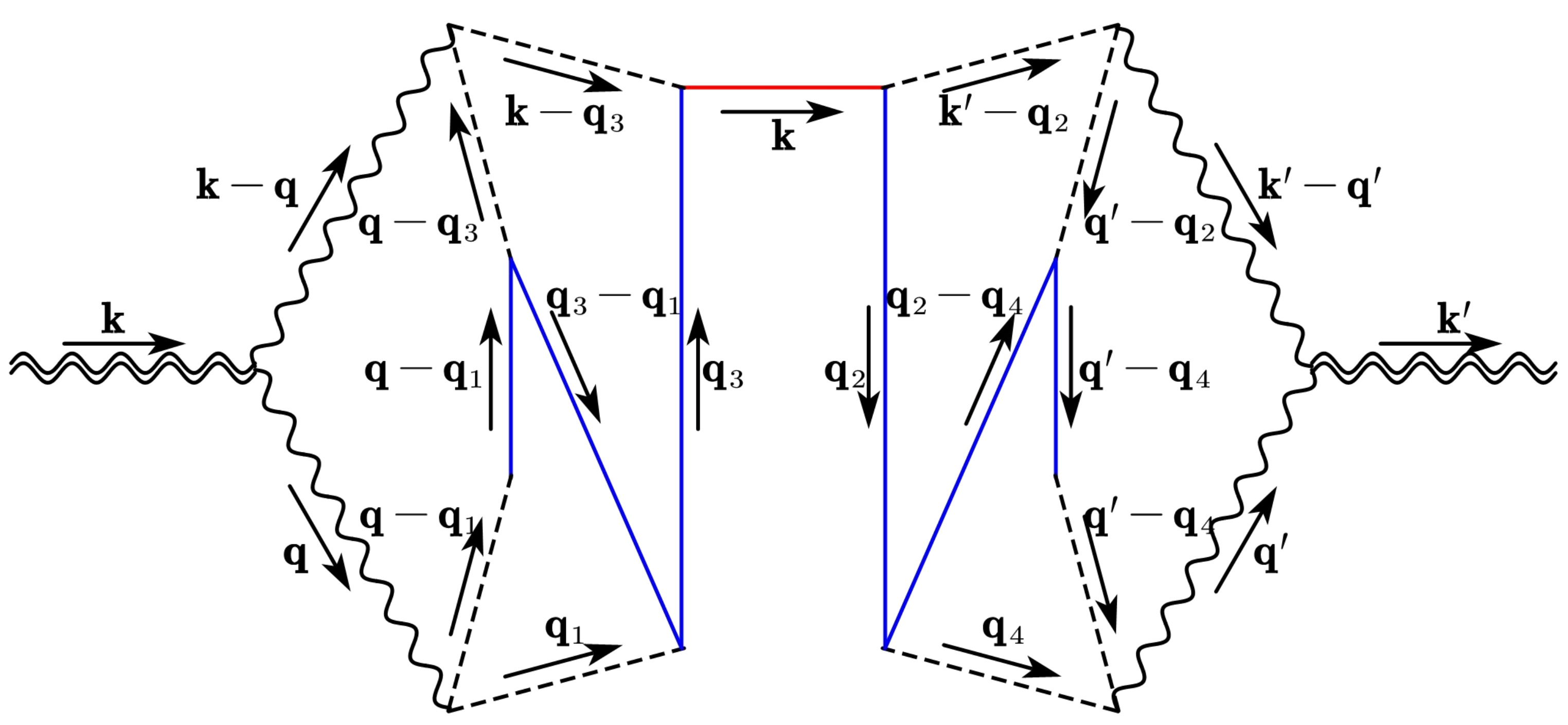}
    \caption{Feynman-like diagrams at $\cO(\Delta^2_L)$ order. For brevity, we omit $\lambda$ for the wavy lines. }\label{fig:APS-all}
\end{figure}

Following the Feynman-like rules in \cref{fig:F_Rules} and \cref{fig:F_Rules_add}, we can obtain all of the nonvanishing Feynman-like diagrams up to linear order in $\Delta_{L}^{2}$.  
However, the disconnected diagrams, which are of zeroth order in $\Delta_{L}^{2}$, correspond to the monopole squared, i.e., $\bar{\omega}_{\uGW}^{2}(\eta,q)$, which has been studied in Eq.~(\ref{eq:omegabar}) and Eq.~(\ref{eq:ogwetaa0qsai}). 
Since they are homogeneous, we disregard them in the following. 
At linear order in $\Delta_L^{2}$, we depict the Feynman-like diagrams in \cref{fig:APS-all}. 
In each panel of \cref{fig:APS-all}, there is an ``$\fnl$ bridge'' that connects the initial inhomogeneities at two different locations separated by a long distance. 
Diagrams at higher order in $\Delta_{L}^{2}$ are negligible due to the assumption of $A_L\ll A_S$.

The Feynman-like diagrams in \cref{fig:APS-all} can be understood as follows. 
On the one hand, there is an ensemble average of the energy density of \acp{SIGW} over a quantity of Hubble horizons in the neighbourhood of $\bx_{\uin}$. 
On the other hand, the energy densities of \acp{SIGW} at $\bx_\uin$ and $\bx_\uin '$ separated by a long distance are connected by the $\fnl$ bridge. 
Therefore, this picture is equivalent to the following mathematical result
\begin{eqnarray}
    \langle\omega_\uGW(\eta_\uin,\bx_\uin,\bq) \omega_\uGW(\eta_\uin,\bx_\uin',\bq')\rangle ^{\mathcal{O}(\Delta_{L}^{2})}
    \sim 
    \langle \langle\omega_\uGW(\eta_\uin,\bk,\bq)\rangle_{\bx_\uin} \langle\omega_\uGW^\ast(\eta_\uin,\bk',\bq')\rangle_{\bx_\uin'}\rangle ^{\mathcal{O}(\Delta_{L}^{2})}\ ,\label{eq:wwinsai}
\end{eqnarray}
where the superscript $^{\mathcal{O}(\Delta_{L}^{2})}$ denotes the linear order in $\Delta_{L}^{2}$. 
During the mathematical derivation, we have approximately take $\bq-\bk\simeq \bq$ because of $k \ll q$. 
Therefore, the eight-point correlator $\langle\zeta^{8}\rangle$, which is used for expressing $\langle\omega_\uGW(\eta_\uin,\bx_\uin,\bq) \omega_\uGW(\eta_\uin,\bx_\uin',\bq')\rangle$, becomes  
\begin{equation}\label{eq:zeta8-contraction}
    \langle \zeta^8 \rangle^{\cO(\Delta^2_L)}
    \propto \left(\frac{3}{5} \fnl\right)^{2}
    \langle\zeta \zeta \zeta \zeta_S\rangle_{\bx_\uin}
    \langle\zeta \zeta \zeta \zeta_S\rangle_{\bx_\uin'} 
    \int \frac{\ud^3 \bk\, \ud^3\bk'}{(2\pi)^3} 
    e^{i\bigl(\bk\cdot\bx_\uin - \bk'\cdot\bx_\uin'\bigr)} 
    \langle \zeta_L(\bk) \zeta_L(-\bk')\rangle \ . 
\end{equation}
Here, $\langle\zeta \zeta \zeta \zeta_S\rangle_{\bx_\uin}$ can be further expressed in terms of the contractions corresponding to the Feynman-like diagrams labeled by $G$, $H$, $C$ and $Z$ in \cref{fig:EDS-G} and \cref{fig:EDS-other}. 
We provide an explicit formula to Eq.~(\ref{eq:wwinsai}) in the following. 
In \cref{fig:APS-all}, we label each diagram with a superscript $^{XY}$ if the $\fnl$ bridge connects two sub-diagrams labeled by $X$ and $Y$ corresponding to diagrams in \cref{fig:EDS-G} and \cref{fig:EDS-other}. 
For the top left panel, we have 
\begin{eqnarray}
    &&\left\langle
        \omega_\uGW (\eta_\uin,\bx_\uin,\bq) \omega_\uGW (\eta_\uin,\bx_\uin',\bq')
    \right\rangle^{GG}\\ 
    &=& 2^6 \Bigl(\frac{3}{5} \fnl\Bigr)^2  
        \bar{\omega}_\uGW^{G} (\eta_\uin,q) 
        \bar{\omega}_\uGW^{G} (\eta_\uin,q')
        \int \frac{\ud^3 \bk}{(2\pi)^{3}}  
        e^{i \bk\cdot(\bx_\uin-\bx_\uin')} P_L (k) \ ,\nonumber
\end{eqnarray}
where the constant $2^{6}$ is an additional symmetric factor due to the $\fnl$ bridge. 
This diagram has been first evaluated in Ref.~\cite{Bartolo:2019zvb}, but used different convention (the authors of Ref.~\cite{Bartolo:2019zvb} used $\Gamma$ rather than $\delta\omega_\uGW$, but the two quantities are related with each other via Eq.~(\ref{eq:delta-Gamma})). 
Following Ref.~\cite{Bartolo:2019zvb}, the anisotropies in \acp{SIGW} were further studied in Refs.~\cite{ValbusaDallArmi:2020ifo,Schulze:2023ich,LISACosmologyWorkingGroup:2022kbp,LISACosmologyWorkingGroup:2022jok,Malhotra:2020ket,Dimastrogiovanni:2021mfs,Unal:2020mts,Carr:2020gox}. 
For the top right panel, we have 
\begin{eqnarray}
    & &\left\langle
        \omega_\uGW (\eta_\uin,\bx_\uin,\bq) \omega_\uGW (\eta_\uin,\bx_\uin',\bq')
    \right\rangle^{HG+GH} \\ 
    &=& 2^5 \Bigl(\frac{3}{5} \fnl\Bigr)^2
        \Bigl[\bar{\omega}_\uGW^{H} (\eta_\uin,q) \bar{\omega}_\uGW^{G} (\eta_\uin,q') + \bar{\omega}_\uGW^{G} (\eta_\uin,q) \bar{\omega}_\uGW^{H} (\eta_\uin,q')\Bigr]  
        \int \frac{\ud^3 \bk}{(2\pi)^{3}}  
        e^{i \bk\cdot(\bx_\uin-\bx_\uin')} P_L (k)\ ,\nonumber
\end{eqnarray}
where the constant $2^{5}$ is also an symmetric factor. 
The expressions for the diagrams in other panels can also be obtained in the same way, but the corresponding derivation processes have been neglected here. 
Summing these results, we eventually get the formula as follows 
\begin{eqnarray}\label{eq:wwolsai}
    &&\left\langle
        \omega_\uGW (\eta_\uin,\bx_\uin,\bq) \omega_\uGW (\eta_\uin,\bx_\uin',\bq')
    \right\rangle^{\cO(\Delta^2_L)} \\
    &=&         \frac{\Omega_{\mathrm{ng}} (\eta_\uin,q)}{4\pi} \frac{\Omega_{\mathrm{ng}} (\eta_\uin,q’)}{4\pi} 
    \left(\frac{3}{5} \fnl\right)^2 
        \int \frac{\ud^3 \bk}{(2\pi)^{3}}  
        e^{i \bk\cdot(\bx_\uin-\bx_\uin')} P_L (k) \ ,\nonumber
\end{eqnarray}
where we introduce a new quantity $\Omega_{\mathrm{ng}}$ for concision, defined as 
\begin{equation}
    \Omega_{\mathrm{ng}} (\eta_\uin,q) 
    = 2^3 \bar{\Omega}_\uGW^{G} (\eta_\uin,q)  
        + 2^2 \bar{\Omega}_\uGW^{H} (\eta_\uin,q)  
        + 2^2 \bar{\Omega}_\uGW^{C} (\eta_\uin,q) 
        + 2^2 \bar{\Omega}_\uGW^{Z} (\eta_\uin,q) \ .
\end{equation}

To simplify computation in the following, we equivalently express the initial inhomogeneities $\delta\omega_\uGW (\eta_\uin,\bx_\uin,\bq)$ as follows 
\begin{equation}
    \delta\omega_\uGW (\eta_\uin,\bx_\uin,\bq) 
    =   \frac{\Omega_{\mathrm{ng}} (\eta_\uin,q)}{4\pi} \left(\frac{3}{5} \fnl\right) 
        \int \frac{\ud^3 \bk}{(2\pi)^{3/2}} e^{i\bk\cdot\bx_\uin} \zeta_L (\bk)\ ,
\end{equation}
which can reproduce Eq.~(\ref{eq:wwolsai}). 
Based on \cref{eq:deltaGW-def}, $\delta\omega_\uGW (\eta_\uin,\bx_\uin,\bq)$ can be further transformed into the initial density contrast as 
\begin{equation}\label{eq:deltainitial}
    \delta_{\uGW} (\eta_\uin,\bx_\uin,\bq) 
    = \left(\frac{3}{5} \fnl\right) \frac{\Omega_{\mathrm{ng}} (\eta_\uin,q)}{\bar{\Omega}_\uGW (\eta_\uin,q)} 
        \int \frac{\ud^3 \bk}{(2\pi)^{3/2}} e^{i\bk\cdot\bx_\uin} \zeta_L (\bk)\ ,
\end{equation}
which will be used for computation of the angular power spectrum in the next subsection. 
The factor $\Omega_{\mathrm{ng}}/\bar{\Omega}_{\uGW}$ would be replaced by a constant in the previous work \cite{Bartolo:2019zvb}, but depends on \ac{GW} frequency in our current work.

\subsubsection{Two-point angular correlation functions}

Substituting \cref{eq:deltainitial} and \cref{eq:phisail} into Eq.~(\ref{eq:saidelta}), we have the observed density contrast 
\begin{equation}\label{eq:deltaGW}
    \delta_{\uGW} (\bq) 
    = \frac{3}{5} \Biggl\{
        \fnl  \frac{\Omega_{\mathrm{ng}} (\eta_\uin,q)}{\bar{\Omega}_\uGW (\eta_\uin,q)} + [4-n_\uGW(q)] 
    \Biggr\} \int \frac{\ud^3 \bk}{(2\pi)^{3/2}} e^{i\bk\cdot\bx_\uin} \zeta_L (\bk)\ .
\end{equation}
We can express $\langle\delta_{\uGW} (\bq) \delta_{\uGW} (\bq') \rangle$ in terms of the two-point correlator of $\zeta_L$, defined in \cref{eqn:PSL-def}. 
Assuming $\Delta_L^{2}(k)$ to be scale-invariant, we analytically calculate the following integral 
\begin{eqnarray}
    \int \frac{\ud^3 \bk\,\ud^3 \bk'}{(2\pi)^3} 
        e^{i\left(\bk\cdot\bx_\uin - \bk'\cdot\bx_\uin '\right)}
        \langle\zeta_L (\bk) \zeta_L (-\bk')\rangle
    &=& 4 \pi \Delta^2_L \sum_{\ell m} Y_{\ell m} (\bn_0) Y^\ast_{\ell m} (\bn_0')
        \int \ud \ln k\, j_\ell^2 [k \left(\eta_0 - \eta_\uin\right)] \nonumber\\
        &\simeq& \sum_{\ell m} Y_{\ell m} (\bn_0) Y^\ast_{\ell m} (\bn_0') \frac{2\pi}{\ell(\ell+1)}   \times \Delta_{L}^{2}\ ,\label{eq:eq515}
\end{eqnarray}
where we use a relation of $\bx_\uin-\bx_\uin' = (\eta_\uin-\eta_0)(\bn_0-\bn_0')$, the identity of the form
    $e^{i k \mu (\eta_\uin-\eta_0)} 
    = 4 \pi \sum_{\ell m}  (-i)^\ell 
        j_\ell [k (\eta_0-\eta_\uin)] Y^\ast_{\ell m} (\hat{k}) Y_{\ell m} (\bn_0)$, 
and the integral $\int \ud \ln k\, j_{\ell}^2 [k \left(\eta_0 - \eta_\uin\right)]=1/ [2\ell (\ell+1)]$ due to $\eta_0 \gg \eta_\uin$. 
Eventually, combining \cref{eq:delta-lm}, \cref{eq:deltaGW}, and Eq.~(\ref{eq:eq515}), we obtain the reduced angular power spectrum defined in \cref{eq:reduced-angular-power-spectrum-def}, i.e., 
\begin{eqnarray}\label{eq:reduced-APS}
    \widetilde{C}_\ell (q,q') 
    = && 
        \frac{18\pi\Delta^2_L }{25 \ell (\ell+1)} \times 
        \biggl[
            \fnl \frac{\Omega_{\mathrm{ng}} (\eta_\uin,q)}{\bar{\Omega}_\uGW (\eta_\uin,q)}
            + \bigl(4 - n_\uGW (q)\bigr)
        \biggr]\nonumber\\
        &&\hphantom{\frac{18\pi\Delta^2_L }{25 \ell (\ell+1)}} 
        \times \biggl[
            \fnl \frac{\Omega_{\mathrm{ng}} (\eta_\uin,q')}{\bar{\Omega}_\uGW (\eta_\uin,q')}
            + \bigl(4 - n_\uGW (q')\bigr)
        \biggr] \ .\label{eq:rapssai}
\end{eqnarray}
Correspondingly, the angular power spectrum defined in \cref{eq:angular-power-spectrum-def} is given as 
\begin{eqnarray}\label{eq:APS}
    C_\ell (q,q') 
        & = & \frac{9\Delta^2_L \bar{\Omega}_{\uGW} (q)\bar{\Omega}_{\uGW} (q')}{200\pi \ell (\ell+1)} \times
        \biggl[
            \fnl \frac{\Omega_{\mathrm{ng}} (\eta_\uin,q)}{\bar{\Omega}_\uGW (\eta_\uin,q)}
            + \bigl(4 - n_\uGW (q)\bigr)
        \biggr]\nonumber\\
        &&\hphantom{\frac{9\Delta^2_L \bar{\Omega}_{\uGW} (q)\bar{\Omega}_{\uGW} (q')}{200\pi \ell (\ell+1)}}
        \times \biggl[
            \fnl \frac{\Omega_{\mathrm{ng}} (\eta_\uin,q')}{\bar{\Omega}_\uGW (\eta_\uin,q')}
            + \bigl(4 - n_\uGW (q')\bigr)
        \biggr]\ .
\end{eqnarray}
This is the most important formula of this paper. 
Besides the radiation domination, it is so generic that also available during other epochs. 
Eq.~\eqref{eq:APS} indicates that the angular power spectrum of \acp{SIGW} consists of the initial inhomogeneities, the \ac{SW} effect, and the cross terms between them. 
We will evaluate it numerically in the following subsection.

In Eq.~\eqref{eq:reduced-APS}, the initial inhomogeneities are explicitly determined by the parameter $\fnl$ as well as the parameter $A_S\fnl^2$ in $\Omega_{\mathrm{ng}}/\bar{\Omega}_{\uGW}$ (besides $\sigma$ and $q_\ast$), indicating that the sign degeneracy in $\fnl$ is explicitly broken. 
In contrast, the \ac{SW} effect is determined by $A_S\fnl^2$ only (besides $\sigma$ and $q_\ast$). 
These results would lead to interesting theoretical expectations in the next subsection. 
In fact, a ratio between the cross terms that are linear in $\fnl$ and the $\fnl^{2}$ term is roughly proportional to $2(4-n_\uGW)/(\fnl\Omega_\mathrm{ng}/\bar{\Omega}_\uGW)$. 
Since $n_\uGW\sim\mathcal{O}(1)$ and $\Omega_\mathrm{ng}/\bar{\Omega}_\uGW\sim\mathcal{O}(1)$, we have possibilities to get the largest breaking of the sign degeneracy of $\fnl$ when we concern $\fnl\sim\mathcal{O}(1)$. 
In other words, to get the largest breaking of the sign degeneracy of $\fnl$, we require an approximate balance between the $\fnl$ terms and the $n_\uGW$ term, making the cross terms to be roughly equal to other terms, or at least the same order of magnitude. 
To further demonstrate the above issue, we will show some numerical results in the next subsection.

The (reduced) angular power spectrum has multipole dependence and frequency dependence. 
On the one hand, the multiple dependence, i.e., $C_\ell\propto[\ell(\ell+1)]^{-1}$, might be vital for discrimination of \acp{SIGW} from other \ac{GW} sources, e.g., astrophysical foregrounds due to \acp{GW} emitted from \acp{BBH} \cite{Cusin:2018rsq,Wang:2021djr,Bellomo:2021mer} and topological defects such as cosmic string loops \cite{Jenkins:2018nty,LISACosmologyWorkingGroup:2022kbp}. 
For example, in the \ac{LISA} band, the angular power spectrum for inspiralling \acp{BBH} has been shown to roughly scale as $(\ell+1/2)^{-1}$ \cite{Cusin:2018rsq,Wang:2021djr}. 
As a second example, the angular power spectrum for cosmic string loops has been shown to be spectrally white, i.e., $C_\ell\propto \ell^{0}$ \cite{Jenkins:2018nty,LISACosmologyWorkingGroup:2022kbp}. 
On the other hand, Eq.~(\ref{eq:rapssai}) depends on the \ac{GW} frequency band due to a factor $\Omega_{\mathrm{ng}}/\bar{\Omega}_{\uGW}$ in the $\fnl$ term.  
Via the component separation approach, the frequency dependence may be useful for discriminating \acp{SIGW} from other \acp{CGWB} produced by, e.g., the first-order phase transitions in the early universe \cite{Geller:2018mwu,Kumar:2021ffi,LISACosmologyWorkingGroup:2022kbp,Liu:2020mru,Schulze:2023ich}.  

\subsection{Numerical results}\label{sec:APS-Result}

In this subsection, we straightforwardly compute Eq.~\eqref{eq:reduced-APS} and Eq.~\eqref{eq:APS} by utilizing the results of $\bar{\Omega}_{\uGW}^{X}$ obtained in \cref{sec:mono}, where $X = G, H, C, Z, R, P$ and $N$.



\begin{figure}
    \includegraphics[width =1. \columnwidth]{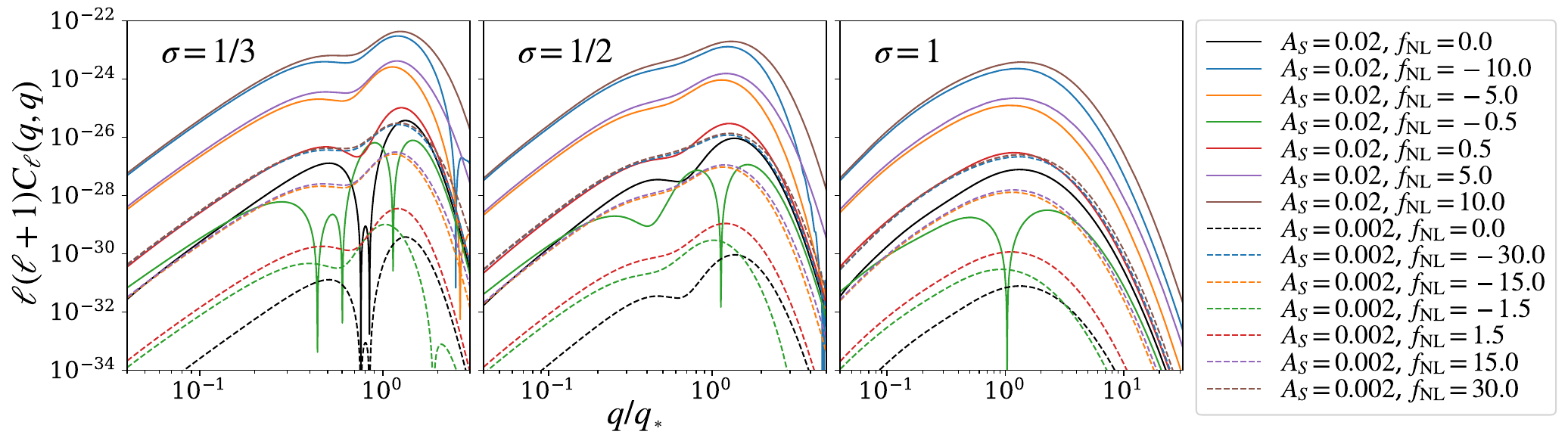}
    \caption{Angular power spectrum for the anisotropies in \acp{SIGW}. 
    The spectra with $A_S=0.02$ are denoted by solid curves, and the spectra with $A_S=0.002$ are denoted by dashed curves. 
    We take $\sigma=1/3,~1/2,~1$ from left to right panels.}\label{fig:Cl_all}
\end{figure}

In \cref{fig:Cl_all}, we show the (auto-correlated) angular power spectrum $\ell(\ell+1)C_{\ell}$ at the same frequency band, i.e., $q'=q$. 
First of all, the sign degeneracy of $\fnl$ is broken obviously in the figure. 
This result can be interpreted by the cross terms in Eq.~\eqref{eq:APS}, because they are linear in $\fnl$. 
In particular, the difference in two spectra with $\pm|\fnl|$ is relatively more significant, when the $\fnl$ term is comparable with the \ac{SW} term in Eq.~\eqref{eq:APS}. 
In addition, we find that $C_\ell$ further depends on $A_S$ and $\sigma$. 
Particularly, it roughly scales in $A_S^{4}$ since $C_\ell$ is approximately proportional to $\bar{\Omega}_{\uGW}^{2}$ in Eq.~\eqref{eq:APS}.

\begin{figure}
    \includegraphics[width =1. \columnwidth]{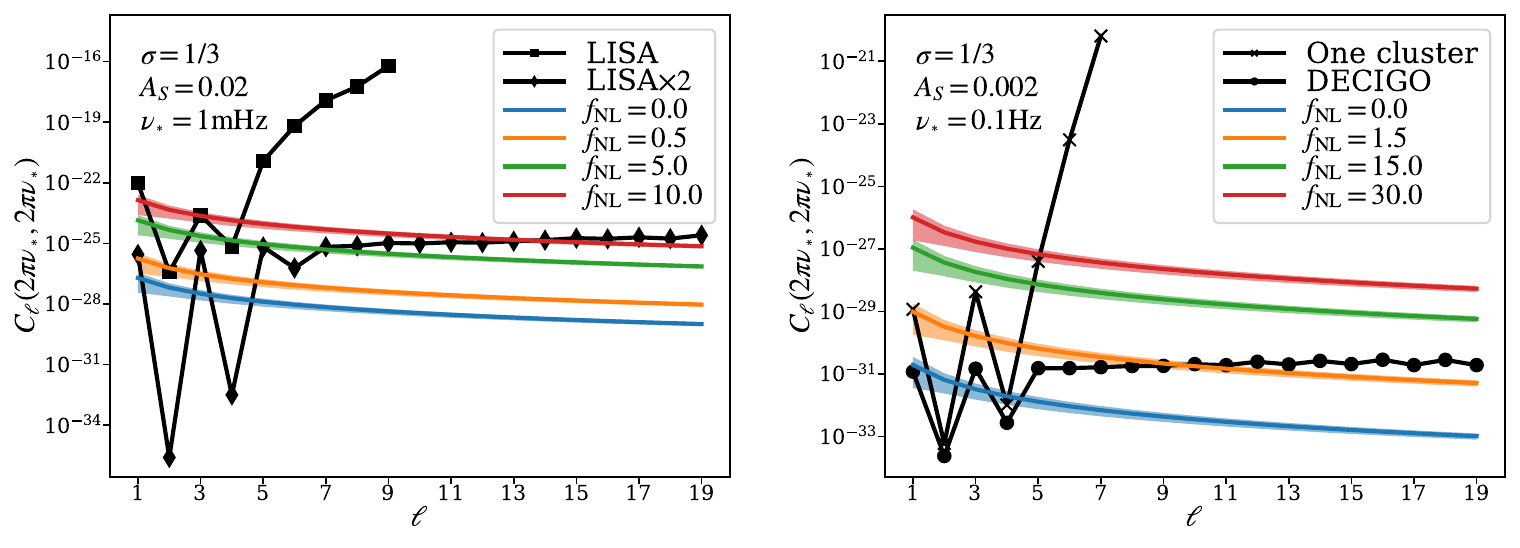}
    \caption{The anticipated angular power spectra versus the noise angular power spectra of \acp{LISA} (at 1 mHz band, left panel) \cite{Capurri:2022lze} and \acp{DECIGO} (at 0.1 Hz band, right panel) \cite{Capurri:2022lze}. The shaded regions stand for the cosmic-variance limits (68\% confidence level).   \label{fig:Cl_sensitivity}}
\end{figure}

As is shown in \cref{fig:Cl_sensitivity}, the angular power spectra with the interested parameter regimes are potentially detectable for \ac{LISA} \cite{Alonso:2020rar} and \ac{DECIGO} \cite{Ishikawa:2020hlo,Kawasaki:2022guk}, particularly on low multipoles. 
For comparison, we plot the shaded regions to stand for the uncertainties at 68\% confidence level due to cosmic variance, which is given as 
\begin{equation}
\frac{\Delta C_{\ell}}{C_{\ell}}=\sqrt{\frac{2}{2\ell+1}}\ .
\end{equation}
A detector network could measure the angular power spectrum for multipoles $\ell=1-19$ with a significantly higher sensitivity than an individual cluster \cite{Capurri:2022lze}. 
This result may also bring new insights to potential developments of the LISA-Taiji network \cite{Wang:2021njt,Cai:2023ywp}.

\begin{figure}
    \includegraphics[width =.9 \columnwidth]{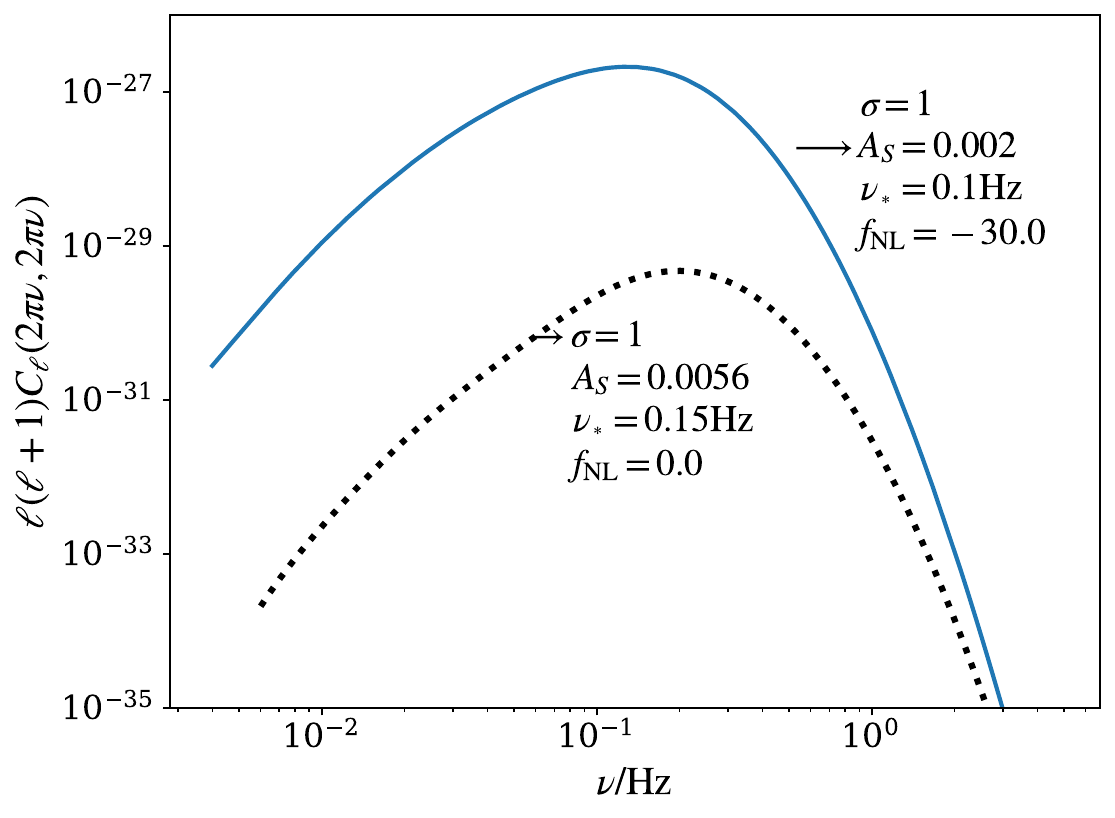}
    \caption{Illustration of the broken degeneracy of model parameters via the angular power spectrum. The degeneracy has been shown in \cref{fig:Omega_sensityvity} for the energy-density fraction spectrum. }\label{fig:compare}
\end{figure}

In \cref{fig:compare}, we show that the degeneracies of model parameters, as have been mentioned in \cref{fig:Omega_sensityvity}, could be explicitly broken by using the angular power spectrum. 
To be specific, corresponding to two curves in \cref{fig:compare}, the two curves with the same labeling in \cref{fig:Omega_sensityvity} almost coincides with each other, indicating degeneracies in these two sets of parameters. 
However, in \cref{fig:compare}, the degeneracies disappear due to an obvious separation of the two curves, with difference of at least two orders of magnitude.

\begin{figure}
    \includegraphics[width =1. \columnwidth]{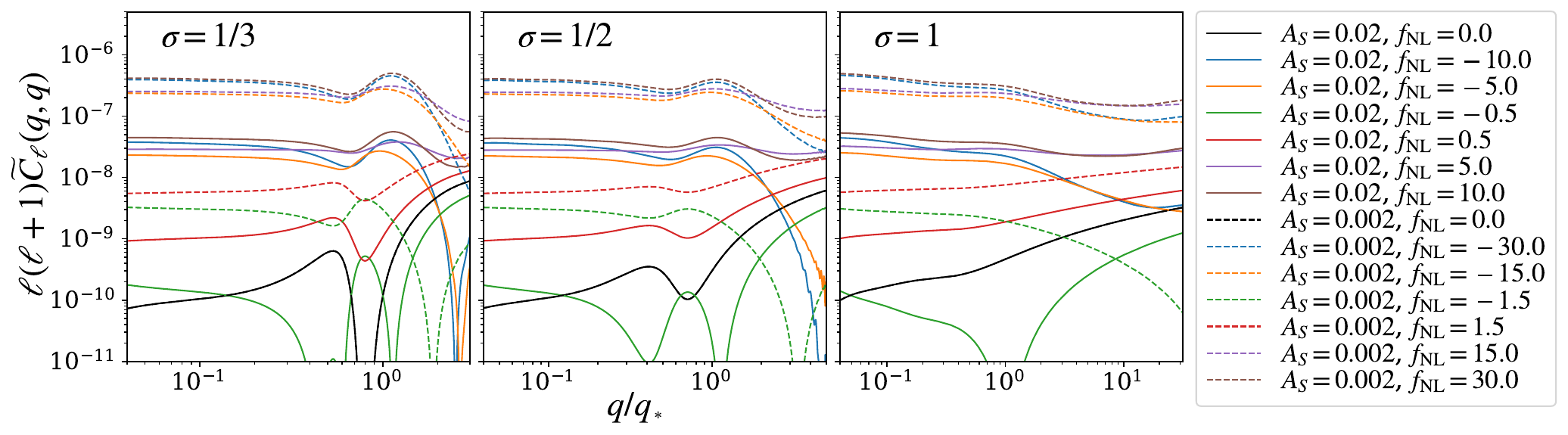}
    \caption{Reduced angular power spectrum as well as its dependence on $\fnl$ and $A_S$, and $\sigma$. 
    The labeling is the same as that of \cref{fig:Cl_all}. }\label{fig:tildeCl_all}
\end{figure}

In \cref{fig:tildeCl_all}, we depict the reduced angular power spectrum to display the difference in parameter dependence between the monopole and multipoles.  
Firstly, the magnitude of $\tilde{C}_{\ell}$ decreases with increase of $A_S$, implying that $C_\ell$ is less dependent on $A_S$ than $\bar{\Omega}_{\uGW}$. 
Secondly, $\tilde{C}_{\ell}$ is roughly red-tilted for a large value of $|\fnl|$, while blue-tilted for a small value. 
The critical value is roughly determined by a balance between the $\fnl$ term and the \ac{SW} term. 
Thirdly, the profiles of  $\tilde{C}_{\ell}$ also vary with values of $\sigma$, implying that $C_\ell$ and $\bar{\Omega}_{\uGW}$ have different dependence on $\sigma$. 
The above theoretical expectations are potentially useful for breaking the degeneracies of model parameters.

\begin{figure}
    \includegraphics[width =1. \columnwidth]{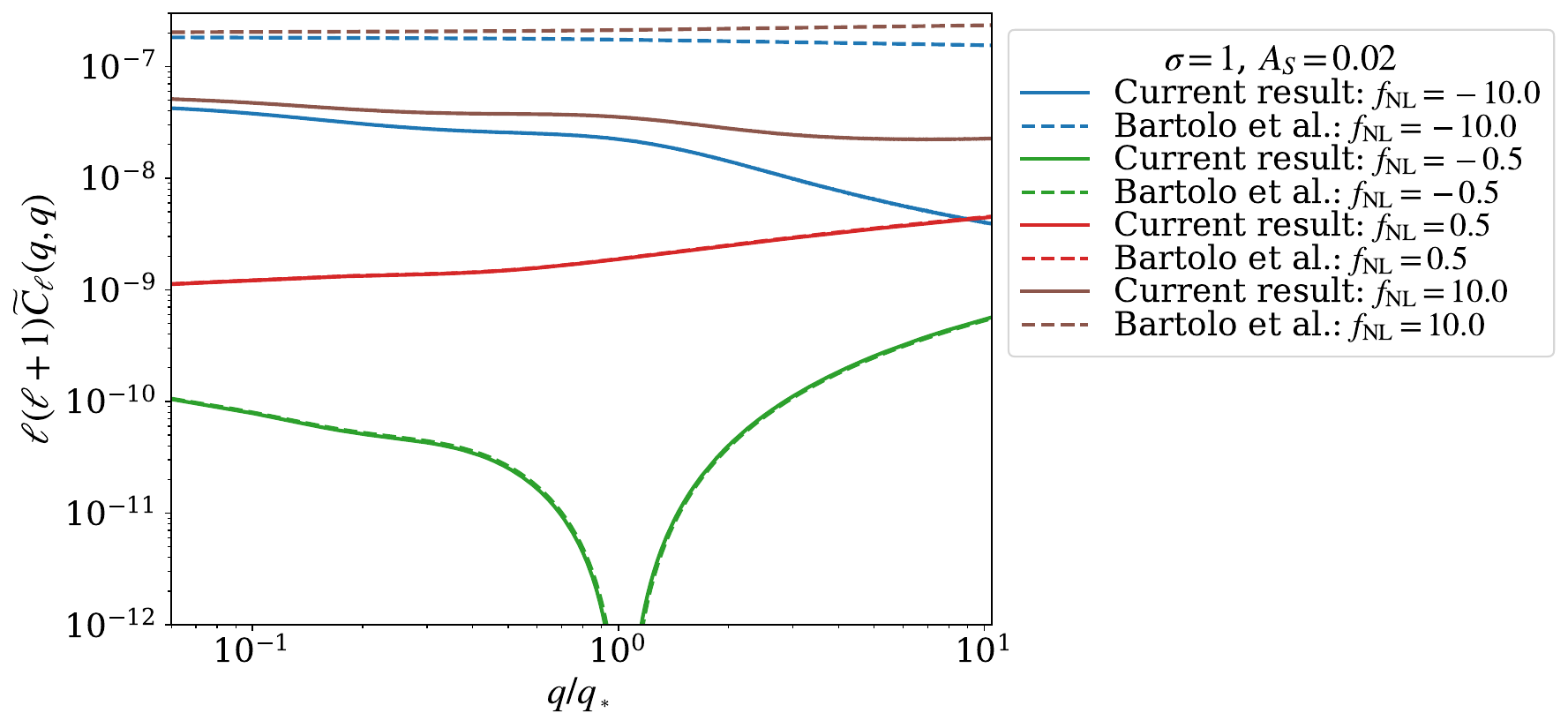}
    \caption{Comparison of the reduced angular power spectra of \acp{SIGW} provided by the current work (solid curves) and Ref.~\cite{Bartolo:2019zvb} (dashed curves). }
    \label{fig:sai}
\end{figure}

In \cref{fig:sai}, we show the results for the reduced angular power spectra anticipated by our current work and then compare them with those of Ref.~\cite{Bartolo:2019zvb}. 
Regarding the frequency dependence, we find that difference between the reduced angular power spectra of the two works is larger, when $|\fnl|$ takes a larger value, given a value of $A_S$. 
This result implies more significant impacts on the anisotropies in \acp{SIGW} with the increase of $|\fnl|$, or more precisely, the combination $A_S\fnl^2$. 
In particular, we find that the difference could be one order of magnitude for a large non-Gaussianity. 
This result can be interpreted as follows.  
On the level of background, i.e., $\bar{\Omega}_{\uGW}$, the authors of Ref.~\cite{Bartolo:2019zvb} considered only the left panel of Fig.~\ref{fig:EDS-G}, implying . 
In contrast, besides this diagram, we take into account the other six diagrams in Fig.~\ref{fig:EDS-other}. 
On the level of fluctuations, only one Feynman-like diagram, i.e., the top left panel of Fig.~\ref{fig:APS-all}, was taken into account in Ref.~\cite{Bartolo:2019oiq}. 
It was shown that the frequency dependence of $\tilde{C}_\ell$ arises from the $n_{\uGW}(q)$ term. 
In contrast, we take into account all of the ten Feynman-like diagrams in Fig.~\ref{fig:APS-all}. 
We show that the frequency dependence of $\tilde{C}_\ell$ arises not only from $n_{\uGW}(q)$, but also from the $\fnl$ term that is now multiplied with a frequency-dependent function of the form $\Omega_\mathrm{ng}/\bar{\Omega}_\mathrm{GW}$. 
In summary, the above two ingredients 
lead to the main difference between our current work and Ref.~\cite{Bartolo:2019oiq}.

\begin{figure}
    \includegraphics[width =1. \columnwidth]{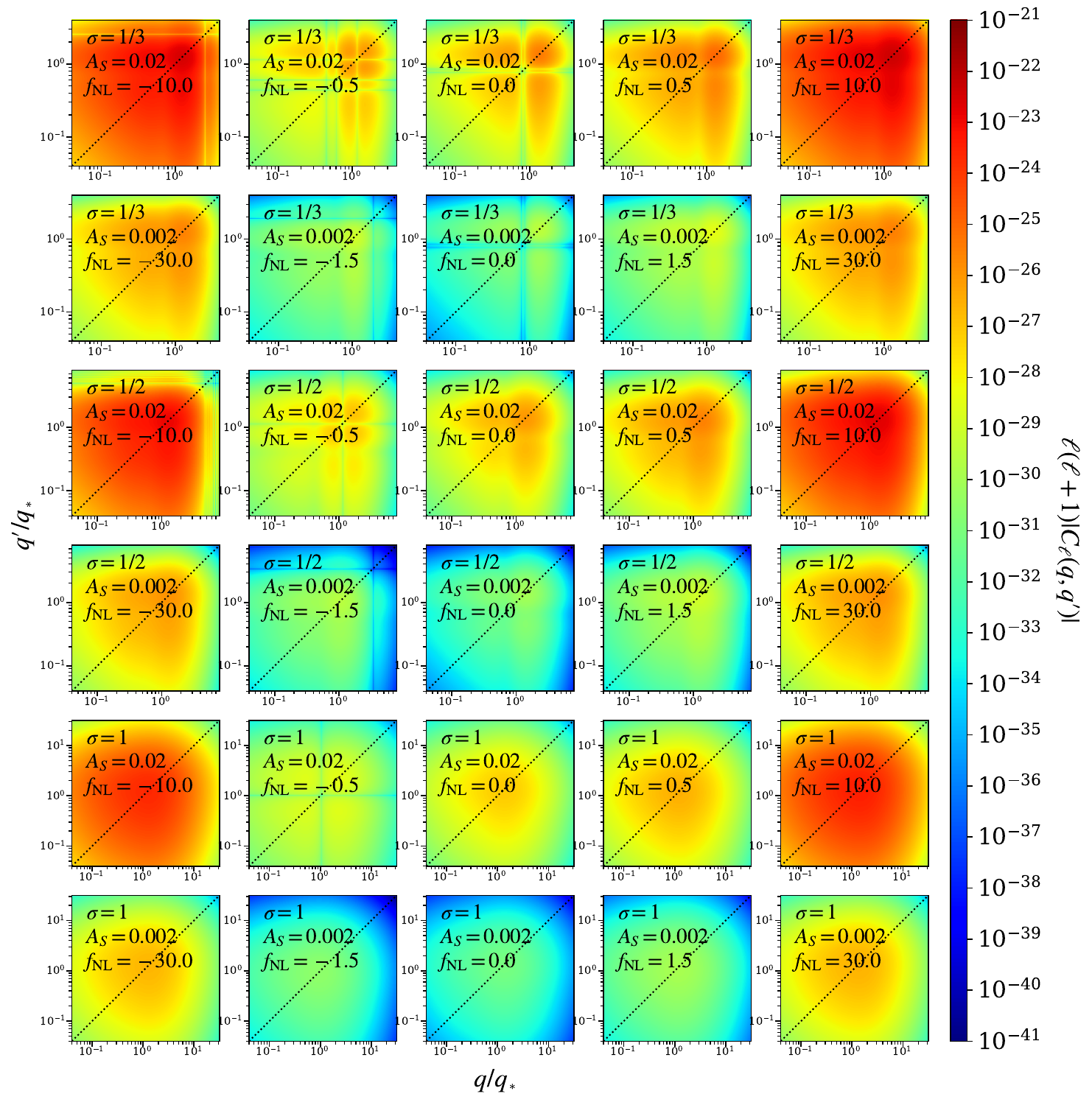}
    \caption{Cross-correlated angular power spectrum with respect to the gravitational-wave frequency band.  
    In each panel, the dotted line refers to the auto-correlated angular power spectrum. 
    }\label{fig:Cl_qq'}
\end{figure}

\begin{figure}
    \includegraphics[width = .9 \columnwidth]{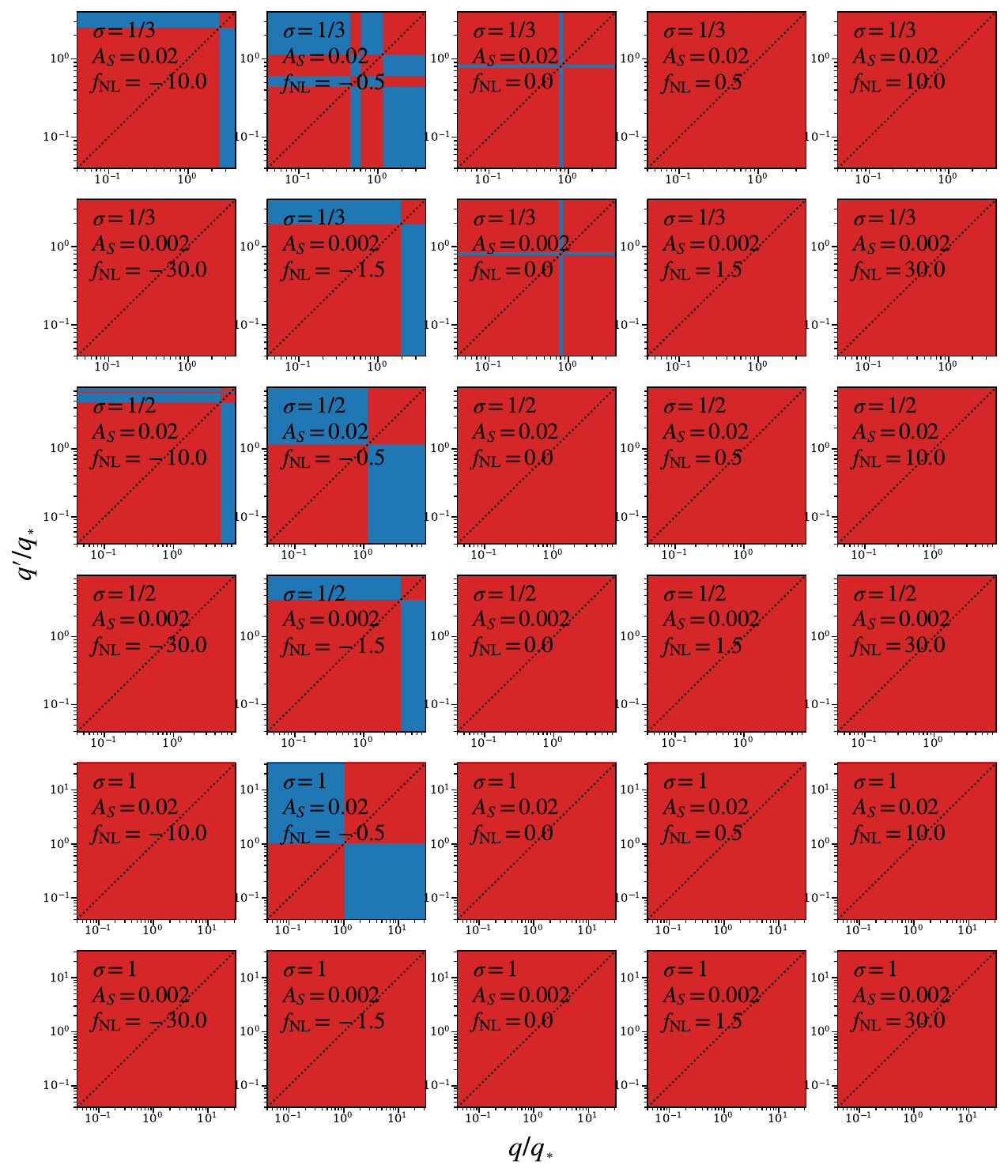}
    \caption{The same as Fig.~\ref{fig:Cl_qq'}, but the correlation factor is shown in red color for $r_{\ell}(q,q')=+1$ while in blue color for $r_{\ell}(q,q')=-1$.      }\label{fig:rellpm}
\end{figure}

In \cref{fig:Cl_qq'}, we depict the (cross-correlated) angular power spectra $\ell(\ell+1)|C_\ell|$ at different frequency bands, i.e., $q'\neq q$. 
Here, hotter colors stand for larger correlations while colder ones denote smaller correlations. 
For comparison, the auto-correlated spectra are also depicted in dotted black lines. 
The cross-correlation might be available to mitigate the stochastic noise that diminishes the anticipated signal. 
A correlation factor is defined as \cite{Schulze:2023ich} 
\begin{equation}\label{eq:r-def}
r_\ell (q,q')=\frac{C_{\ell}(q,q')}{\sqrt{C_\ell(q,q)C_\ell(q',q')}}\ .
\end{equation}
Considering Eq.~\eqref{eq:APS}, we obtain $r_\ell (q,q')=\pm1$. 
Note that we always have $r_\ell (q,q)=+1$. 
As is shown in Fig.~\ref{fig:rellpm}, the changes of the sign depend on the value of the pair $(q,q')$, indicating that \acp{SIGW} encode information in it. 
Therefore, only the sign is important, rather than $C_\ell(q,q')$ itself. 
In contrast, the noise may have different cross-correlation from the signal. 
If so, the cross-correlation would be useful for differentiating the signal from the noise. 
Note that this prediction is
to some extent speculative. 
However, we would like to point out such a possibility, which may be useful to future related studies. 
In \cref{fig:Cl_qq'}, we still depict the absolute value of $C_{\ell}(q,q')$, with dotted lines standing for the auto-correlated spectra. 
However, $r_\ell (q,q')$ could be straightforwardly computed in practice.

\section{Conclusion}\label{sec:Conclusion}

In this work, we proposed the anisotropies in \acp{SIGW} as a powerful probe to the local-type \acl{PNG} in the cosmological curvature perturbations. 
For the energy-density fraction spectrum of \acp{SIGW}, we reproduced the existing results in the literature and showed the degeneracies between the non-Gaussian parameter and other model parameters, that bring challenges to determination of the \acl{PNG}. 
For the first time, we provided the complete analysis to the (reduced) angular power spectrum of anisotropies in \acp{SIGW}, particularly, the contributions from the \acl{PNG}. 
In Eq.~\eqref{eq:reduced-APS}, we showed that such a spectrum is explicitly determined by $\fnl$, $A_S\fnl^2$, $\sigma$, and $q_\ast$, indicating that the degeneracies of model parameters can be broken. 
The spectrum was also shown to have multipole dependence, i.e., $C_\ell\sim[\ell(\ell+1)]^{-1}$, and be dependent on \ac{GW} frequency. 
In particular, the initial inhomogeneities were shown to be dependent on \ac{GW} frequency. 
These properties may be useful for the component separation and foreground removal. 
Despite challenges for breaking the sign degeneracy of $\fnl$ in the angular power spectrum for large $|\fnl|$, probing \acp{PBH} may provide a promising way to further break this degeneracy. The presence of primordial non-Gaussianity has substantial impacts on the abundance and mass distribution of \acp{PBH}, as their formation threshold is influenced by levels of this non-Gaussianity \cite{Byrnes:2012yx,Young:2013oia,Nakama:2016gzw,Bartolo:2019zvb,Meng:2022ixx,Atal:2019erb,Escriva:2022pnz}. Notably, a sizable negative $\fnl$ would be incompatible with detection of \acp{PBH} \cite{Byrnes:2012yx,Young:2013oia}, since the abundance of \acp{PBH} is expected to be suppressed significantly. Conversely, it is expected that a sizable positive $\fnl$ could significantly enhance the abundance of \acp{PBH}. 
Therefore, measuring the anisotropies in \acp{SIGW} and probing \acp{PBH} can serve as complementary approaches to break the sign degeneracy of $\fnl$. 
In addition, the theoretical formalism could be straightforwardly generalized to study \acp{SIGW} produced during other epochs \cite{Malhotra:2022ply} or other \acp{CGWB}. 
The theoretical predictions of this work may be tested by space-borne \ac{GW} detectors or networks in future.

\acknowledgments

We acknowledge Dr. Bin Gong and Dr. Tao Liu for useful suggestions on the \texttt{vegas} \cite{Lepage:2020tgj} package. We would also like to thank Dr. Siyu Li and Dr. Yi Wang for helpful discussions on the anisotropies in cosmic microwave background and inflationary non-Gaussianity, respectively. 
S.W. and J.P.L. are supported by the National Natural Science Foundation of China (Grant No. 12175243). Z.C.Z. is supported by the National Natural Science Foundation of China (Grant NO. 12005016). K.K. is supported by KAKENHI Grants No. JP17H01131, No. JP19H05114, No. JP20H04750 and No. JP22H05270.








\bibliography{biblio}
\bibliographystyle{JHEP}
\end{document}